\newcommand\beq{\begin{eqnarray}}
\newcommand\eeq{\end{eqnarray}}

\newcommand\missET{E_T^{\rm miss}}

\newcommand\epsu{\epsilon_{\scriptscriptstyle U}}
\newcommand\epsup{\epsilon_{\scriptscriptstyle U}'}
\newcommand\epsd{\epsilon_{\scriptscriptstyle D}}
\newcommand\epsdp{\epsilon_{\scriptscriptstyle D}'}
\newcommand\epsn{\epsilon_{\scriptscriptstyle N}}
\newcommand\epse{\epsilon_{\scriptscriptstyle E}}
\newcommand\AppendixA{Appendix A}
\newcommand\AppendixB{Appendix B}

\def\lsim{\mathrel{\rlap{\lower4pt\hbox{$\sim$}}
    \raise1pt\hbox{$<$}}}                
\def\gsim{\mathrel{\rlap{\lower4pt\hbox{$\sim$}}
    \raise1pt\hbox{$>$}}}

\documentclass[
amsmath,
prd,nofootinbib,floatfix,11pt,
]{revtex4}

\allowdisplaybreaks
\usepackage{graphicx}
\usepackage{setspace}

\begin{document}
\renewcommand{\theequation}{\arabic{section}.\arabic{equation}}

\title{\Large%
\baselineskip=21pt
Extra vector-like matter and the lightest Higgs scalar boson mass 
in low-energy supersymmetry
}

\author{Stephen P. Martin}
\affiliation{
{\it Department of Physics, Northern Illinois University, DeKalb IL 60115,} 
and 
\\
{\it Fermi National Accelerator Laboratory, P.O. Box 500, Batavia IL 60510.}
}

\begin{abstract}\normalsize \baselineskip=15pt 
The lightest Higgs scalar boson mass in supersymmetry can be raised 
significantly by extra vector-like quark and lepton supermultiplets with 
large Yukawa couplings but dominantly electroweak-singlet masses. I 
consider models of this type that maintain perturbative gauge coupling 
unification. The impact of the new particles on precision electroweak 
observables is found to be moderate, with the fit to Z-pole data as good 
or better than that of the Standard Model even if the new Yukawa 
couplings are as large as their fixed-point values and the extra 
vector-like quark masses are as light as 400 GeV. I study the size of 
corrections to the lightest Higgs boson mass, taking into account the 
fixed-point behavior of the scalar trilinear couplings. I also discuss 
the decay branchings ratios of the lightest new quarks and leptons and 
general features of the resulting collider signatures.
\end{abstract}


\maketitle

\baselineskip=14.9pt
\tableofcontents

\vfill\eject
\baselineskip=14.5pt

\setcounter{footnote}{1}
\setcounter{page}{2}
\setcounter{figure}{0}
\setcounter{table}{0}

\section{Introduction}
\label{sec:intro}
\setcounter{equation}{0}
\setcounter{footnote}{1}

The Minimal Supersymmetric Standard Model \cite{primer} (MSSM)
predicts that the lightest 
neutral Higgs boson, $h^0$, has a mass that can only exceed that of the 
$Z^0$ boson by virtue of radiative corrections. If the superpartners are 
not too heavy, then it becomes a challenge to evade the constraints on 
$h^0$ set by CERN LEPII $e^+e^-$ collider searches. On the other hand, 
larger superpartner masses tend to require some tuning in order to 
accommodate the electroweak symmetry breaking scale. In recent years 
this has motivated an exploration of models that extend the MSSM and can 
raise the prediction for $m_{h^0}$.

In the MSSM, the largest radiative corrections to $m_{h^0}$ come from 
loop diagrams involving top quarks and squarks, and are proportional to 
the fourth power of the top Yukawa coupling. This suggests that one can 
further raise the Higgs mass by introducing new heavy supermultiplets 
with associated large Yukawa couplings. In recent years there has been
renewed interest \cite{Frampton:1999xi,Maltoni:1999ta,He:2001tp,Novikov:2002tk,Holdom:2006mr,Kribs:2007nz,Hung:2007ak,Ozcan:2008qk,Fok:2008yg,Murdock:2008rx,Ibrahim:2008gg,Burdman:2008qh,Dobrescu:2009vz,Bobrowski:2009ng,Chanowitz:2009mz,Novikov:2009kc,Holdom:2009rf,Liu:2009cc,Litsey:2009rp}
in the possibility of a fourth family of quarks
and leptons, which can be reconciled with precision electroweak
constraints with or without supersymmetry.
However, within the context of supersymmetry, if the new heavy supermultiplets 
are chiral (e.g. a sequential fourth family), then 
in order to evade discovery at the Fermilab Tevatron $p\bar p$ collider the Yukawa couplings 
would have to be so large that perturbation theory 
would break down not far above the electroweak scale. This would negate the 
success of apparent gauge coupling unification in the MSSM. Furthermore, 
the corrections to precision electroweak physics would rule out such 
models without some fine tuning. 

These problems can be avoided if 
the extra supermultiplets are instead vector-like, as proposed in 
\cite{Moroi:1991mg,Moroi:1992zk,Babu:2004xg,Babu:2008ge}. If the 
scalar members of the new supermultiplets are heavier than the fermions, 
then there is a positive correction to $m_{h^0}$. As I will show below, 
the corrections to precision electroweak parameters decouple fast enough 
to render them benign.

To illustrate the general structure of such models, suppose that the 
new left-handed chiral supermultiplets include an $SU(2)_L$ doublet 
$\Phi$ with weak hypercharge $Y$ and an $SU(2)_L$ singlet $\overline 
\phi$ with weak hypercharge $-Y -1/2$, 
and $\overline{\Phi}$ and $\phi$ with the opposite gauge quantum numbers. 
The fields $\Phi$ and $\phi$ transform as the same representation of $SU(3)_C$
(either a singlet, a fundamental, or an 
anti-fundamental), and $\overline{\Phi}$ and ${\overline 
\phi}$ transform appropriately as the opposite. The superpotential allows 
the terms:
\beq
W = M_\Phi \Phi \overline{\Phi} + M_\phi \phi \overline{\phi} 
    + k H_u \Phi \overline{\phi} - h H_d \overline{\Phi} \phi,
\label{eq:genW}
\eeq
where $M_\Phi$ and $M_\phi$ are vector-like (gauge-singlet) masses, and 
$k$ and $h$ are Yukawa couplings to the weak hypercharge $+1/2$ and 
$-1/2$ MSSM Higgs fields $H_u$ and $H_d$, respectively. 
In the following, 
I will consistently use the letter $k$ for Yukawa couplings of new fields 
to $H_u$, and $h$ for couplings to $H_d$. 
Products of weak isospin doublet fields implicitly have
their $SU(2)_L$ indices contracted with an antisymmetric tensor 
$\epsilon^{12} = -\epsilon^{21} = 1$, with the first component of every 
doublet
having weak isospin $T_3 = 1/2$ and the second $T_3 = -1/2$. 
So, for example, $\Phi \overline \Phi = \Phi_1 \overline\Phi_2 -
\Phi_2 \overline\Phi_1$, with the components 
$\Phi_1$, $\Phi_2$, $\overline \Phi_1$, and $\overline \Phi_2$ 
having electric charges $Y+1/2$, $Y - 1/2$,
$-Y+1/2$, and $-Y-1/2$ respectively.

The scalar members 
of the new chiral supermultiplets participate in soft supersymmetry 
breaking Lagrangian terms:
\beq
-{\cal L}_{\rm soft} = 
\Bigl (
b_\Phi \Phi \overline{\Phi} + b_\phi \phi \overline{\phi} +
a_k H_u \Phi \overline{\phi} - a_h H_d \overline{\Phi} \phi \Bigr ) + {\rm c.c.}
+ m_{\Phi}^2 |\Phi|^2 + m_{\phi}^2 |\phi|^2
,
\label{eq:gensoft}
\eeq
where I use the same name for each chiral superfield and its scalar component. 

The fermion content of this model consists of two Dirac 
fermion-anti-fermion pairs with electric charges $\pm(Y+1/2)$
and one Dirac fermion-anti-fermion pair with electric charges $\pm(Y-1/2)$.
The doubly degenerate squared-mass eigenvalues of the fermions with 
charge $\pm(Y+1/2)$ are obtained at tree-level by 
diagonalizing the matrix
\beq
m_F^2 = \begin{pmatrix}
        {\cal M}_F {\cal M}_F^\dagger & 0 \cr
        0 & {\cal M}_F^\dagger {\cal M}_F
        \end{pmatrix}
\label{eq:m2Fmatrix}
\eeq 
with 
\beq
{\cal M}_F = \begin{pmatrix}
             M_\Phi & k v_u \cr
             h v_d & M_\phi
             \end{pmatrix},
\eeq
which is assumed to be
dominated by the $M_\Phi$ and $M_\phi$ entries on the diagonal. 
Here $v_u = v 
\sin\beta$ and $v_d = v \cos\beta$
are the vacuum expectation values (VEVs) of the MSSM Higgs fields 
$H_u$ and $H_d$, in a normalization where $v \approx 175$ GeV. 
The scalar partners of these have a 
squared-mass matrix given by, in the basis
$(\Phi, \phi, \overline \Phi^*, \overline \phi^*)$: 
\beq
m_S^2 = m_F^2 + \begin{pmatrix} 
m_\Phi^2 + \Delta_{\frac{1}{2},Y+ \frac{1}{2}} & 0 & 
b^*_\Phi & a^*_k v_u -k \mu v_d \cr
0 & m_\phi^2 + \Delta_{0,Y+\frac{1}{2}} & a^*_h v_d - h \mu v_u & 
b^*_\phi 
\cr
b_\Phi & a_h v_d - h \mu^* v_u & 
m_{\overline \Phi}^2 + \Delta_{-\frac{1}{2},-Y- \frac{1}{2}} & 0 
\cr
a_k v_u - k \mu^* v_d & b_\phi & 0 & m_{\overline \phi}^2 + 
\Delta_{0,-Y- \frac{1}{2}} \end{pmatrix}
\label{eq:m2Smatrix}
\eeq
where the
$\Delta_{T_3, q} = [T_3 - q \sin^2\theta_W] \cos(2\beta) m_Z^2$ 
are electroweak $D$-terms, with $T_3$ and $q$ the weak isospin and
electric charge.
The scalar particle squared-mass eigenvalues of 
eq.~(\ref{eq:m2Smatrix}) are presumably 
larger than those of their fermionic partners because
of the effects of
$m^2_\Phi$, $m^2_\phi$, $m^2_{\overline \Phi}$ and $m^2_{\overline \phi}$,
inducing a significant positive one-loop correction
to $m_{h^0}^2$. 
If $\tan\beta$ is not too small, the corrections to $m_{h^0}^2$ are 
largest if the $k$-type Yukawa coupling
is as large as possible, i.e. near its infrared quasi-fixed point.

The fermions of charge $\pm(Y-1/2)$ have squared mass 
$M_\Phi^2$, and their scalar partners have a squared-mass matrix 
\beq
\begin{pmatrix} 
|M_\Phi|^2 + m_\Phi^2 + \Delta_{-\frac{1}{2}, Y - \frac{1}{2}}
& -b_\Phi^* \cr
-b_\Phi &
|M_\Phi|^2 + m_{\overline \Phi}^2 + \Delta_{\frac{1}{2}, -Y + \frac{1}{2}}
\end{pmatrix}
.
\label{eq:othermassmat}
\eeq
These particles do not contribute to $m^2_{h^0}$ except through the small
electroweak $D$-terms, since they do not have Yukawa couplings to the
neutral Higgs boson.  Since that contribution is therefore parametrically
suppressed, it will be neglected in the following. 

With the phases of $H_u$ and $H_d$ chosen so that their vacuum 
expectation values (VEVs) are real, then in complete generality only 
three of the new parameters $M_\Phi$, $M_\phi$, $k$ and $h$ can be 
simultaneously chosen real and positive by convention. Nevertheless, I 
will take all four to be real and positive below. (I will usually be 
assuming that the magnitude of at least one of the new Yukawa couplings 
is small, so that the potential CP-violating effects are negligible 
anyway.)

In the MSSM, the running gauge couplings extrapolated to very high mass
scales appear to approximately 
unify near $Q = M_{\rm unif} = 2.4 \times 10^{16}$ GeV. In
order to maintain this success, it is necessary to include additional
chiral supermultiplets, besides the ones just mentioned. These other
fields again do not have Yukawa couplings to the Higgs boson, so
their contribution to $\Delta m^2_{h^0}$ will be neglected below. 

I will be assuming that the superpotential vector-like mass terms
are not much larger than the TeV scale. This can be accomplished by 
whatever mechanism also generates the $\mu$ term in the MSSM. For 
example, it may be that the terms $M_\Phi$ and $M_\phi$ are 
forbidden at tree-level in the renormalizable Lagrangian, and arise from
non-renormalizable terms in the superpotential of the form:
\beq
W = \frac{\lambda}{M_{\rm Pl}} S \overline S \Phi \overline \Phi
+ \frac{\lambda'}{M_{\rm Pl}} S \overline S \phi \overline \phi,
\eeq
after the scalar components of singlet supermultiplets 
$S$ and $\overline S$ obtain vacuum
expectation values of order the geometric mean of the Planck and
soft supersymmetry-breaking scales. 
Then $M_\Phi, M_\phi \lsim 1$ TeV can be natural, just 
as for $\mu$ in the MSSM.

In the remainder of 
this paper, I will discuss aspects of the phenomenology of models of 
this type, concentrating on the particle content and renormalization 
group running (section 2), corrections to $m_{h^0}$ (section 3), 
precision 
electroweak corrections (section 4), and branching ratios and signatures 
for the lightest of the new fermions in each model (section 5).

\section{Supersymmetric models with new vector-like fields }
\label{sec:models}
\setcounter{equation}{0}
\setcounter{footnote}{1}

\subsection{Field and particle content}

To construct and describe models, consider the following possible fields 
defined by their
transformation properties under $SU(3)_C \times SU(2)_L \times U(1)_Y$:
\beq
&&
Q = ({\bf 3}, {\bf 2}, 1/6), 
\qquad
\overline{Q} = ({\bf \overline{3}}, {\bf 2}, -1/6), 
\qquad
U = ({\bf 3}, {\bf 1}, 2/3), 
\qquad
\overline{U} = ({\bf \overline{3}}, {\bf 1}, -2/3), 
\nonumber
\\
&&D = ({\bf 3}, {\bf 1}, -1/3), 
\qquad
\overline{D} = ({\bf \overline{3}}, {\bf 1}, 1/3), 
\qquad
L = ({\bf 1}, {\bf 2}, -1/2), 
\qquad
\overline{L} = ({\bf 1}, {\bf 2}, 1/2), 
\nonumber
\\
&&
E = ({\bf 1}, {\bf 1}, -1), 
\qquad
\overline{E} = ({\bf 1}, {\bf 1}, 1), 
\qquad
N = ({\bf 1}, {\bf 1}, 0), 
\qquad
\overline{N} = ({\bf 1}, {\bf 1}, 0).
\label{eq:extraslist}
\eeq
Restricting the new supermultiplets to this list assures that 
small mixings with the MSSM fields can eliminate stable exotic particles
which could be disastrous relics from the early universe.
In this paper, I will reserve the above capital letters for new extra 
chiral supermultiplets,
and use lowercase letters for the MSSM quark and lepton supermultiplets:
\beq
&&
q_i = ({\bf 3}, {\bf 2}, 1/6), 
\qquad
\overline{u}_i = ({\bf \overline{3}}, {\bf 1}, -2/3), 
\qquad
\overline{d}_i = ({\bf \overline{3}}, {\bf 1}, 1/3), 
\qquad
\nonumber \\ &&
\ell_i = ({\bf 1}, {\bf 2}, -1/2), 
\qquad
\overline{e}_i = ({\bf 1}, {\bf 1}, 1), 
\nonumber \\ &&
H_u = ({\bf 1}, {\bf 2}, 1/2), 
\qquad
H_d = ({\bf 1}, {\bf 2}, -1/2).
\eeq
with $i = 1,2,3$ denoting the three families.
So the MSSM superpotential, in the approximation that only third-family 
Yukawa couplings are included, is:
\beq
W = \mu H_u H_d + y_t H_u q_3 \overline u_3 - y_b H_d q_3 \overline d_3 
- y_\tau H_d \ell_3 \overline e_3.
\label{eq:WMSSM}
\eeq
It is well-known that gauge coupling unification is maintained
if the new fields taken together transform as complete $SU(5)$ 
multiplets. However, this is not a necessary condition.
There are three types of models that can successfully maintain 
perturbative gauge 
coupling unification with the masses of new extra chiral supermultiplets
at the TeV scale. 

First, there is a model to be called
the ``LND model" in this paper, consisting of chiral supermultiplets 
$L, \overline L, N, \overline N, D, \overline D$, with a superpotential
\beq
W = M_L L \overline L + M_N N \overline N + M_D D \overline D +
k_N H_u L \overline N - h_N H_d \overline L N .
\label{eq:WLND}
\eeq
Here $L, \overline{L}$ play the role of $\Phi, \overline{\Phi}$ and 
$N,\overline{N}$ 
the role of $\phi, \overline{\phi}$ in 
eqs.~(\ref{eq:genW})-(\ref{eq:othermassmat}).
In most of the following, I will consider only the case that the multiplicity of
each of these fields is 1, although 1, 2, or 3 copies of each would be 
consistent with perturbative gauge coupling unification. These fields
consist of a ${\bf 5} + {\bf \overline 5}$ of $SU(5)$, plus a 
pair\footnote{Here I choose the minimal model of this type that includes Yukawa
couplings of the kind mentioned in the Introduction while 
not violating lepton number.
It is also possible to identify the fields $N$ and 
$\overline N$, since they are gauge singlets, 
or to eliminate them
(and their Yukawa couplings) entirely.} of 
singlet fields. 
The non-MSSM mass eigenstate fermions consist of
a charged lepton $\tau'$, a pair of neutral fermions $\nu'_{1,2}$,
and a charge $-1/3$ quark $b'$. Their superpartners are
complex scalars $\tilde \tau'_{1,2}$, $\tilde \nu'_{1,2,3,4}$, and
$\tilde b'_{1,2}$. The primes are used to distinguish these states from those
of the usual MSSM that have the same charges.

Second, one has a model consisting of a ${\bf 10} + {\bf \overline{10}}$
of $SU(5)$, to be called the ``QUE model" below, consisting of fields
$Q, \overline Q, U, \overline U, E, \overline E$ with a superpotential
\beq
W = M_Q Q \overline Q + M_U U \overline U + M_E E \overline E +
k_U H_u Q \overline U - h_U H_d \overline Q U .
\label{eq:WQUE}
\eeq
The non-MSSM particles in this case consist of charge $+2/3$ quarks
$t'_{1,2}$, a charge $-1/3$ quark $b'$, and a charged lepton $\tau'$,
and their scalar partners
$\tilde t'_{1,2,3,4}$, $\tilde b'_{1,2}$ and $\tilde \tau'_{1,2}$.

Third, one has a ``QDEE model" 
consisting of fields 
$Q, \overline Q, D, \overline D, E_i, \overline E_i$ ($i=1,2$) with a 
superpotential
\beq
W = M_Q Q \overline Q + M_U D \overline D + M_{E_i} E_i \overline E_i +
k_D H_u \overline Q D - h_D H_d Q \overline D .
\label{eq:WQDEE}
\eeq
Although this particle content does not happen to contain complete 
multiplets of $SU(5)$, it still gives perturbative gauge coupling 
unification. 
The non-MSSM particles in this model consist of charge $-1/3$ quarks
$b'_{1,2}$, a charge $+2/3$ quark $t'$, and two charged leptons $\tau'_{1,2}$,
and their scalar partners
$\tilde b'_{1,2,3,4}$, $\tilde t'_{1,2}$ and $\tilde \tau'_{1,2,3,4}$.

The field and particle content of these three models is summarized in 
Table \ref{table:models}.
\renewcommand{\arraystretch}{1.25}
\begin{table}
\begin{tabular}[c]{|c||c|c|c||c|c|}
\hline
Model & \multicolumn{3}{|c||}{New supermultiplets}
& \multicolumn{2}{|c|}{New particles}
\\
& ~~$\Phi, \overline \Phi$~~ & ~~$\phi, \overline \phi$~~ &
~~~~others~~~~ &  ~~~~spin $1/2$~~~~ & ~~~~~~~~~~spin $0$~~~~~~~~~~
\\ 
\hline\hline
LND & $L,\overline L$ & $N, \overline N$ & $D, \overline D$ &
$\nu_{1,2}'$~ $\tau'$~ $b'$ & 
$\tilde \nu_{1,2,3,4}'$~ $\tilde \tau_{1,2}'$~ $ \tilde b_{1,2}'$
\\
\hline
QUE & $Q,\overline Q$ & $U, \overline U$ & $E, \overline E$ &
$t_{1,2}'$~ $b'$~ $\tau'$ & 
$\tilde t_{1,2,3,4}'$~ $\tilde b_{1,2}'$~ $ \tilde \tau_{1,2}'$
\\
\hline
~~~QDEE~~~& $\overline Q, Q$ & $\overline D, D$ & 
$E_{1,2}, \overline E_{1,2}$ &
$b_{1,2}'$~ $t'$~ $\tau_{1,2}'$ & 
$\tilde b_{1,2,3,4}'$~ $\tilde t_{1,2}'$~ $ \tilde \tau_{1,2,3,4}'$
\\
\hline
\end{tabular}
\caption{The new chiral supermultiplets and the new particle
content of the models discussed in this paper. The notation
for $\Phi, \overline \Phi, \phi, \overline \phi$ follows that of the
Introduction.
\label{table:models}}
\end{table}

In reference \cite{Babu:2008ge}, it is suggested that
a model with extra chiral supermultiplets in  
${\bf 5} + {\bf \overline{5}} + {\bf 10} + {\bf \overline{10}}$
of $SU(5)$, or equivalently (if a pair of singlets is added) 
${\bf 16} + {\bf \overline{16}}$ of 
$SO(10)$, will also result in gauge coupling unification. 
However,  
the multi-loop running of gauge couplings actually renders them 
non-perturbative below the putative unification scale,
unless the new particles have masses well above the 1 TeV scale. 
For example, working to three-loop order, if one requires that the unified coupling (defined
to be the common value of $\alpha_1$ and $\alpha_2$ at their
meeting point) satisfies the perturbativity condition $\alpha_{\rm unif} < 0.35$, then the average threshold of the new particles must exceed 5 TeV 
if the MSSM particles are treated as having a common threshold at or below 1 TeV
as suggested by naturalness and the little hierarchy problem. 
In that case, the new particles will certainly decouple from LHC phenomenology.
Even if one allows the MSSM soft mass scale to be as heavy as the new particles,
treating all non-Standard Model particles as having a common threshold, I find that
this threshold must be at least 2.8 TeV if the new Yukawa couplings vanish
and at least 2.1 TeV if the new Yukawa couplings are as large as their
fixed-point values. While such heavy mass spectra are possible, they go
directly against the motivation provided by the little hierarchy problem.
Furthermore, at the scale of apparent unification of $\alpha_1$ and $\alpha_2$
in such models, the value of $\alpha_3$ is 
considerably smaller,
rendering the apparent unification of gauge couplings at best completely accidental, dependent on the whim of out-of-control high-scale
threshold corrections. 
I will therefore not consider that model further here, although it could be
viable if one accepts the loss of perturbative unification and control at
high scales. 
The collider phenomenology should be qualitatively similar to that of the
LND and QUE models, since the particle content is just the union of them.
 
\subsection{Renormalization group running}

\begin{figure}[!tp]
\begin{minipage}[]{0.47\linewidth}
\caption{\label{fig:unification}
Gauge coupling unification in the MSSM, LND and QUE models.
The running is performed with 3-loop beta functions, with all
particles beyond the Standard Model taken to decouple at $Q = 600$ GeV,
and $m_t = 173.1$ GeV with $\tan\beta = 10$.}
\end{minipage}
\begin{minipage}[]{0.52\linewidth}
\begin{flushright}
\includegraphics[width=8.0cm,angle=0]{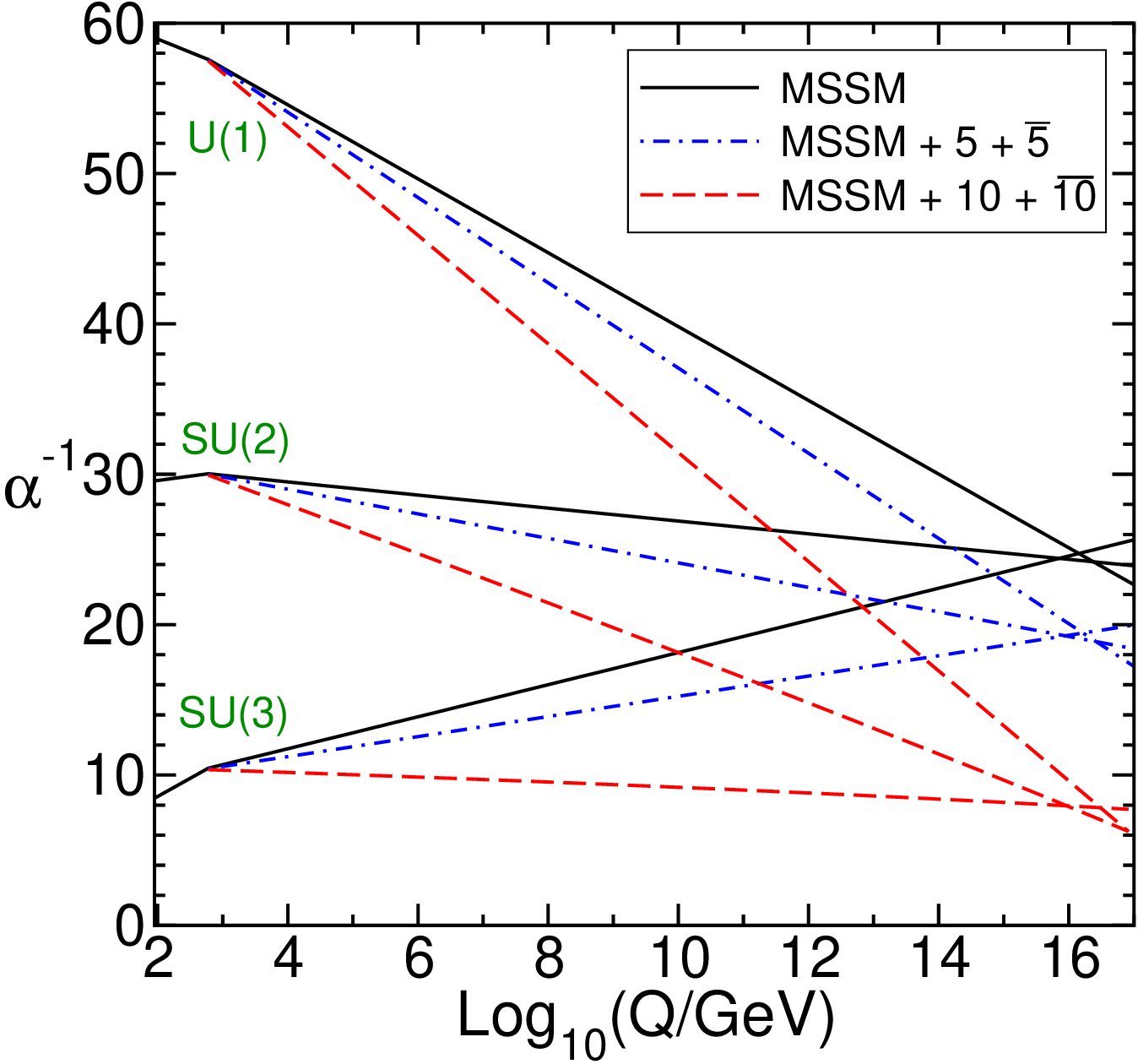}
\end{flushright}
\end{minipage}
\end{figure}

The unification of running gauge couplings in the MSSM, LND, and QUE 
models  is shown in Figure \ref{fig:unification}.
In this graph, 3-loop beta functions
are used for 
the MSSM gauge
couplings, and $m_t = 173.1$ GeV and $\tan\beta = 10$, and all
non-Standard-Model particles are taken to decouple at $Q = 600$ GeV. (The 
Yukawa
couplings $k_N$ and $h_N$ in the LND model and $k_U$ and $h_U$ in the QUE 
model are set to 0 here for simplicity; they do not have a dramatic effect on 
the results as long as they are at or below their fixed-point 
trajectories.) 
The running for the QDEE model is not shown, because it is
very similar to that for 
the QUE model. 
Indeed, it will turn out that many features of the
QUE and QDEE models are similar, insofar as the $U + \overline{U}$
fields can be interchanged with the 
$D + \overline{D} + E + \overline{E}$ 
fields. This 
similarity does not extend, however, to the collider phenomenology
as discussed in section 5.
Note that the unification scale, defined as the 
renormalization scale $Q$ at which $\alpha_1 = \alpha_2$, is somewhat 
higher with the extra chiral supermultiplets in place; in the MSSM,
$M_{\rm unif} \approx 2.4 \times 10^{16}$ GeV, but
$M_{\rm unif} \approx 2.65 \times 10^{16}$ GeV in the LND model, and
$M_{\rm unif} \approx 8.3 \times 10^{16}$ GeV in the QUE and QDEE models.
The strong coupling $\alpha_3$ misses the unified $\alpha_1$ and $\alpha_2$,
but by a small amount that can be reasonably ascribed to threshold corrections
of whatever new physics occurs at $M_{\rm unif}$.

\begin{figure}
\begin{minipage}[]{0.49\linewidth}
\begin{flushleft}
\includegraphics[width=7.8cm,angle=0]{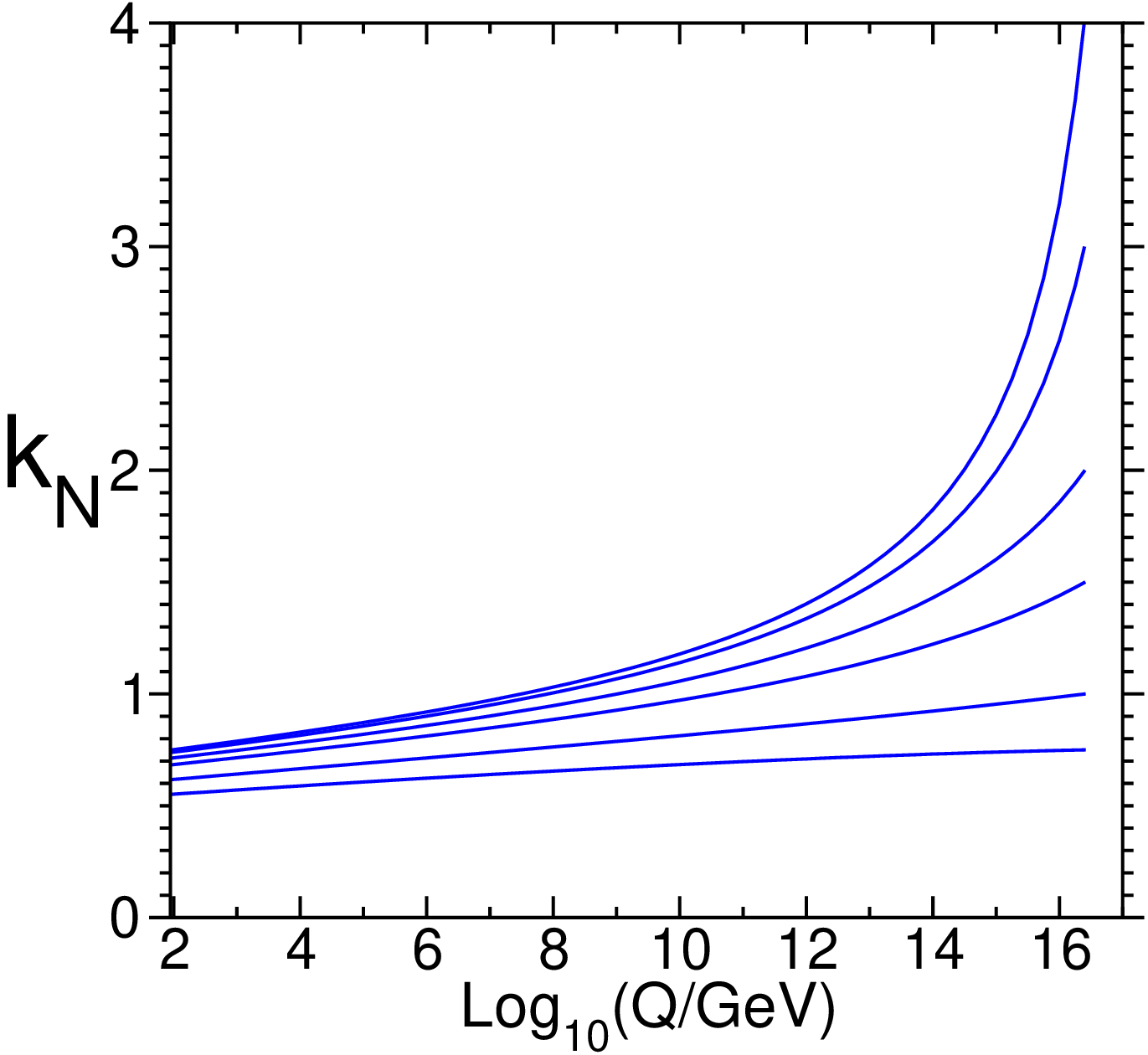}
\end{flushleft}
\end{minipage}
\begin{minipage}[]{0.49\linewidth}
\begin{flushright}
\includegraphics[width=7.8cm,angle=0]{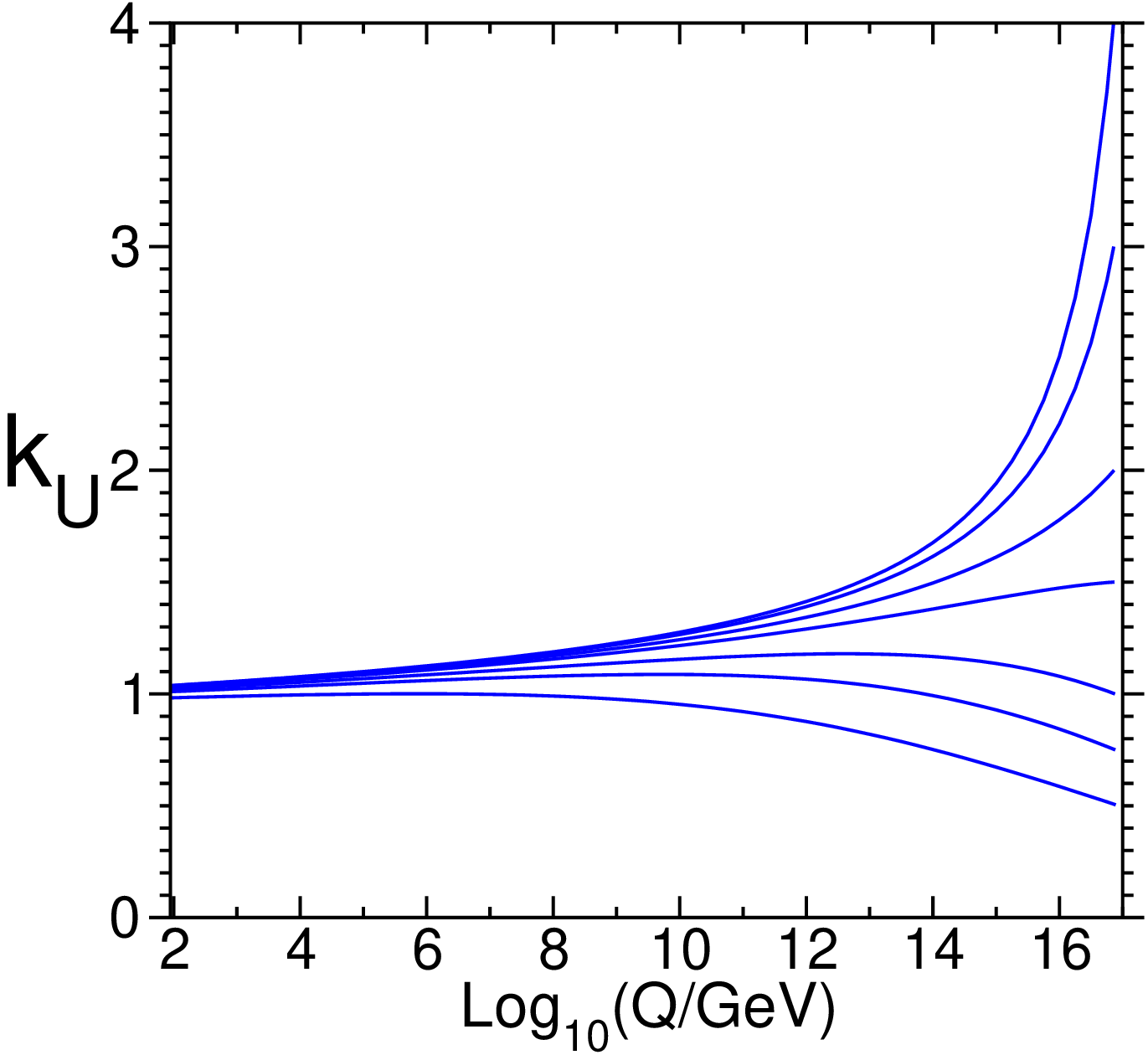}
\end{flushright}
\end{minipage}
\caption{\label{fig:fixedrun}
Renormalization group trajectories near the fixed point for 
$k_N$ in the LND model (left panel) and
$k_U$ in the QUE model (right panel), showing the infrared-stable
quasi-fixed point behaviors.
Here $m_t = 173.1$ GeV and $\tan\beta = 10$ are assumed.
}
\end{figure}

\begin{figure}[!tp]
\begin{minipage}[]{0.47\linewidth}
\caption{\label{fig:fixedrects}
The contours represent the infrared-stable quasi-fixed points of the
2-loop renormalization group equations in the plane of Yukawa couplings
$(k_i, h_i)$ evaluated at $Q = 500$ GeV. The allowed perturbative regions
(defined by $k_i, h_i < 3$ at $Q = M_{\rm unif}$) are to the left and below
the contours. The long dashed (blue) line corresponds to $k_N, h_N$ in the
LND model. The solid (black) line corresponds to $k_U, h_U$ in the QUE
model, and the nearly overlapping short dashed (red) line corresponds to
$k_D, h_D$ in the QDEE model. Here $m_t = 173.1$ GeV and $\tan\beta = 10$
are assumed. The very nearly rectangular shape of these contours reflects
the absence of direct coupling between the Yukawa couplings in the one-loop 
$\beta$ functions.}
\end{minipage}
\begin{minipage}[]{0.52\linewidth}
\begin{flushright}
\includegraphics[width=8.0cm,angle=0]{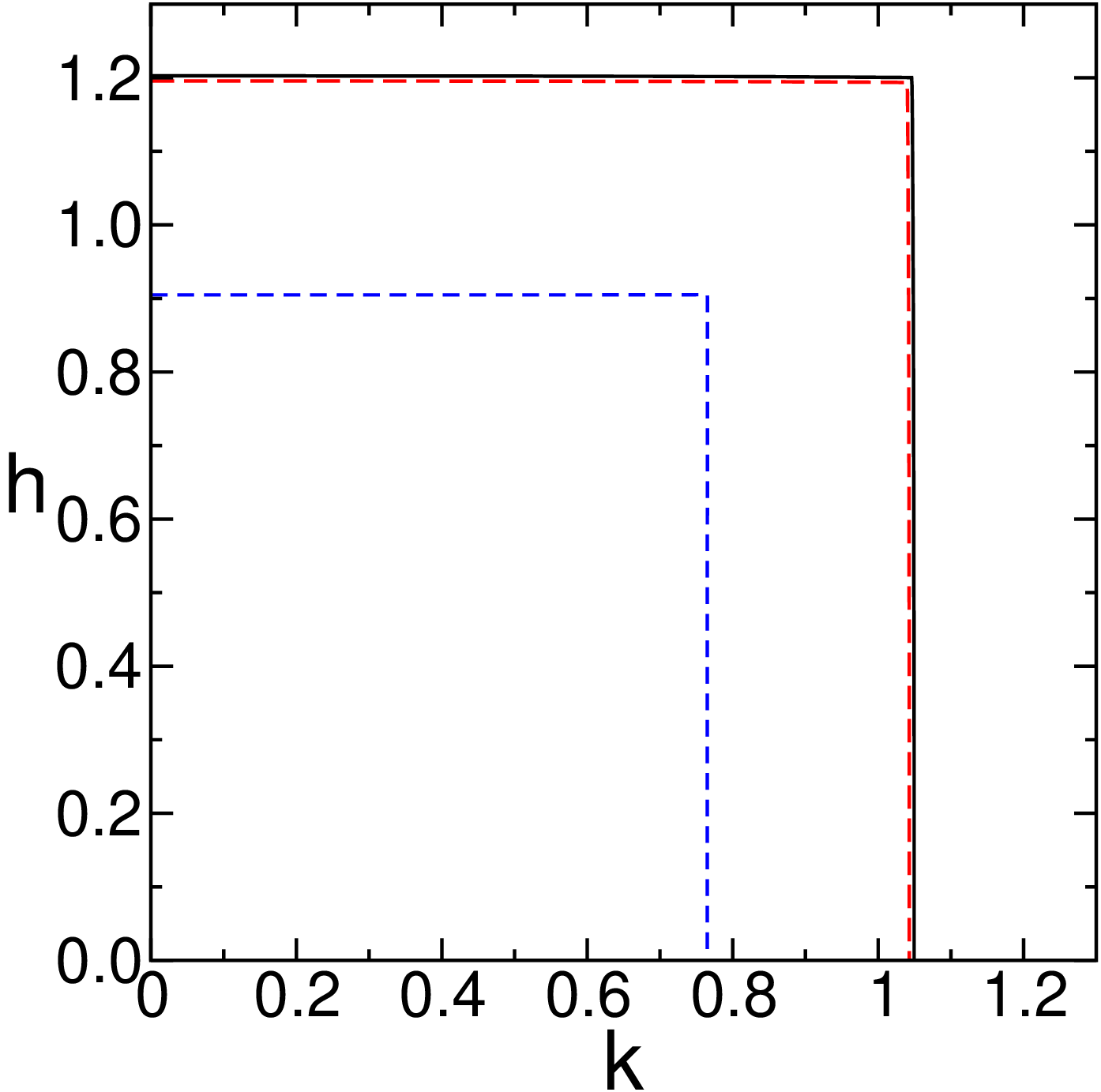}
\end{flushright}
\end{minipage}
\end{figure}

The largest corrections to $m_{h^0}$ are obtained when the
new Yukawa couplings of the type $k_N$, $k_U$, or $k_D$ are as large as 
possible in the LND, QUE, and QDEE models respectively.
These new Yukawa couplings have infrared quasi-fixed point 
behavior, which limits how large they can be at the TeV scale while
staying consistent with perturbative unification.
This is illustrated in Figure \ref{fig:fixedrun}, which shows the
renormalization group running\footnote{In this paper,
I use 3-loop beta functions for the gauge couplings and gaugino masses, 
and 2-loop beta functions for the Yukawa couplings,
soft scalar trilinear couplings, and soft scalar squared masses. 
These can be obtained quite
straightforwardly from the general results listed in 
\cite{betas:1,betas:2,betas:3}, and so are not given explicitly here.}
of the $k_N$ coupling in the LND model 
and $k_U$ in the QUE model.%
The running of $k_D$ in the QDEE model
is very similar to the latter (and so is not shown).
In this paper, I will somewhat arbitrarily define the fixed-point 
trajectories to be those for which the extreme Yukawa couplings are equal 
to\footnote{Formally, 
it turns out that the 2-loop and 3-loop beta functions for 
these Yukawa couplings have ultraviolet-stable fixed points,
although these occur at such large values $(>5)$ that they cannot be 
trusted to reflect the true behavior. Simply requiring 
the high-scale value of
the Yukawa couplings to be somewhat smaller avoids this issue.}
3 at the scale $M_{\rm unif}$ where $\alpha_1$ and $\alpha_2$ unify.
Then, assuming that only one of the new Yukawa couplings is turned on
at a time, and that $\tan\beta = 10$ with $m_t = 173.1$ GeV and with
all new particle thresholds taken to be at $Q = 600$ GeV, the
fixed point values also evaluated at $Q = 600$ GeV are
\beq
&&
\mbox{LND model:}\qquad\>\, k_N = 0.765\quad\mbox{or}\quad h_N = 0.905,
\\
&&
\mbox{QUE model:}\qquad\>\, k_U = 1.050\quad\mbox{or}\quad h_U = 1.203,
\\
&&
\mbox{QDEE model:}\qquad\! k_D = 1.043\quad\mbox{or}\quad h_D = 1.196.
\eeq
Turning on both Yukawa couplings at the same time in
each model hardly affects the results at all, because 
$k_i, h_i$ decouple from each other's beta functions at one loop order 
for each of $i = N,U,D$.  This is illustrated by the very nearly
rectangular shape of the fixed-line contours in Figure
\ref{fig:fixedrects}. 

\begin{figure}[!tp]
\begin{minipage}[]{0.47\linewidth}
\caption{\label{fig:gauginosQUE}
Running of gaugino masses in the QUE model, assuming a unified value 
$m_{1/2}$ at $M_{\rm unif}$. The gluino mass parameter $M_3$ is evolved 
according to the 1, 2, and 3 loop beta functions in the top three lines. 
The 1 and 3 loop beta functions are shown for the parameters $M_1$ 
(bottom two lines) and $M_2$, for which the 2-loop and 3-loop results are not 
visually distinguishable. Note the significant running of $M_3$ due to 
multi-loop effects.}
\end{minipage}
\begin{minipage}[]{0.52\linewidth}
\begin{flushright}
\includegraphics[width=8.0cm,angle=0]{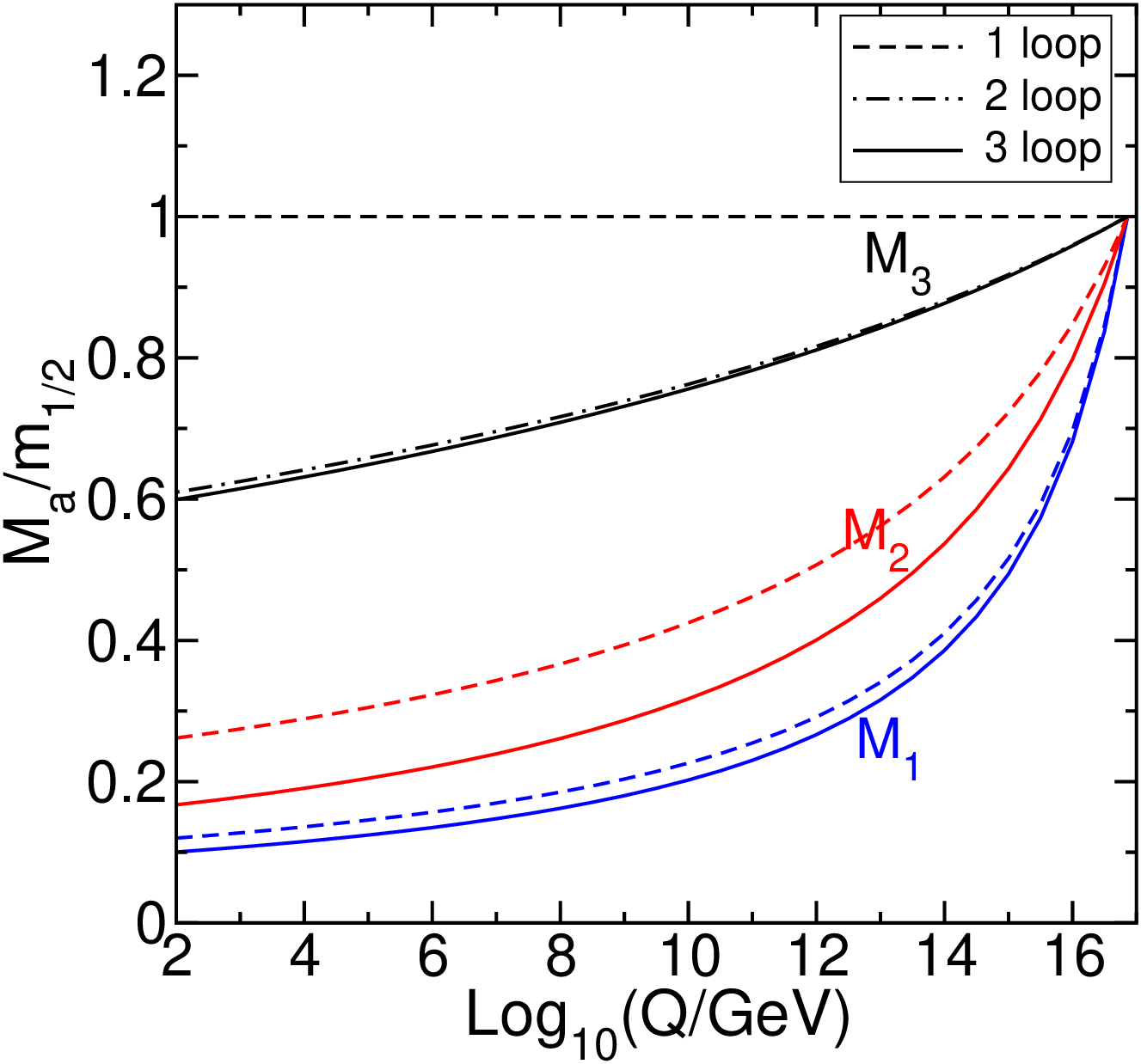}
\end{flushright}
\end{minipage}
\end{figure}

The phenomenology of supersymmetric models is crucially dependent on the
ratios of gaugino masses. In the MSSM, it is will known that if the
gaugino masses unify at $M_{\rm unif}$, then working to one-loop order they
obey $M_1/\alpha_1 = M_2/\alpha_2 = M_3/\alpha_3 = m_{1/2}/\alpha_{\rm
unif}$, and this relation has only moderate corrections from higher-loop
contributions to the beta functions. The presence of extra matter
particles strongly affects this prediction, however. In Table
\ref{table:gauginomasses}, the predictions for $M_1$, $M_2$, and $M_3$ at
$Q = 1$ TeV are given for the MSSM, the LND model, the QUE model, and the
QDEE model.%
\begin{table}
\begin{tabular}[c]{|l|c|c|c|c|c|}
\hline
\phantom{XXXX} & \phantom{i}$M_1/m_{1/2}$ \phantom{i}& \phantom{i}$M_2/m_{1/2}$ \phantom{i}& \phantom{i}$M_3/m_{1/2}$\phantom{i} & 
\phantom{i}$M_2/M_1$\phantom{i} & \phantom{i}$M_3/M_1$ \phantom{i}\\ \hline\hline
MSSM & 0.41 & 0.77 & 2.28 & 1.88 & 5.53 \\ \hline 
LND 
    & 0.32 & 0.59 & 1.75 & 1.86 & 5.52 \\ 
\hline
QUE $(k_U = 0)$ & 0.097 & 0.147 & 0.571 & 1.52 & 5.90 \\
\phantom{QUE} 
$(k_U(M_{\rm unif}) = 3)$ 
& 0.109 & 0.176 & 0.617 & 1.61 & 5.66 \\ \hline
QDEE $(k_D = 0)$ & 0.094 & 0.153 & 0.572 & 1.62 & 6.08 \\ 
\phantom{QDEE} 
$(k_D(M_{\rm unif}) = 3)$ 
& 0.107 & 0.178 & 0.615 & 1.66 & 5.72 \\ \hline
\end{tabular}
\caption{Gaugino masses $M_a/m_{1/2}$ for $(a = 1,2,3)$ and ratios of gaugino masses $M_2/M_1$ and $M_3/M_1$,
evaluated at $Q = 1$ TeV in the models described in the text,
assuming unified gaugino masses $m_{1/2}$ at $M_{\rm unif}$.
\label{table:gauginomasses}}
\end{table}
In the latter two cases, I distinguish between the cases of vanishing 
extra Yukawa couplings and the fixed-point trajectories with $(k_U, h_U) 
= (3,0)$ and $(k_D, h_D) = (3,0)$, respectively, at $Q = M_{\rm unif}$. As 
before, I have used $\tan\beta = 10$ and $m_t = 173.1$ GeV, and taken all 
new particle thresholds to be at $Q = 600$ GeV, and assumed for 
simplicity that the new scalar trilinear couplings vanish at $Q = M_{\rm 
unif}$. The results will change slightly if these assumptions are 
modified, but there are a couple of striking and robust features to be 
pointed out about the gaugino masses in these models. First, because the 
unified gauge couplings are so much larger in the models with extra 
matter than in the MSSM, the gaugino masses at the TeV scale are 
suppressed relative to $m_{1/2}$ by a significant amount compared to the 
MSSM. Secondly, the one-loop prediction for the ratios of gaugino masses 
is very strongly violated by two-loop effects\footnote{Similar effects have been 
noted long ago in the context of ``semi-perturbative unification"
\cite{Kolda:1996ea}.}
in the extended models, 
which were evidently neglected in \cite{Babu:2008ge}. For example, in 
both the QUE and QDEE models, the one-loop prediction is that $M_3 = 
m_{1/2}$, independent of $Q$, since the one-loop beta function for $M_3$ 
happens to vanish. However, the correct result is that $M_3$ does run 
significantly, with $M_3/m_{1/2}$ reduced by some 40\% from 
unity, 
depending on the Yukawa coupling value. This reflects, in part, the accidental 
vanishing of the one-loop beta function; in contrast, 
the three-loop contribution to the 
running is quite small compared to the two-loop one. This is illustrated 
for the QUE model in Figure \ref{fig:gauginosQUE}, which shows the 
renormalization-scale dependence of the running gaugino mass parameters 
$M_1$, $M_2$, and $M_3$ in the QUE model [with $(k_U, h_U) = (3,0)$ at $Q 
= M_{\rm unif}$], evolved according to the 1, 2, and 3-loop beta 
functions.

Another 
notable feature of the extended models is that they permit 
gaugino mass 
domination for the soft supersymmetry breaking terms at the unification 
scale, according to which all soft scalar masses and scalar trilinear 
couplings are assumed negligible compared to the gaugino masses, or $A_0 
= 0$, $m_0^2 = 0$ in the usual mSUGRA language. In the MSSM, this 
``no-scale" boundary condition is problematic if applied strictly, 
because it predicts that the lightest supersymmetric particle (LSP) is not a neutralino. However, in the QUE 
and QDEE models, the increased size of the gauge couplings at high scales 
gives extra gaugino-mediated renormalization group contributions to the 
scalar squared masses, so that they are safely heavier than the bino-like 
LSP. For the squarks and sleptons of the first two families, this is 
illustrated in Table \ref{table:MSSMsfermionmasses} (for the same models 
as in Table \ref{table:gauginomasses}), by giving the ratios of the 
running masses to the unified gaugino mass parameter $m_{1/2}$.
\begin{table}
\begin{tabular}[c]{|l|c|c|c|c|c|}
\hline
\phantom{XXXX} & \phantom{x}$m_{\tilde q}$\phantom{x} & \phantom{x}$m_{\tilde{\bar{u}}}$\phantom{x} & 
\phantom{x}$m_{\tilde{\bar{d}}}$\phantom{x} & \phantom{x}$m_{\tilde \ell}$\phantom{x} & \phantom{x}$m_{\tilde{\bar{e}}}$\phantom{x} 
\\ \hline\hline
MSSM & 2.08 & 2.01 & 2.00 & 0.67 & 0.37 \\ \hline 
LND  & 1.89 & 1.82 & 1.81 & 0.63 & 0.35 \\ \hline
QUE $(k_U = 0)$ & 1.24 & 1.20 & 1.19 & 0.45 & 0.28 \\
\phantom{QUE} 
$(k_U(M_{\rm unif}) = 3)$ 
& 1.29 & 1.24 & 1.24 & 0.47 & 0.30 \\ \hline
QDEE $(k_D = 0)$ & 1.24 & 1.20 & 1.20 & 0.45 & 0.28\\ 
\phantom{QDEE} 
$(k_D(M_{\rm unif}) = 3)$ 
& 1.30 & 1.25 & 1.24 & 0.47 & 0.30 \\ \hline
\end{tabular}
\caption{Ratios of first- and second-family MSSM squark and slepton 
mass parameters to $m_{1/2}$,
evaluated at $Q = 1$ TeV, assuming unified gaugino mass dominance 
at $Q = M_{\rm unif}$
($m_0^2 = 0$ and $A_0 = 0$).\label{table:MSSMsfermionmasses}}
\end{table}
The contributions of the gaugino masses to the new extra squarks and
sleptons in the LND, QUE, and QDEE models are listed below: 
\beq
&&
\mbox{LND:}\quad (m_{D}, m_{\overline{D}}, m_{L}, m_{\overline{L}}, 
m_{N}, m_{\overline N}) = 
(1.80, 1.80, 0.63, 0.63, 0, 0),
\\
&&
\mbox{QUE:}\quad (m_{Q}, m_{\overline{Q}}, m_{U}, m_{\overline{U}}, m_{E}, m_{\overline E}) = 
(1.17, 1.29, 1.25, 0.94, 0.267, 0.299),
\label{eq:QUEsfermions}
\\
&&
\mbox{QDEE:}\quad (m_{Q}, m_{\overline{Q}}, m_{D}, m_{\overline{D}}, m_{E}, m_{\overline E}) =
(1.30,  1.18,  0.94,  1.24,  0.266,  0.304).
\label{eq:QDEEsfermions}
%
\eeq
Here I have chosen to display the 
results for boundary conditions at $M_{\rm unif}$ of
$k_N = h_N = 0$ and for $k_U = 3, h_U = 0$ and for $k_D = 3, h_D = 0$,
respectively.
It should be noted that these are all running mass parameters, and the physical mass parameters will be different.
Also, if there are non-zero contributions to the running scalar squared masses and scalar trilinear couplings
at $M_{\rm unif}$, the results will of course change. For example, including a
non-zero common $m_0^2$, as in mSUGRA, will raise all of the scalar 
squared masses, yielding a more degenerate scalar mass spectrum.

For the QUE and QDEE models, we see from Tables 
\ref{table:gauginomasses} and
\ref{table:MSSMsfermionmasses} and eqs. (\ref{eq:QUEsfermions}) and 
(\ref{eq:QDEEsfermions}) that the bino mass parameter is well over a 
factor of 2 smaller than the lightest slepton mass, for unified, dominant 
gaugino masses. Since neutralino mixing only decreases 
the LSP mass compared to the bino mass parameter, the LSP will be a 
neutralino. In contrast, for the LND models, the gaugino mass dominance 
boundary condition would predict that the scalar component of $N$ or 
$\overline N$ (a non-MSSM sneutrino) should be the LSP, and should be 
nearly massless. In fact, including a non-zero Yukawa coupling $h_N$ or 
$k_N$ would give the corresponding scalar a negative squared mass. If 
there is an additional positive contribution to that sneutrino mass, 
then it can be the LSP, and it might be interesting to consider it as a 
possible dark matter candidate. 

\begin{figure}[!tp]
\begin{minipage}[]{0.47\linewidth}
\begin{flushright}
\includegraphics[width=8.0cm,angle=0]{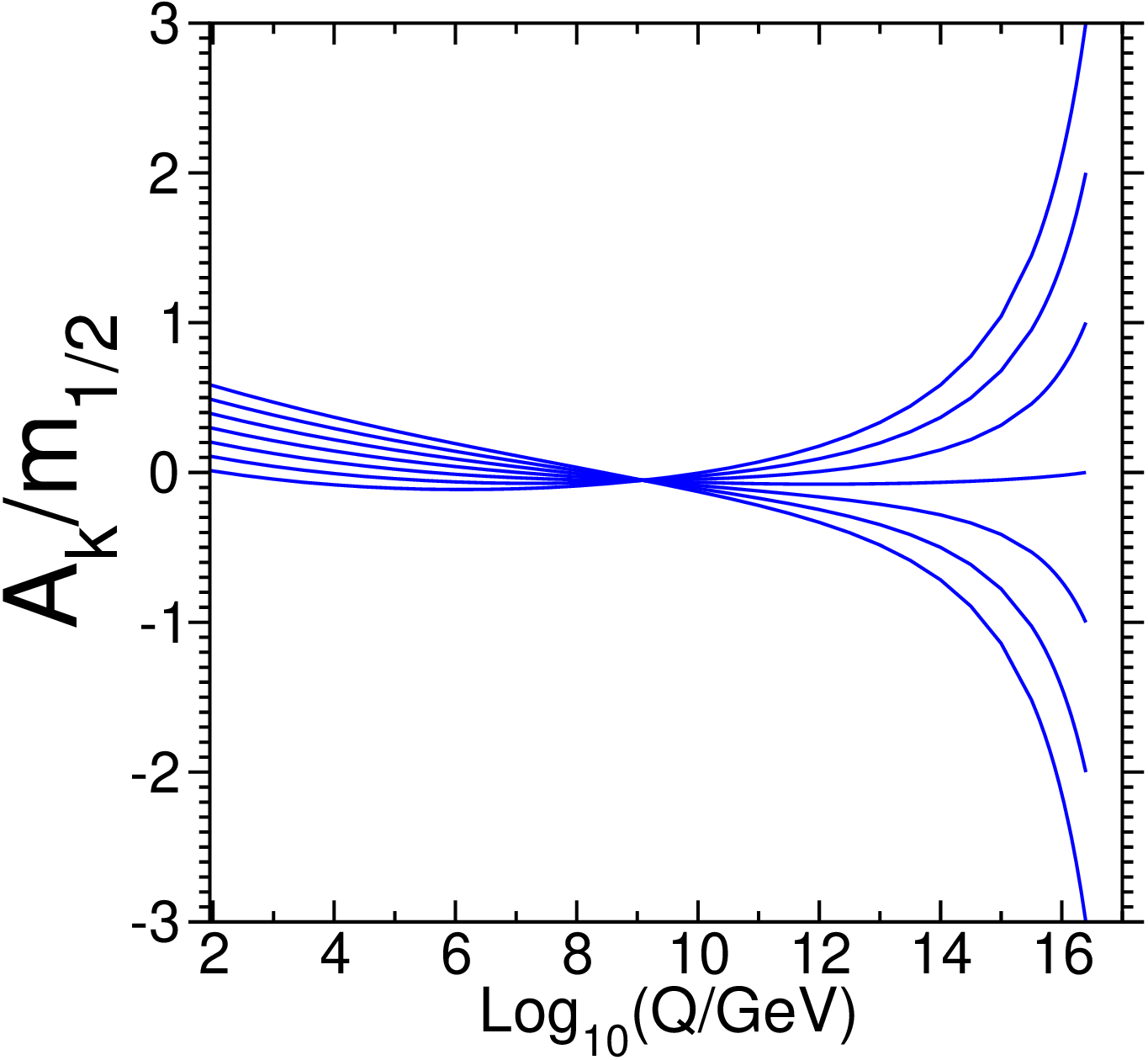}
\end{flushright}
\end{minipage}
\begin{minipage}[]{0.52\linewidth}
\begin{flushright}
\includegraphics[width=8.0cm,angle=0]{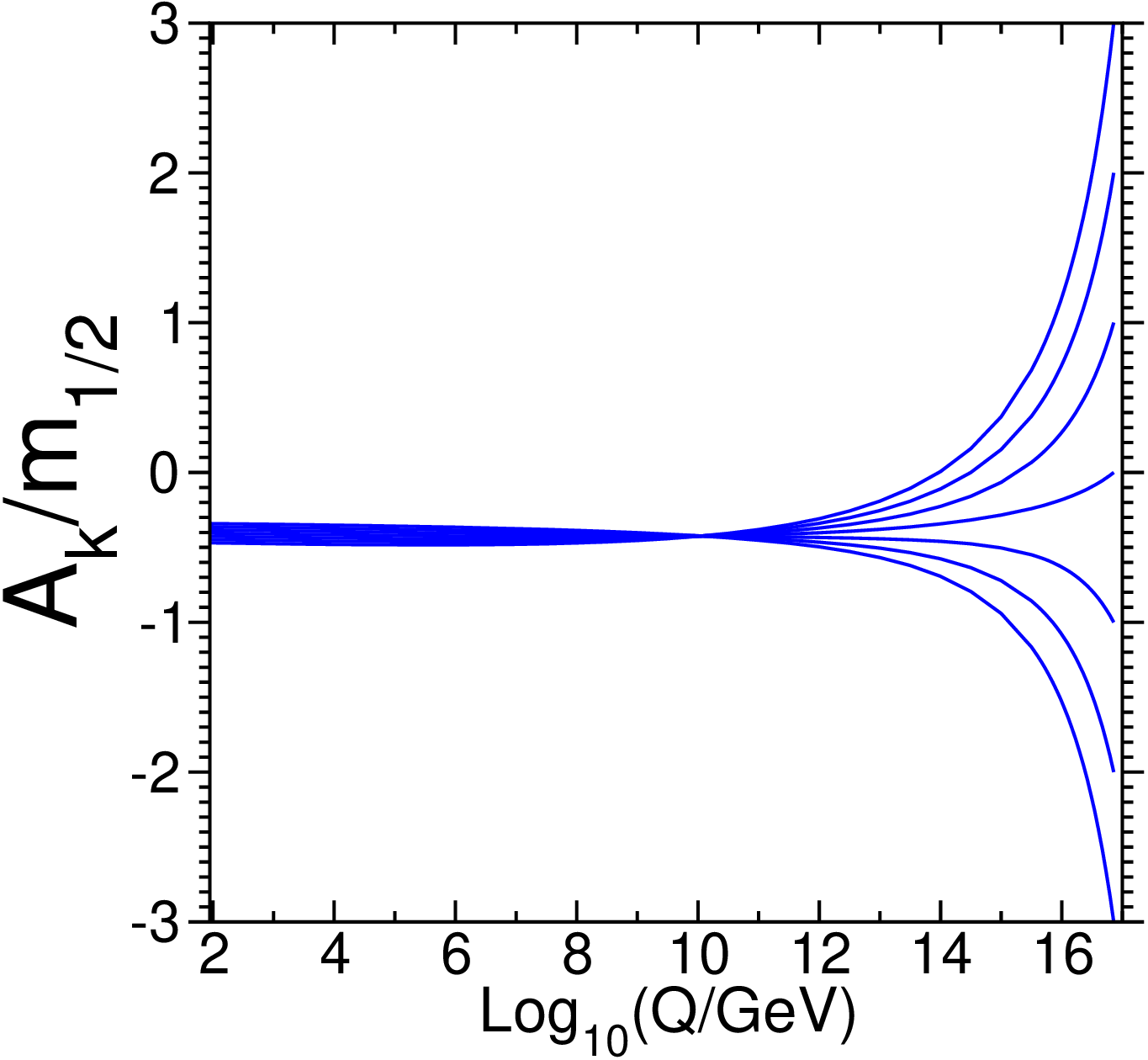}
\end{flushright}
\end{minipage}
\caption{\label{fig:Arunning}
Renormalization group running of scalar trilinear couplings $A_{k_N}$ 
in the LND
model (left panel) and $A_{k_U}$ in the QUE model (right panel), 
normalized to
$m_{1/2}$, the common gaugino mass parameter at the unification 
scale $M_{\rm unif}$. 
The different lines correspond to different boundary conditions at
$M_{\rm unif}$.
The corresponding Yukawa couplings $k_N$ and $k_U$ are taken to be near their
fixed-point trajectories, with $k_N = 3$ and $k_U = 3$ at $M_{\rm unif}$.
The running of $A_{k_D}$ in the QDEE model is very similar to that shown here
for $A_{k_U}$.}
\end{figure}

The corrections to the lightest Higgs squared mass considered in the next
section depend on the scalar trilinear coupling $a_{k_N}$, $a_{k_U}$,
or $a_{k_D}$ of the type appearing in
eq.~(\ref{eq:gensoft}). It is therefore useful to note that these
couplings have a strongly attractive fixed-point behavior in the infrared
when the corresponding superpotential couplings $k_N$, $k_U$ and $k_D$ 
are near their fixed points.  To illustrate this, consider the quantities
\beq
A_{k_N} \equiv a_{k_N}/k_N, \quad
A_{k_U} \equiv a_{k_U}/k_U, \quad
A_{k_D} \equiv a_{k_D}/k_D,
\label{eq:defAs}
\eeq
for the LND, QUE, and QDEE models, respectively. The renormalization group 
runnings
of $A_{k_N}$ and $A_{k_U}$ (each normalized to $m_{1/2}$) are shown in Figure 
\ref{fig:Arunning}, for various input values at the unification scale.
The running of $A_{k_N}$ in the LND model is seen to have a mild focusing
behavior, leading to values at the weak scale of $-0.1 \lsim
A_{k_N}/m_{1/2} \lsim 0.6$ for input values at $M_{\rm unif}$ in the range
$-3 \lsim A_{k_N}/m_{1/2} \lsim 3$. In the case of $A_{k_U}$ in the QUE
model, one finds an even 
stronger focusing behavior leading to $-0.5 \lsim
A_{k_U}/m_{1/2} \lsim -0.3$ at the weak scale. 
The running of $A_{k_D}$ in the QDEE model is
very similar (and so is not shown). It is useful to note that in the cases
of $A_{k_U}$ in the QUE model and $A_{k_D}$ in the QDEE model, most of the
contribution to the running comes from the gluino mass parameter. This
will still be true if one does not assume gaugino mass unification,
provided only that the gluino mass parameter $M_3$ is not very small
compared to the bino and wino mass parameters $M_1$ and $M_2$. Therefore,
the previous results concerning the fixed-point behavior of $A_{k_D}$ and
$A_{k_U}$ remain approximately valid if $m_{1/2}$ is replaced by the 
value of $M_3$ at the unification scale. 

\subsection{Fine-tuning considerations}

One of the primary model-building motivations in recent years is the
supersymmetric little hierarchy problem, which concerns the tuning required
to obtain the electroweak scale, given the large supersymmetry breaking
effects needed to avoid a light Higgs boson that should have been seen at
LEP and to evade direct searches for superpartners at LEP and the
Tevatron. One way to express this problem is to note that the $Z$ boson
mass is related to the parameters $|\mu|$ and $m^2_{H_u}$ near the weak
scale by: 
\beq
-\frac{1}{2} m_Z^2 = |\mu|^2 + m^2_{H_u} + \frac{1}{2 v_u} 
\frac{\partial}{\partial v_u} \Delta V + {\cal O}(1/\tan^2\beta) ,
\eeq
where $\Delta V$ is the radiative part of the effective potential.
Although there can be no such thing as an objective measure of fine tuning
in parameter space, the cancellation needed between $|\mu|^2$ and
$m^2_{H_u}$ can be taken as an indication of how ``difficult" it is to
achieve the observed weak scale. Large values of $-m^2_{H_u}$ require more
tuning in this sense. 

In the MSSM, 
with the gauge and Yukawa couplings taken to be the values of the infamous
benchmark point SPS$1\rm{a}'$ \cite{AguilarSaavedra:2005pw}
for a concrete example, 
one finds
\beq
-m^2_{H_u} &=& 
1.82 \hat M_3^2 
- 0.212 \hat M_2^2 
+ 0.156 \hat M_3 \hat M_2 
+ 0.023 \hat M_1 \hat M_3 
-0.32 \hat A_t \hat M_3 - 0.07 \hat A_t \hat M_2 
\nonumber \\ &&
+ 0.11 \hat A_t^2  
-0.64 \hat m^2_{H_u} 
+ 0.36 \hat m^2_{\tilde q_3} 
+ 0.28 \hat m^2_{\tilde u_3}
+ \ldots .
\label{eq:mHuMSSM}
\eeq
Here $m_{H_u}^2$ on the left side is evaluated at the scale $Q = 600$ GeV,
where corrections to $\Delta V$ are presumably not too large.
The non-MSSM particle thresholds are also taken to be
at the same scale. The hats on the parameters on the right side 
denote that they are 
inputs at the apparent unification scale
$M_{\rm unif} = 2.4 \times 10^{16}$ GeV. They 
consist of gaugino masses $\hat M_{1,2,3}$,
scalar squared masses 
$\hat m^2_{H_u}$, $\hat m^2_{\tilde q_3}$ and $\hat m^2_{\tilde u_3}$, 
and $\hat A_t \equiv a_t/y_t$. 
I have neglected to write other contributions with small coefficients.
Note that the gaugino masses and scalar 
squared masses are not assumed to be unified here. 
The essence of the supersymmetric little hierarchy problem is that
after constraints from non-observation of the lightest Higgs 
boson, the charged supersymmetric particles, and from the relic abundance 
of dark matter are taken into account, the remaining parameter space 
tends to yield $-m_{H_u}^2 \gg m_Z^2/2$, so that some fine adjustment is
needed between $-m_{H_u}^2$ and $|\mu|^2$.
It was noted long ago in ref.~\cite{Kane:1998im} 
that the gluino mass parameter $M_3$ is actually mostly responsible for 
the tuning needed in $m^2_{H_u}$, because of its large coefficient
as seen in eq.~(\ref{eq:mHuMSSM}), and this problem can be ameliorated 
significantly by taking $|\hat M_3/\hat M_2|$ 
smaller than unity at $M_{\rm unif}$. 
This can easily be achieved in non-mSUGRA models. For example, taking 
$|\hat M_3/\hat M_2| \sim 1/3$ produces near cancellation between the 
$\hat M_3^2$ 
and $\hat M_2^2$ terms with opposite signs in eq.~(\ref{eq:mHuMSSM}), 
yielding a smaller value for 
$m^2_{H_u}$.

Now let us compare to the corresponding formulas in the LND, QUE and 
QDEE models under study here. For the QUE model, I find near 
the fixed 
point $k_U = 1.05$ with $h_U = 0$ that
the most significant contributions are approximately:
\beq
-m^2_{H_u} &=&
2.10 \hat M_3^2 
+ 0.035 \hat M_2^2 
+ 0.019 \hat M_1^2 
-0.014 \hat M_3 \hat M_2 
-0.075 \hat A_t \hat M_3 - 0.016 \hat A_t \hat M_2 
\nonumber \\ &&
+ 0.022 \hat A_{k_U} \hat M_3 + 0.014 \hat A_{k_U} \hat M_2 
+ 0.057 \hat A_t^2 - 0.015 \hat A_t \hat A_{k_U} + 0.25 \hat A_{k_U}^2 
\nonumber \\ &&
-0.17 \hat m^2_{H_u} + 0.34 \hat m^2_{\tilde q_3} 
+ 0.27 \hat m^2_{\tilde u_3}
+0.47 m^2_Q + 0.40 m^2_{\overline U} + \ldots
.
\label{eq:tuningQUEfixed}
\eeq
Again the hats on parameters on the right side denote their status as
input values at
$M_{\rm unif}$, and $m^2_{H_u}$ on the left side
is evaluated at $Q = 600$ GeV, which is also
where the new particle thresholds are placed, and $\tan\beta = 10$. 
The result of eq.~(\ref{eq:tuningQUEfixed}) seems to reflect a worsening 
of the little 
hierarchy problem, 
since the
contribution to $-m_{H_u}^2$ proportional to $\hat M_3^2$ is even larger 
than 
in the MSSM
case, while the physical MSSM superpartner masses are actually lower for
fixed values of the input soft parameters, as can be seen from Tables
\ref{table:gauginomasses} and \ref{table:MSSMsfermionmasses}.
This implies that for a given scale of physical superpartner masses,
including notably the top squarks that contribute strongly to $m_{h^0}^2$,
one will need larger $-m_{H_u}^2$, and thus larger $|\mu|^2$, and 
so a more
delicate cancellation between the two. Note also that since the 
contribution proportional to $\hat M_2^2$ is positive (and quite small), there cannot be
a cancellation as in the MSSM for large $|\hat M_2/\hat M_3|$. 
Counteracting these considerations, there is the fact that there 
are 
large positive corrections
to $m_{h^0}^2$ from the new particles, as discussed in the following 
section, so that the top squark masses need not be so large. 
 
It is interesting to compare with the corresponding result when
the new Yukawa coupling $k_U$ is instead taken to vanish:
\beq
-m^2_{H_u} &=& 
1.14 \hat M_3^2 
- 0.107 \hat M_2^2 
+ 0.153 \hat M_3 \hat M_2 
+ 0.022 \hat M_1 \hat M_3
-0.436 \hat A_t \hat M_3 
- 0.090 \hat A_t \hat M_2
\nonumber \\ &&
+ 0.125 \hat A_t^2
-0.70 \hat m^2_{H_u} + 0.30 \hat m^2_{\tilde q_3}
+ 0.21 \hat m^2_{\tilde u_3}
+ \ldots
\eeq
for $k_U = 0$. Here the impact on fine-tuning is less 
because the coefficient of $\hat M_3^2$ is reduced, there is
no large positive contribution from the new scalar soft masses, and the
possibility of significant cancellation between the gluino and wino mass 
contributions (if $|\hat M_2/\hat M_3| > 1$) is restored. But, 
counteracting 
this, there is no large positive
contribution to $m_{h^0}^2$ from the extra vector-like sector when $k_U = 
0$.

Results for the QDEE model are quite similar. 
At the fixed point with $k_D = 1.043$, I find
\beq
-m^2_{H_u} &=&
2.12 \hat M_3^2 
+ 0.034 \hat M_2^2 
+ 0.006 \hat M_1^2 
-0.013 \hat M_3 \hat M_2 
-0.085 \hat A_t \hat M_3 - 0.017 \hat A_t \hat M_2 
\nonumber \\ &&
+ 0.029 \hat A_{k_D} \hat M_3 + 0.014 \hat A_{k_D} \hat M_2 
+ 0.054 \hat A_t^2 - 0.027 \hat A_t \hat A_{k_D} + 0.12 \hat A_{k_D}^2 
\nonumber \\ &&
-0.22 \hat m^2_{H_u} + 0.33 \hat m^2_{\tilde q_3} 
+ 0.26 \hat m^2_{\tilde u_3}
+0.37 m^2_{\overline Q} + 0.39 m^2_{D} + \ldots
,
\label{eq:tuningQDEEfixed}
\eeq
and for $k_D = 0$,
\beq
-m^2_{H_u} &=&
1.15 \hat M_3^2
- 0.106 \hat M_2^2
+ 0.154 \hat M_3 \hat M_2
+ 0.024 \hat M_1 \hat M_3
-0.439 \hat A_t \hat M_3
- 0.090 \hat A_t \hat M_2
\nonumber \\ &&
+ 0.125 \hat A_t^2
-0.70 \hat m^2_{H_u} + 0.30 \hat m^2_{\tilde q_3}
+ 0.21 \hat m^2_{\tilde u_3}
+ \ldots
.
\eeq
The same general
comments therefore apply for the QDEE model as for the QUE model.

Treating the LND model in the same way, I find for $k_N = 0.765$:
\beq
-m^2_{H_u} &=& 
1.74 \hat M_3^2 
- 0.166 \hat M_2^2 
+ 0.131 \hat M_3 \hat M_2 
+ 0.020 \hat M_3 \hat M_1
-0.33 \hat A_t \hat M_3 
- 0.06 \hat A_t \hat M_2 
\nonumber \\ &&
+ 0.07 \hat A_{k_N} \hat M_3 
+ 0.11 \hat A_t^2 
- 0.04 \hat A_t \hat A_{k_N} 
+ 0.05 \hat A_{k_N}^2
\nonumber \\ &&
-0.62 \hat m^2_{H_u} + 0.38 \hat m^2_{\tilde q_3} + 0.30 \hat m^2_{\tilde u_3}
+ \ldots .
\eeq
The dependence on the soft parameters in the sector of new
extra particles is very slight, due to the fact that the fixed-point
Yukawa coupling is not too large. Here we see that even at its fixed 
point
the LND model is
qualitatively quite similar to the MSSM, in that a ratio of 
$|\hat M_2/\hat M_3|$ larger than 1
at the unification scale can reduce $-m^2_{H_u}$ and therefore mitigate
the amount of tuning required with $|\mu|^2$. 
For comparison, the result with $k_N = 0$ is
\beq
-m^2_{H_u} &=& 
1.68 \hat M_3^2 
- 0.178 \hat M_2^2 
+ 0.164 \hat M_3 \hat M_2 
+ 0.020 \hat M_3 \hat M_1
-0.36 \hat A_t \hat M_3 
- 0.08 \hat A_t \hat M_2 
\nonumber \\ &&
+ 0.12 \hat A_t^2 
- 0.66 \hat m^2_{H_u} 
+ 0.34 \hat m^2_{\tilde q_3} 
+ 0.26 \hat m^2_{\tilde u_3}
+ \ldots ,
\eeq
which shows quite similar characteristics.

Summarizing the preceding discussion, there are two general counteracting effects
on the little hierarchy problem from introducing vector-like supermultiplets
with large Yukawa couplings. The impact of contributions to $-m_{H_u}^2$
generally tends to worsen the problem, but the additional correction to $m^2_{h^0}$ discussed
in the next section works to mitigate the problem.
(Ref.~\cite{Babu:2008ge} obtained 
qualitatively similar results, but with quite different numerical 
details, presumably due to neglect of higher-loop contributions to the
running of gaugino masses, as noted above.)
I will make 
no attempt to further quantify the competition between
these two competing and opposite impacts on the little hierarchy problem, 
because there is simply no such thing as an objective measure on 
parameter space, and because there is great latitude in 
choosing the remaining parameters anyway.

\section{Corrections to the lightest Higgs scalar boson mass}
\label{sec:deltamh}
\setcounter{equation}{0}
\setcounter{footnote}{1}

The contributions of the new supermultiplets to the lightest Higgs scalar
boson mass can be computed using the effective potential approximation,
which amounts to neglecting non-zero external momentum effects in
$h^0$ self-energy diagrams. Since $m_{h^0}^2$ is much smaller than any of 
the new particle masses, this approximation is quite good for these
contributions. The one-loop contribution to the effective potential due to
the supermultiplets in eqs.~(\ref{eq:genW})-(\ref{eq:m2Smatrix}) is: 
\beq
\Delta V = 2 N_c \sum_{i=1}^4 [F(M^2_{S_i}) - F(M^2_{F_i})]
,
\eeq
where $N_c$ is the number of colors of $\Phi$, and $M^2_{S_i}$ and 
$M^2_{F_i}$ are the squared-mass eigenvalues of
eqs.~(\ref{eq:m2Fmatrix}) and (\ref{eq:m2Smatrix}), and
$
F(x) = x^2 [\ln(x/Q^2) - 3/2]/64 \pi^2.
$
Here $Q$ is the renormalization scale. I will assume the decoupling approximation that the neutral Higgs mixing angle
is $\alpha \approx \beta - \pi/2$, which is valid
if $m_{A^0}^2 \gg m^2_{h^0}$. Then the correction to $m^2_{h^0}$
is
\beq
\Delta m_{h^0}^2 = 
\left \lbrace \frac{\sin^2\beta}{2} \Bigl [ 
\frac{\partial^2}{\partial v_u^2} - \frac{1}{v_u}\frac{\partial}{\partial v_u}\Bigr ]
               + \frac{\cos^2\beta}{2} \Bigl [ 
\frac{\partial^2}{\partial v_d^2} - \frac{1}{v_d}\frac{\partial}{\partial v_d}\Bigr ]
               + \sin\beta\cos\beta 
\frac{\partial^2}{\partial v_u \partial v_d} \right \rbrace \Delta V
.
\eeq

\begin{figure}[!tp]
\begin{minipage}[]{0.52\linewidth}
\caption{\label{fig:fx}
The functions $f(x)$ and $f_{\rm max}(x)$ described in the text, graphed as a function of $\sqrt{x} = $ the average ratio of scalar to fermion masses.}
\end{minipage}
\begin{minipage}[]{0.47\linewidth}
\begin{flushright}
\includegraphics[width=5.80cm,angle=0]{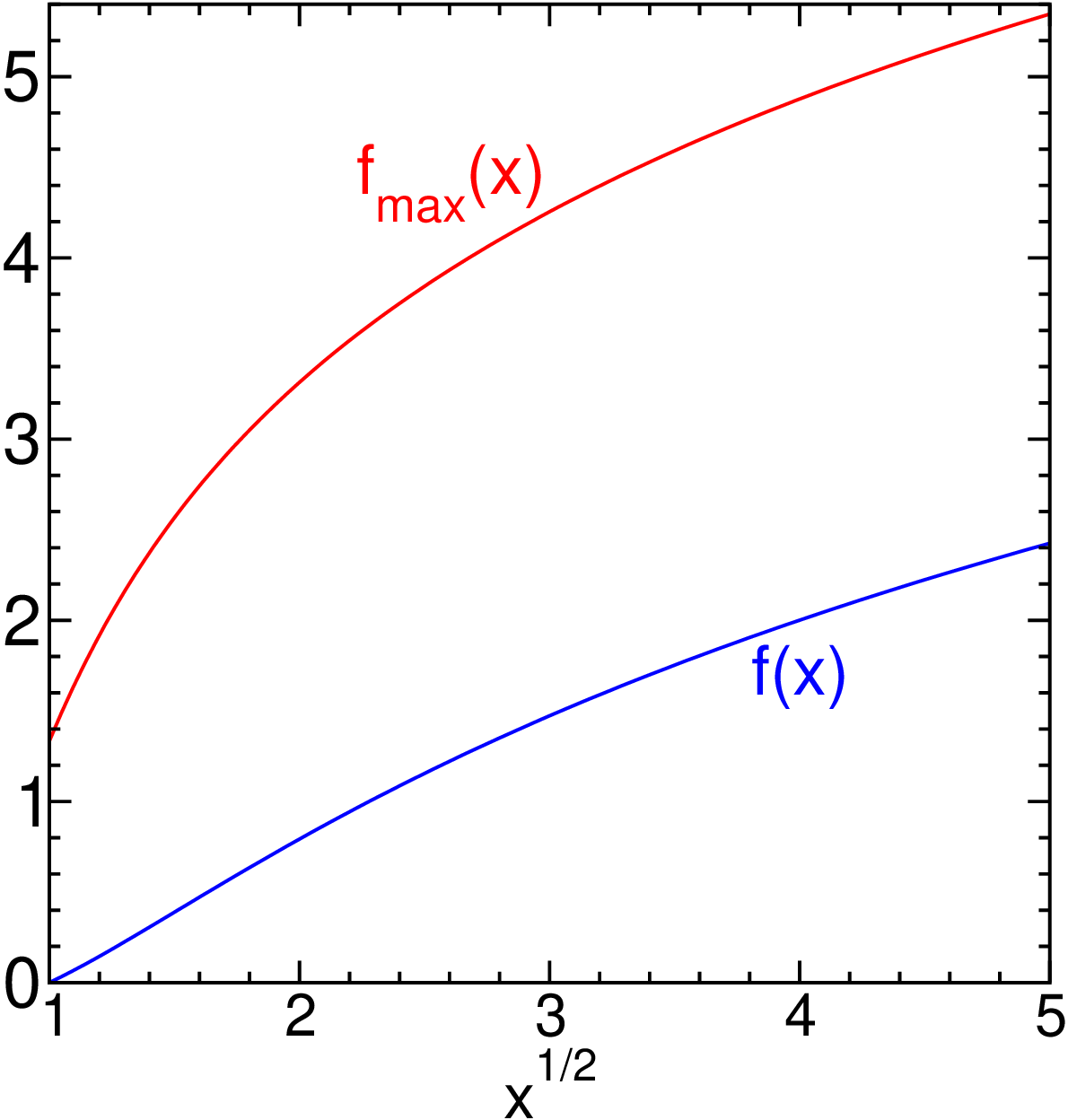}
\end{flushright}
\end{minipage}
\end{figure}

\begin{figure}[!tp]
\begin{minipage}[]{0.52\linewidth}
\caption{\label{fig:deltamhsimple}
Estimates for the corrections to the Higgs mass as a function of 
$\sqrt{x}$, where $x = (M^2 + m^2)/M^2$
is the ratio of the mean scalar squared mass to the mean 
fermion squared mass, in the simplified model
framework used in eq.~(\ref{eq:deltamhsimple}) of the text, using
$N_c = 3$ and $k^4 v^2 \sin^4\beta =$ (190 GeV)$^2$,
corresponding roughly 
to the QUE or QDEE model near the fixed point with reasonably large 
$\tan\beta$. The lower line is the no-mixing case $X_k = 0$, and the upper line
is the maximal mixing case $X_k = 2 M^2 (3 x-1)$. The Higgs mass before the 
correction is taken to be 110 GeV.}
\end{minipage}
\begin{minipage}[]{0.47\linewidth}
\begin{flushright}
\includegraphics[width=6.5cm,angle=0]{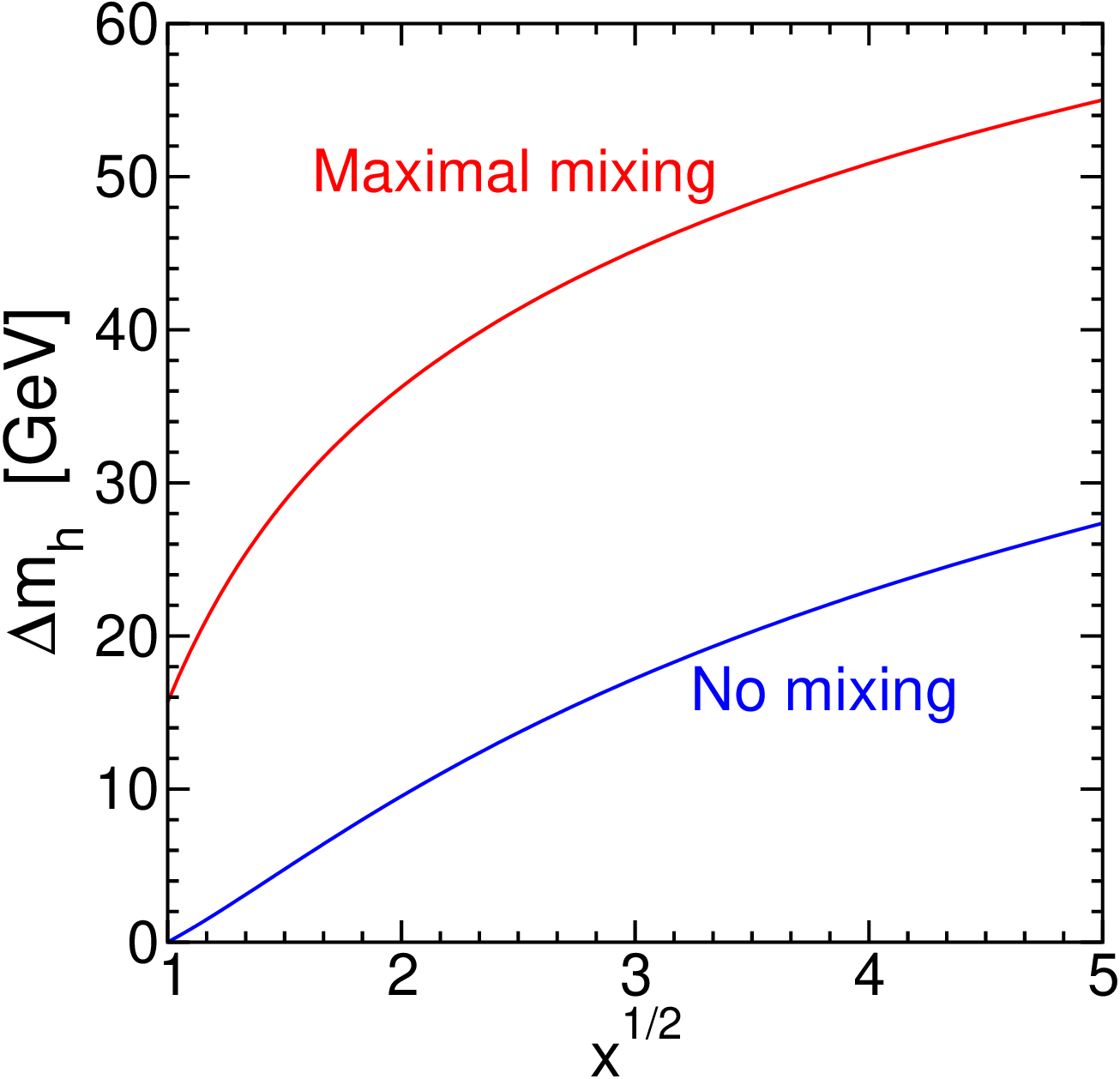}
\end{flushright}
\end{minipage}
\end{figure}

Before presenting some numerical results, it is useful to note a 
relatively simple analytical result that can be obtained if the 
superpotential vector-like fermion masses are taken to be
equal ($M_\Phi = M_\phi 
\equiv M_F$) and the soft supersymmetry-breaking non-holomorphic masses 
are equal ($m^2_\Phi = m^2_{\overline\Phi} = m^2_{\phi} = m^2_{\overline 
\phi} \equiv m^2$), and the small electroweak $D$-terms and the 
holomorphic soft mass terms $b_\Phi$ and $b_\phi$ are neglected. Then, 
writing
\beq
&& 
M_S^2 = M_F^2 + m^2 = \mbox{average scalar mass}
\\
&&
x \equiv M_S^2/M^2_F
\\
&&
\bar k \equiv k \sin\beta,\qquad
\bar h \equiv h \cos\beta,
\\
&&
X_k \equiv A_k - \mu \cot\beta,\qquad 
X_h \equiv A_h - \mu \tan\beta,
\eeq
and expanding to leading order in the normalized Yukawa couplings $\bar 
k$ and $\bar h$, one obtains:
\beq
\Delta m_{h^0}^2 &= &\frac{N_c v^2}{4 \pi^2} \biggl (
\bar k^4 \left [f(x) + \frac{X_k^2}{x M^2} (1 - \frac{1}{3 x})
                - \frac{X_k^4}{12 x^2 M^4} \right ]
\nonumber \\ &&
+ \bar k^3 \bar h \left [-\frac{2}{3}(2 - \frac{1}{x})(1 - \frac{1}{x}) - X_k (2 X_k + X_h)/(3 x^2 M^2) \right ]
\nonumber \\ &&
+ \bar k^2 \bar h^2 \left [-(1 - \frac{1}{x})^2 - (X_k + X_h)^2/(3 x^2 M^2) \right ]
\nonumber \\ &&
+ \bar k \bar h^3 \left [-\frac{2}{3}(2 - \frac{1}{x})(1 - \frac{1}{x}) - X_h (2 X_h + X_k)/(3 x^2 M^2) \right ]
\nonumber \\ &&
+ \bar h^4 \left [f(x) + \frac{X_h^2}{x M^2} (1 - \frac{1}{3 x})
                - \frac{X_h^4}{12 x^2 M^4} \right ]
\biggr ).
\label{eq:approxm2h}
\eeq
where 
\beq
f(x) \equiv \ln(x) -\frac{1}{6}(5 - \frac{1}{x})(1 - \frac{1}{x}) .
\eeq
It is often a good approximation to keep only the contribution proportional to $\bar k^4$, corresponding to the
case where $k \tan\beta \gg h$. 
In that limit, 
eq.~(\ref{eq:approxm2h}) agrees with the result given in \cite{Babu:2008ge},
which can be rewritten as simply:
\beq
\Delta m_{h^0}^2 &=&\frac{N_c}{4 \pi^2} k^4 v^2 \sin^4\beta 
\left[ f(x) + \frac{X_k^2}{x M^2} (1 - \frac{1}{3 x})
                - \frac{X_k^4}{12 x^2 M^4} \right ]
.
\label{eq:deltamhsimple}
\eeq
Note that $x$ is, to first approximation, the ratio of the mean 
squared masses of the scalars to the fermions. 
A key feature of the result for $\Delta m_{h^0}^2$ is that the 
contribution of the vector-like particles does not decouple with the 
overall extra particle mass scale, 
provided that there is a hierarchy $x$ maintained between the 
scalar and fermion squared masses.
To get an idea of the impact of this hierarchy, the 
function $f(x)$ is depicted in Figure~\ref{fig:fx}.
In the limit of unbroken supersymmetry, $f(1) = 0$, and 
$f(x)$ monotonically increases for scalars heavier than fermions ($x>1$).
The other significant feature that could lead to enhanced $\Delta m^2_{h^0}$ is
the mixing parameterized by $X_k$.  The maximum possible value of the
$X_k$ contribution in eq.~(\ref{eq:deltamhsimple}) is obtained when $X_k^2
= 2 M^2 (3 x-1)$, leading to a ``maximal mixing" result $\Delta m^2_{h^0}
= \frac{N_c}{4 \pi^2} k^4 v^2 \sin^4\beta f_{\rm max}(x)$ where $f_{\rm
max} = f(x) + (3 - 1/x)^2/3$. This function is also graphed in Figure
\ref{fig:fx} to show the maximal effects of mixing from the new fermion
sector. In Figure \ref{fig:deltamhsimple}, I show an estimate of the
corresponding corrections to $\Delta m_{h^0}$, taking $N_c = 3$ and
$k^4 v^2 \sin^4\beta =$ (190 GeV)$^2$ (corresponding roughly to the QUE or
QDEE model near the fixed point with reasonably large $\tan\beta$) and
assuming that the predicted
Higgs mass before the correction is 110 GeV, so 
that
$\Delta m_{h^0} = \sqrt{(110\>{\rm GeV})^2 + \Delta m^2_{h^0}} - 110\>{\rm
GeV}$. 

\begin{figure}[!tp]
\begin{minipage}[]{0.47\linewidth}
\caption{\label{fig:deltamh}
Corrections to $m_{h^0}$ in the QUE model with $k_U = 1.05$, for 
varying $m_{1/2}$ with other 
parameters described in the text.
Here $M_F = 400$, 600, and 800 GeV is the vector-like superpotential 
fermion mass term,
and $M_S$ is the geometric mean of the new up-type scalar masses.
The upper and lower lines in each case correspond to $A_k = -0.5 m_{1/2}$
and $A_k = -0.3 m_{1/2}$, respectively, at the TeV scale.
The value of $m_{h^0}$ before these corrections is assumed to be 110 GeV.}
\end{minipage}
\begin{minipage}[]{0.52\linewidth}
\begin{flushright}
\includegraphics[width=8.0cm,angle=0]{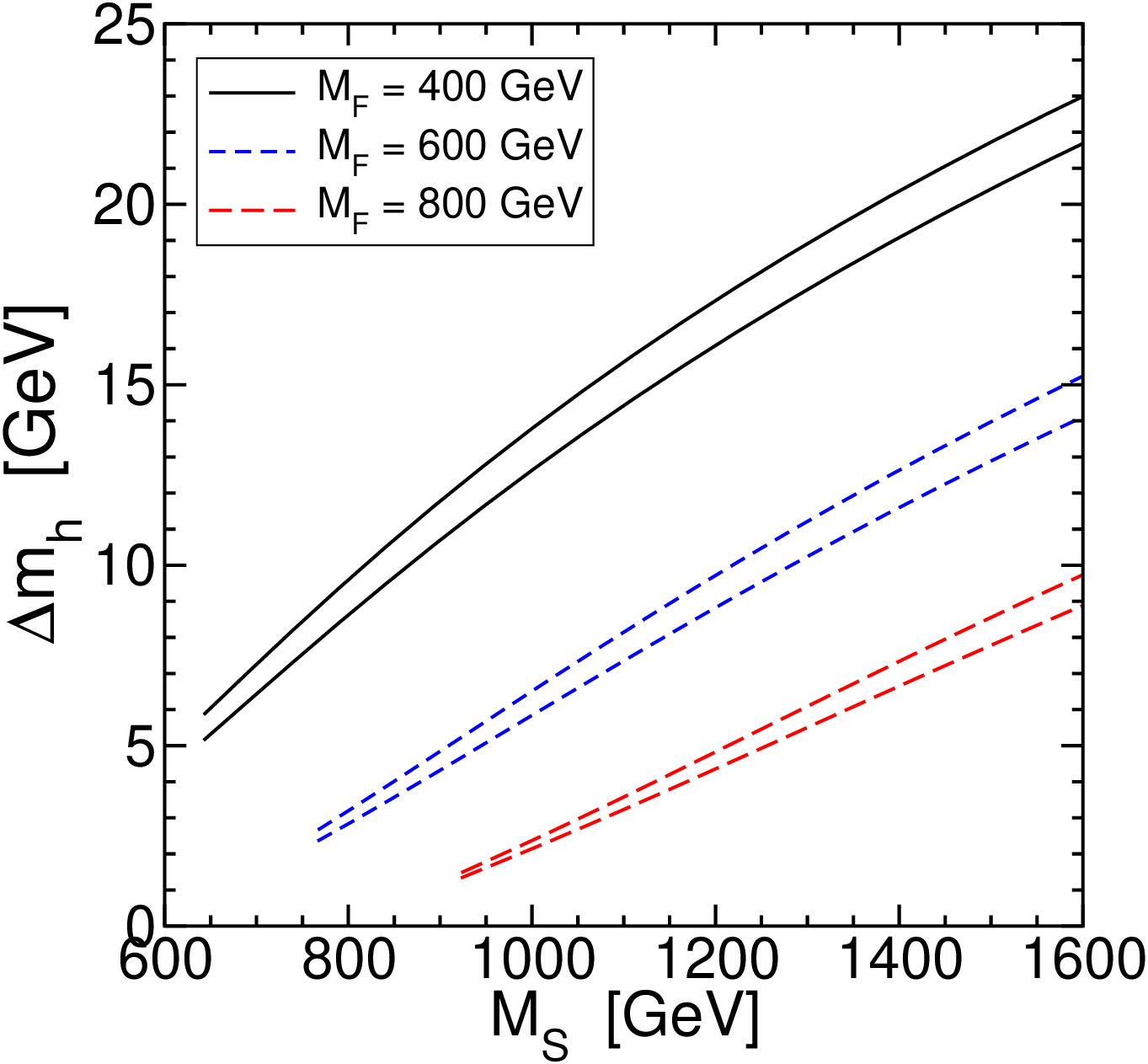}
\end{flushright}
\end{minipage}
\end{figure}

The previous depiction may be too simplistic, 
since the superpotential and soft supersymmetry-breaking 
masses need not have the simple degeneracies that were assumed. 
Also, as found in the previous section,
the scalar trilinear coupling has a fixed point behavior that 
implies that the mixing is neither maximal
nor zero (but closer to the latter). A 
more realistic estimate is therefore as depicted in Figure \ref{fig:deltamh}.
Here, I take scalar masses inspired by the renormalization group 
solutions of the previous section for the QUE model.
In particular, I take three cases for the vector-like superpotential 
masses at the TeV scale, $M_Q = M_U \equiv M_F = 400$, 600, and 800 
GeV.
The Yukawa couplings are taken to be at the fixed-point values 
$k_U = 1.05$ and $h_U = 0$.
The soft supersymmetry-breaking terms are parameterized by
$(m_Q, m_{\overline Q}, m_U, m_{\overline U}) = 
(1.17, 1.29, 1.25, 0.94) m_{1/2}$ and
$(b_Q, b_U) = -(M_Q, M_U) m_{1/2}$, and $A_k = -0.3 m_{1/2}$ and $-0.5 m_{1/2}$, all at a renormalization scale of 1 TeV.
The results turn out to be not very sensitive to $b_Q$ or $b_U$, 
or to the MSSM supersymmetric Higgs mass parameter $\mu$ 
(taken to be $800$ GeV here), or to
$\tan\beta$ as long as it is not too small ($\tan\beta = 10$ was used here).
Figure \ref{fig:deltamh} shows the results for 
$\Delta m_{h^0} \equiv \sqrt{(110\>{\rm GeV})^2 + \Delta m^2_{h^0}}
- 110\>{\rm GeV}$, as a function of $M_S$, 
the geometric mean of the scalar masses. Quite similar results obtain 
for the QDEE model at the fixed point with $k_D = 1.043$, $h_D = 0$.

Figure \ref{fig:deltamh} illustrates that the contribution of the new
extra particles to $m_{h^0}$ is probably much less than the ``maximal
mixing" scenario, if one assumes that the TeV-scale parameters
(particularly the scalar trilinear coupling $a_k$) can be obtained by
renormalization group running from the unification scale. Note that the
models illustrated in Figure \ref{fig:deltamh} represent gaugino-mass
dominated examples. If one assumes that the soft scalar squared masses at
$M_{\rm unif}$ actually have significant positive values (from, for
example, running between $M_{\rm unif}$ and $M_{\rm Planck}$), then the
low-scale model will be even closer to the no-mixing scenario,
since the diagonal entries in the scalar mass matrix will be enhanced,
while the mixing terms are still subject to the strong focusing behavior
seen in Figure \ref{fig:Arunning}. 

In the case of the minimal LND model, one expects the maximum
contributions to $\Delta m^2_{h^0}$ to be suppressed by a factor of
roughly $(k_N/k_U)^4/N_c \approx (0.765/1.05)^4/3 \approx 0.094$. This
leads to corrections that are typically not large compared to the inherent
uncertainties in the total prediction. This counts against
the minimal LND model as a way of significantly increasing the Higgs mass.
One can also consider $n>1$ copies of the LND model, with each $k_N$ Yukawa
coupling near a common fixed point to maximize $\Delta m^2_{h^0}$.
However, then the common fixed point value is even smaller, with
$k_N = 0.695$ for $n=2$, and $k_N = 0.650$ for $n=3$. (A much more significant
correction to $m^2_{h^0}$ can occur
if one enhances the model with
several copies of the extra fields connected by the ``lateral" gauge
group idea of \cite{Babu:2004xg}.) 

\section{Precision electroweak effects}
\label{sec:PEW}
\setcounter{equation}{0}
\setcounter{footnote}{1}

Because the Yukawa couplings responsible for large effects on $m^2_{h^0}$
break the custodial symmetry of the Higgs sector, it is necessary to
consider the possibility of constraints due to precision electroweak
observables arising from virtual corrections to electroweak vector boson
self-energies. 
In this section, I will show that these corrections are
actually benign (and much smaller than previously estimated), 
at least if one uses only $M_t$, $M_W$, and $Z$-peak
observables as in the LEP Electroweak Working Group analyses \cite{LEPEWWG,LEPEWWG2} rather than including also
low-energy observables as in \cite{RPP}. The essential reason for this is
that the corrections decouple with larger vector-like masses, even if the
Yukawa couplings are large and soft supersymmetry breaking effects produce
a large scalar-fermion hierarchy. Indeed, they decouple even when the
corrections to $m^2_{h^0}$ do not.

The most important new physics 
contributions to precision electroweak observables can 
be summarized in terms of the Peskin-Takeuchi $S$ and $T$ parameters 
\cite{Peskin:1991sw}. 
For the measurements of Standard Model observables, I use the updated values:
\beq
s^2_{\rm eff} &=& 0.23153 \pm 0.00016 \quad \mbox{ref.~\cite{LEPEWWG}}
\\
M_W &=& 80.399 \pm 0.025\>{\rm GeV} \quad \mbox{ref.~\cite{Wmass,LEPEWWG2}}
\\
\Gamma_\ell &=& 83.985 \pm 0.086\>{\rm MeV} \quad \mbox{ref.~\cite{LEPEWWG}}
\\
\Delta \alpha_h^{(5)}(M_Z) &=& 0.02758 \pm 0.00035 \quad \mbox{ref.~\cite{LEPEWWG}}
\\
M_t &=& 173.1 \pm 1.3\>{\rm GeV} \quad \mbox{ref.~\cite{Tevatrontop}}
\\
\alpha_s(M_Z) &=& 0.1187 \pm 0.0020 \quad \mbox{ref.~\cite{RPP}}
\eeq
with $M_Z = 91.1875$ GeV held fixed. For the Standard Model predictions 
for $s^2_{\rm eff}$, $M_W$, and $\Gamma_\ell$ in terms of the other 
parameters, I use refs.~\cite{Awramik:2006uz}, \cite{Awramik:2003rn}, and 
\cite{Ferroglia:2002rg}, respectively. These values are then used to 
determine the best experimental fit values and the 68\% and 95\% confidence level (CL) ellipses for $S$ and $T$, relative to a 
Standard Model template with $M_t = 173.1$ GeV and $M_h = 115$ GeV, using
\beq
\frac{s_{\rm eff}^2}{(s_{\rm eff}^2)_{\rm SM}} &=& 
1 + \frac{\alpha}{4 s_W^2 c_{2W}} S - \frac{\alpha c_W^2}{c_{2W}} T
,
\\
\frac{M_W^2}{(M_W^2)_{\rm SM}} &=& 
1 - \frac{\alpha}{2 c_{2W}} S + \frac{\alpha c_W^2}{c_{2W}} T
,
\\
\frac{\Gamma_\ell}{(\Gamma_\ell)_{\rm SM}} &=& 
1 - \alpha d_W S + \alpha (1 + s_{2W}^2 d_W) T
,
\eeq
where $s_W = \sin\theta_W$, $c_W = \cos\theta_W$, $s_{2W} = 
\sin(2 \theta_W)$, $c_{2W} = \cos(2 \theta_W)$, and
$d_W = (1 - 4 s_W^2)/[(1 - 4 s_W^2 + 8 s_W^4) c_{2 W}]$.
The best fit turns out to be $S = 0.057$ and $T = 
0.080$.

The new physics contributions to $S$ and $T$ are given in terms of
one-loop corrections to the electroweak vector boson self-energies
$\Pi_{WW}$, $\Pi_{ZZ}$, $\Pi_{Z\gamma}$ and $\Pi_{\gamma\gamma}$, which
are computed for each of the LND, QUE and QDEE models in \AppendixA. They
are dominated by the contributions from the fermions when the soft
supersymmetry-breaking scalar masses are large. It is useful and
instructive to consider the simplified example that occurs when, in the
notation of the Introduction, $M_\Phi = M_\phi = M_F$ with an expansion in
small $m_u \equiv k v_u$, $m_d \equiv h v_d$, and $M_W$. Then one finds for the new fermion contributions: 
\beq
\Delta T &=& \frac{N_c}{480 \pi s_W^2 M_W^2 M_F^2}
\left [13 (m_u^4 + m_d^4) + 2 (m_u^3 m_d + m_d^3 m_u) + 18 m_u^2 m_d^2 \right ]
\label{eq:simpT}
\\
\Delta S &=& \frac{N_c}{30 \pi M_F^2}
\left [ 4 (m_u^2 + m_d^2) + m_u m_d (3 + 20 Y_\Phi) \right ]
,
\label{eq:simpS}
\eeq
where 
$Y_\Phi$ is the weak hypercharge of the left-handed fermion doublet, 
denoted $\Phi$ in the Introduction, that
has a Yukawa coupling to $H_u$
(so that $Y_\Phi = -1/2$, $1/6$, and $-1/6$ for the LND, QUE, and QDEE models
respectively).
Equations (\ref{eq:simpT}), (\ref{eq:simpS})  
agree\footnote{However, note that
the result for $\Delta T$ quoted in ref.~\cite{Babu:2008ge}
actually corresponds to the improbable case $h v_d = k v_u$, rather 
than $h=0$. 
So, for small $h$,
the actual correction
to $\Delta T$ is almost a factor of 4 smaller than their estimate.
As a result, much smaller values for $M_F$ are admissible 
than would be indicated by ref.~\cite{Babu:2008ge}.}
with the results found in 
\cite{Lavoura:1992np,Maekawa:1995ha}.
An important feature of this is that the corrections decouple 
quadratically with increasing $M_F$, regardless of the soft supersymmetry
breaking terms. This is 
in contrast to the contributions to $\Delta m^2_{h^0}$, 
which do not decouple as long as there is a hierarchy between the
scalar and fermion masses within a heavy supermultiplet. 
It also contrasts with the situation for chiral fermions (as in a 
sequential fourth family), which yields much larger $\Delta S$, $\Delta T$.

If $h = 0$, then the results of eqs.~(\ref{eq:simpT}), (\ref{eq:simpS})
from the fermions become, numerically:
\beq
\Delta T &=& 
0.54 N_c\, k^4 \sin^4(\beta) \left (\frac{\mbox{100 GeV}}{M_F} \right)^2 
,
\label{eq:delTnum}
\\
\Delta S &=& 
0.13 N_c\, k^2 \sin^2(\beta) \left (\frac{\mbox{100 GeV}}{M_F} \right)^2
.
\label{eq:delSnum}
\eeq
These rough formulas show that it is not too hard to obtain agreement with 
the precision electroweak data, provided that $M_F$ is not too small,
but it should be noted that especially for light new fermions with mass
of order 100 GeV, the expansion in large $M_F$ is not very accurate,
with eqs.~(\ref{eq:delTnum}) and (\ref{eq:delSnum})
overestimating the actual corrections.

\begin{figure}[!tp]
\begin{minipage}[]{0.49\linewidth}
\caption{\label{fig:STforLND}
Corrections to electroweak precision observables
$S,T$ from the LND model at the fixed point
$(k_N, h_N) = (0.765, 0)$, for varying $M_L = M_N = m_{\tau'} > 100$ GeV,
in the limit of heavy scalar superpartners. The seven dots on the line segment
correspond to $m_{\tau'} = 100, 120, 150, 200, 250, 400$ GeV and $\infty$,
from top to bottom. The experimental best fit is shown as the $\times$ at
$(\Delta S, \Delta T) = (0.057, 0.080)$. Also shown are the $68\%$ and $95\%$
CL ellipses, obtained as described in the text.
The point $\Delta S= \Delta T = 0$ is defined to be the Standard Model prediction
for $m_t = 173.1$ GeV and $m_{h^0} = 115$ GeV.}
\end{minipage}
\begin{minipage}[]{0.49\linewidth}
\begin{flushright}
\includegraphics[width=7.7cm,angle=0]{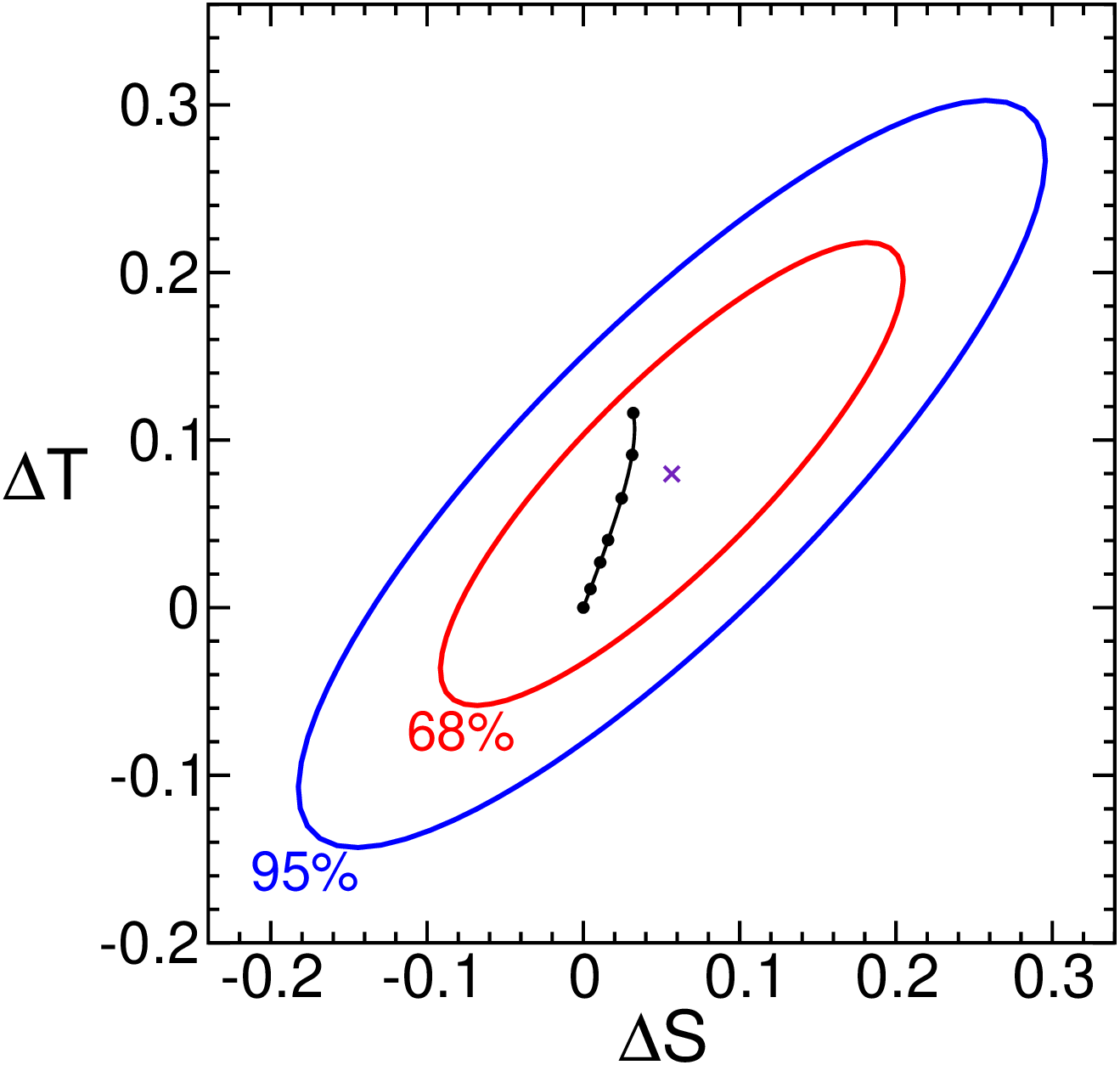}
\end{flushright}
\end{minipage}
\end{figure}

\begin{figure}[!tp]
\begin{minipage}[]{0.49\linewidth}
\caption{\label{fig:STforQUE}
Corrections to electroweak precision observables
$S,T$ from the QUE model at the fixed point
$(k_U, h_U) = (1.050, 0)$, for varying $M_Q = M_U = m_{b'}$,
with $m_{1/2} = 600$ GeV and $A_{k_U} = -0.4 m_{1/2}$ and $m_0 = 0$ and $b_Q = 
b_U = -m_{1/2} M_Q$, using eqs.~(\ref{eq:QUEsfermions}) and 
(\ref{eq:defAs}). The eight dots on the line segment
correspond to $m_{t'_1} = 275, 300, 350, 400, 500, 700, 1000$ GeV and 
$\infty$,
from top to bottom. The experimental best fit is shown as the $\times$ at
$(\Delta S, \Delta T) = (0.057, 0.080)$. Also shown are the $68\%$ and $95\%$
CL ellipses, obtained as described in the text.
The point $\Delta S= \Delta T = 0$ is defined to be the Standard Model prediction
for $m_t = 173.1$ GeV and $m_{h^0} = 115$ GeV.
Results for the QDEE model are very similar.}
\end{minipage}
\begin{minipage}[]{0.49\linewidth}
\begin{flushright}
\includegraphics[width=7.7cm,angle=0]{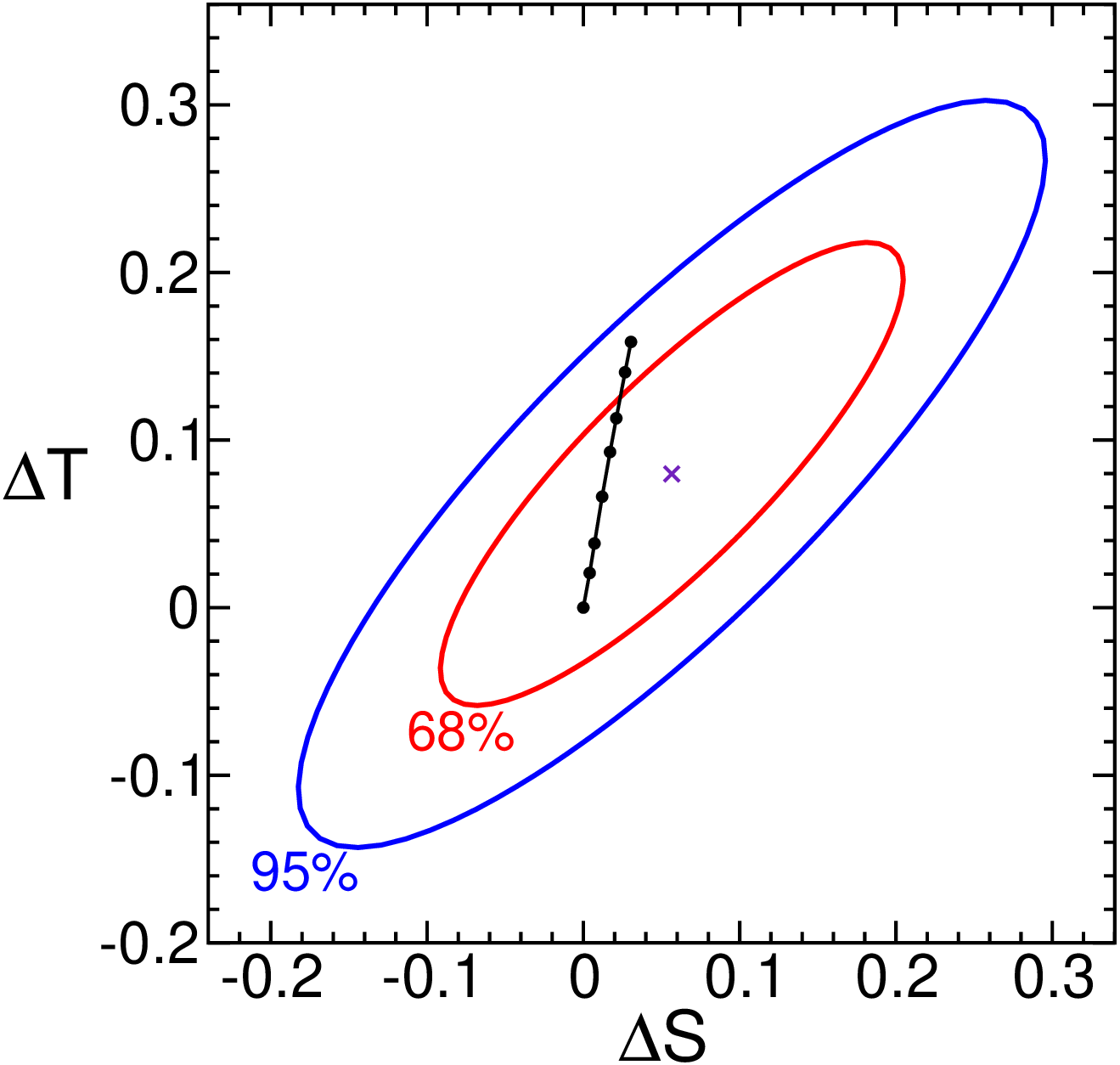}
\end{flushright}
\end{minipage}
\end{figure}

A more precise evaluation, using the formulas of \AppendixA,
is shown in figures \ref{fig:STforLND} and \ref{fig:STforQUE}, which
compares the experimental best fit and 68\% and 95\% CL ellipses
to the predictions from the models.
Note that in these figures
I do not include the contributions
from the ordinary MSSM superpartners, which are typically not very large
and which become small quadratically with large soft supersymmetry 
breaking masses. 
Figure \ref{fig:STforLND} shows the corrections for the LND model at the 
Yukawa coupling
fixed point $(k_N, h_N) = (0.765, 0)$, 
for varying $M_N = M_L = m_{\tau'} > 100$ GeV
as a line segment with dots at 
$m_{\tau'} = 100, 120, 150, 200, 250, 400$ GeV and $\infty$.
These contributions are due to the fermions $\nu_{1,2}', \tau'$, with
their scalar superpartners assumed heavy enough to decouple. Note that
in the LND model
$b'$ and $\tilde b_{1,2}'$ do not contribute to $S,T$ as defined above,
since they do not have Yukawa couplings to the Higgs sector. Figure
\ref{fig:STforLND} shows that even for $m_{\tau'}$ as small as 100 GeV,
the $S$ and $T$ parameters remain within the 68$\%$ CL ellipse, and can
even give a slightly better fit to the experimental results provided that
$m_{\tau'}\gsim 120$ GeV. If the Yukawa coupling $k_N$ is less than the
fixed point value, or if $M_L < M_N$, then the corrections to $S$ and $T$
are smaller, for a given $m_{\tau'}$. 

Figure \ref{fig:STforQUE} shows the corrections for the QUE model at the 
Yukawa coupling
fixed point $(k_U, h_U) = (1.050, 0)$, 
for varying $M_U = M_Q = m_{b'}$
as a line segment with dots at 
$m_{t'_1} = 275, 300, 350, 400, 500, 700, 1000$ GeV and $\infty$.
[For a comparison to the approximate formulas (\ref{eq:delTnum}) and
(\ref{eq:delSnum}), the appropriate values are
$M_F = m_{b'} \approx 355$, 381, 432, 483, 584, 786, 1088 GeV and $\infty$,
respectively.]
Here, I have included the contributions from the scalar
states $\tilde t_{1,2,3,4}$ and $\tilde b_{1,2}$, obtained for 
$m_{1/2} = 600$ GeV and $A_{k_U} = -0.4 m_{1/2}$ and $m_0 = 0$ and $b_Q = b_U =
-m_{1/2} M_Q$, using eqs.~(\ref{eq:QUEsfermions}) and 
(\ref{eq:defAs}). Smaller values of $m_{1/2}$ would imply a chargino lighter
than the LEP2 bound; see Table~\ref{table:gauginomasses}. 
From Figure \ref{fig:STforQUE} we see that a slightly
better fit than the Standard Model 
can be obtained for $m_{t_1'} \gsim 400$ GeV, but even for $m_{t_1'}$
as light as 275 GeV, the corrections remain within the 95\% CL 
ellipse. The corrections to $S$ and $T$ for a given $m_{t_1'}$ are even 
smaller (and so the fit is even better) if
any of the following conditions apply: the Yukawa coupling $k_U$ is below
its fixed point value, $m_{1/2}$ or $m_0$ is larger so that the new squarks
are heavier, or $M_Q \not= M_U$. For $t_1'$ masses less than about 400 GeV, the fit also improves slightly if $m_{h^0}$ is larger than 115 GeV.

I have also looked at the QDEE model at its fixed point 
$(k_D, h_D) = (1.043, 0)$,
with scalar squared-mass soft terms given by eq.~(\ref{eq:QDEEsfermions}) 
with $m_{1/2} = 600$ GeV. The results are nearly identical to those found in
the QUE model in figure \ref{fig:STforQUE} with the
values for $m_{t_1'}$ replaced by $m_{b_1'}$, and so are not depicted.

If one also included lower energy data as used in \cite{RPP}, the fits to 
$S$ and $T$ would be somewhat worse, so it is important to keep in mind 
that the results above are sensitive to the choice of following the LEP 
Electroweak Working Group \cite{LEPEWWG,LEPEWWG2} in using fits based on 
the $Z$-pole data and $m_t$ and $m_W$. With this caveat, one may conclude 
that the models considered here fit at least as well as the Standard Model, 
provided that the new quarks with large Yukawa couplings are heavier than 
roughly 400 GeV, and can do even better if the new squarks are heavy enough 
to decouple.

\section{Collider phenomenology of the extra fermions}
\label{sec:BR}
\setcounter{equation}{0}
\setcounter{footnote}{1}

The extra particles in the 
models discussed above will add considerable richness to the already 
complicated LHC phenomenology of the MSSM. A full discussion of the 
different signals, and how to disentangle them, is beyond the scope of 
the present paper, but it is likely that the most important 
distinguishing collider 
signals will arise from production of the new fermions, 
especially the new quarks. This is simply because of the relatively large 
production cross-section compared to the scalars, which are presumably
much heavier due to the effects of soft supersymmetry breaking masses. 
One can therefore expect signals from
direction pair production of the lightest new quark,
and possibly also from cascade decays of somewhat heavier fermions down
to them. 
For concreteness, I will concentrate on only the final states
from decays of the lightest new quark in each model.
In general, the lightest new quark and the lightest new lepton 
would be stable, were it not for mixing with the 
Standard Model fermions. At least some small 
such mixing is necessary to avoid a cosmological
disaster from unwanted heavy relics. If the mixing is very small, then 
the new fermions could be quasi-stable, with decay lengths on the scale
of collider detectors.
Then the collider signatures will involve 
particles that leave highly-ionizing and slow tracks in the detectors,
or feature macroscopic decay kinks or charge-changing tracks. 
These can be either the new charged leptons or hadronic bound states of 
the new quarks. Such signals have been discussed before in a variety of 
different model-building contexts; for some reviews, see 
refs.~\cite{Culbertson:2000am, Kraan:2004tz, Fairbairn:2006gg, 
Aad:2009wy}. 

In the following, I will assume that the mixing of the new fermions with
Standard Model fermions is large enough to provide for prompt decays.
Mixing of the new fermions with the first and second family Standard 
Model
fermions is highly constrained by flavor-changing neutral currents,
since the vector-like gauge quantum number assignments 
eliminate the GIM-type 
suppression.
Therefore, I will assume that the mixing is with the third 
Standard Model family, for which the constraints are much easier to 
satisfy.
Then the final states of the decays will always involve a single 
third-family quark or lepton, together with a $W$, $Z$, or $h^0$ boson. 
Below, I will discuss the possibilities for the branching ratios of the 
new quarks and leptons, and their dependence on the type of mixing. 

There are existing limits on the extra quarks coming from Tevatron, 
although these 
have mostly been found with assumed 100\% branching ratios for 
particular decay modes (which as we will see below is not 
necessarily likely). The current limits are, for prompt decays:
\begin{itemize} 
\item 
$m_{t'} > 311$ GeV for BR$(t' \rightarrow Wq) = 1$,
based on 2.8 fb$^{-1}$ \cite{CDF9446}
\item
$m_{b'} > 325$ GeV for BR$(b' \rightarrow Wt) = 1$,
based on 2.7 fb$^{-1}$ \cite{CDF9759}
\item
$m_{b'} > 268$ GeV for BR$(b' \rightarrow Zb) = 1$,
based on 1.06 fb$^{-1}$ \cite{CDF8737}
\item
$m_{b'} > 295$ GeV for BR$(b' \rightarrow Wt, Zb, h^0b) = 0.5, 0.25, 0.25$,
based on 1.2 fb$^{-1}$ \cite{CDFbp211}
\end{itemize}
and for quasi-stable quarks:
\begin{itemize}
\item $m_{t'} > 220$ GeV, based on $dE/dx$ for 90 pb$^{-1}$ 
at $\sqrt{s} = 1.8$ TeV \cite{Acosta:2002ju}
\item $m_{b'} > 190$ GeV, based on $dE/dx$ for 90 pb$^{-1}$ 
at $\sqrt{s} = 1.8$ TeV \cite{Acosta:2002ju}
\item $m_{b'} > 170$ GeV for 3mm $ < c\tau_{b'} < 20$mm,
based on 163 pb$^{-1}$ \cite{CDF7244}
\end{itemize}
Also, if the cross-section upper bound 
found from 
time-of-flight measurements with 1.0 fb$^{-1}$ 
in ref.~\cite{Aaltonen:2009kea} for stable top squarks also applies
to stable $t'$ quarks with no change
in efficiency, then I estimate a bound $m_{t'} \gsim 360$ GeV should be
obtainable, with a somewhat weaker bound for stable
$b'$ due to a lower detector efficiency.

\begin{figure}[!tp]
\begin{minipage}[]{0.42\linewidth}
\caption{\label{fig:cross}
Production cross-section for new quarks as a function of the mass,
for the Tevatron $p\bar p$ collisions at $\sqrt{s} = 1.96$ TeV, and for the LHC $pp$ collisions with
$\sqrt{s} = 7, 10, 12$, and 14 TeV. The graph was made at leading order using 
CTEQ5LO parton
distribution functions \cite{CTEQ5} with $Q = m_{q'}$ and 
applying a $K$ factor of
1.5 for LHC and 1.25 for Tevatron.}
\end{minipage}
\begin{minipage}[]{0.56\linewidth}
\begin{flushright}
\includegraphics[width=9.0cm,angle=0]{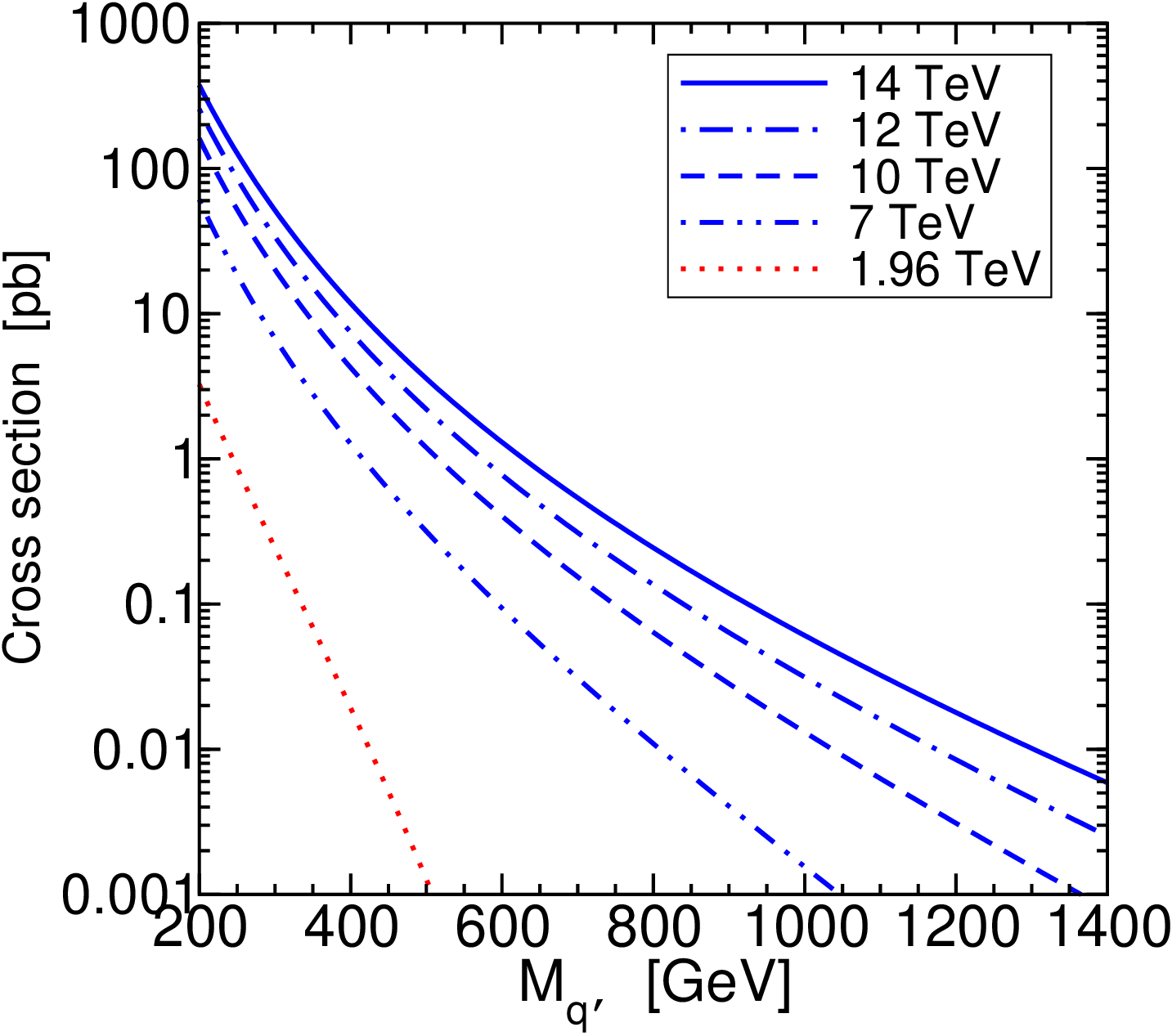}
\end{flushright}
\end{minipage}
\end{figure}

At hadron colliders, the production cross-section of the new quarks
is due to $gg$ and $q\overline q$ initial 
states and is mediated by the strong interactions, and so is nearly 
model-independent when expressed as
a function of the mass. The leading order cross-section is shown 
in figure \ref{fig:cross} for the Tevatron $p\overline p$ collider at
$\sqrt{s} = 1.96$ TeV and for the LHC $pp$ collider with 
$\sqrt{s} = 7, 10, 12$, and 14 TeV.%
Note the Tevatron will probably be unable to strengthen the existing 
constraints very significantly, at least for promptly decaying new 
quarks, due to the rather steep fall of the production cross-section with 
mass. At the LHC pair production of mostly vector-like quarks should 
provide a robust signal; see for example studies (in diverse other model
contexts) in refs.~\cite{Arik:1996qd, AguilarSaavedra:2005pv, Mehdiyev:2006tz, 
Skiba:2007fw, AguilarSaavedra:2009es, Holdom:2006mr}.
(Note that in the models under study here, there is no reason  
that the 
flavor-violating charged current couplings should be large enough to enable
a viable signal from 
single $q'$ production in association with a Standard Model fermion
through $t$-channel $W$ exchange, unlike in other
model contexts as studied in 
refs.~\cite{Han:2003wu, Perelstein:2003wd, Azuelos:2004dm, Dennis:2007tv, 
Contino:2008hi, Atre:2008iu, Mrazek:2009yu}.) The branching ratios and 
possible signals for the LND, QUE, and QDEE models are examined below.

\subsection{The LND model}
\label{subsec:LNDdecays}

In the LND model, the fermions consist of a $b'$, $\tau'$, and two
neutral fermions $\nu_1'$ and $\nu_2'$. The $\nu_1'$ is
always lighter than the $\tau'$. 
The fermions $b'$ and $\nu_1'$ can therefore decay only through their
mixing with the Standard Model fermions from the superpotential
\beq
W = 
-\epsd H_d q_3 \overline D 
+ \epsn H_u \ell_3 \overline N 
-\epse H_d L \overline e_3,
\label{eq:WLNDmix}
\eeq
where $\epsd$, $\epsn$, and $\epse$ are new Yukawa couplings that are assumed here to be small enough to 
provide mass mixings that can
be treated as perturbations compared to the other entries in the mass matrices.

\begin{figure}[!tp]
\begin{minipage}[]{0.49\linewidth}
\begin{flushright}
\includegraphics[width=8.0cm,angle=0]{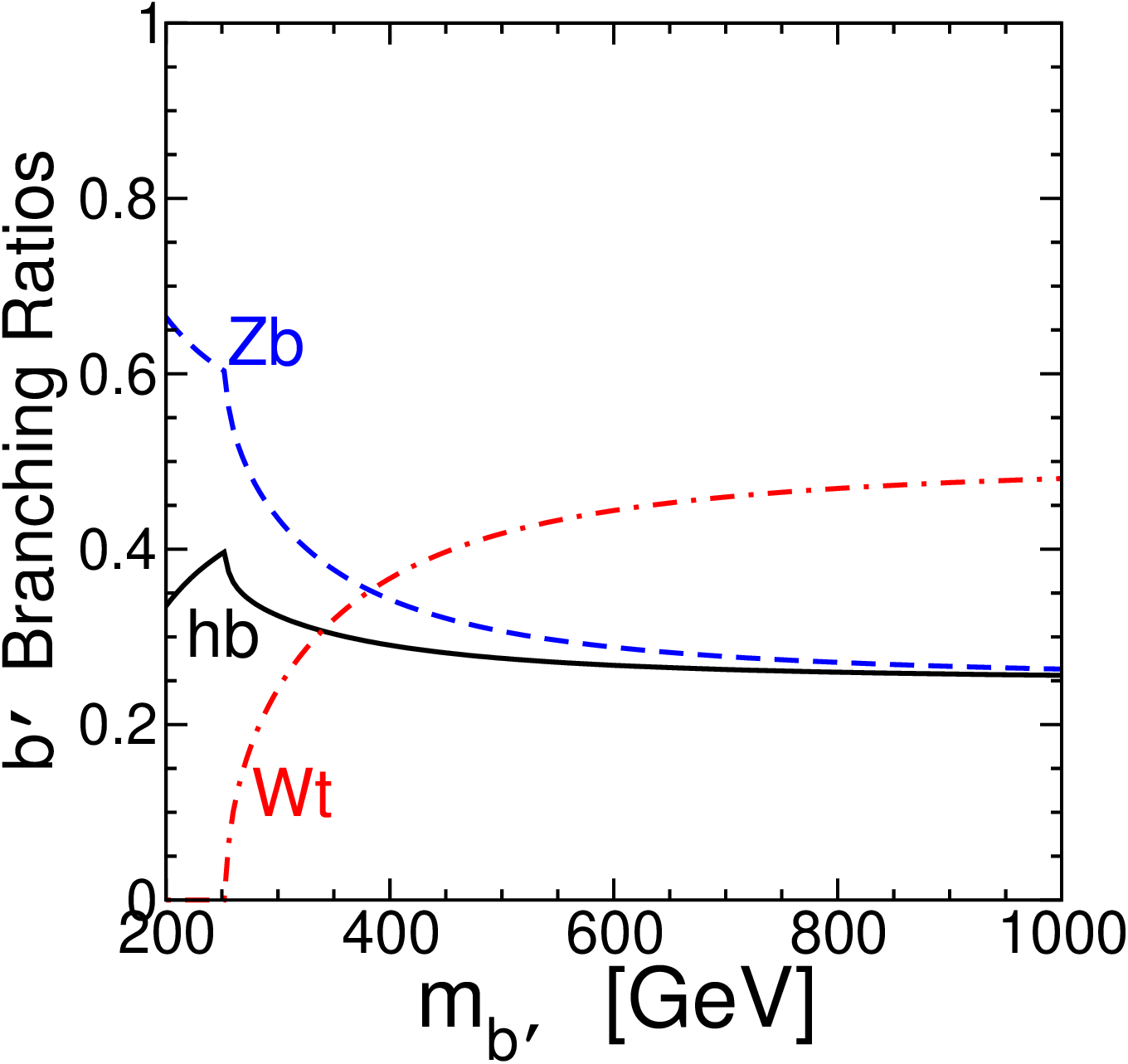}
\end{flushright}
\end{minipage}
\begin{minipage}[]{0.49\linewidth}
\begin{flushright}
\includegraphics[width=8.0cm,angle=0]{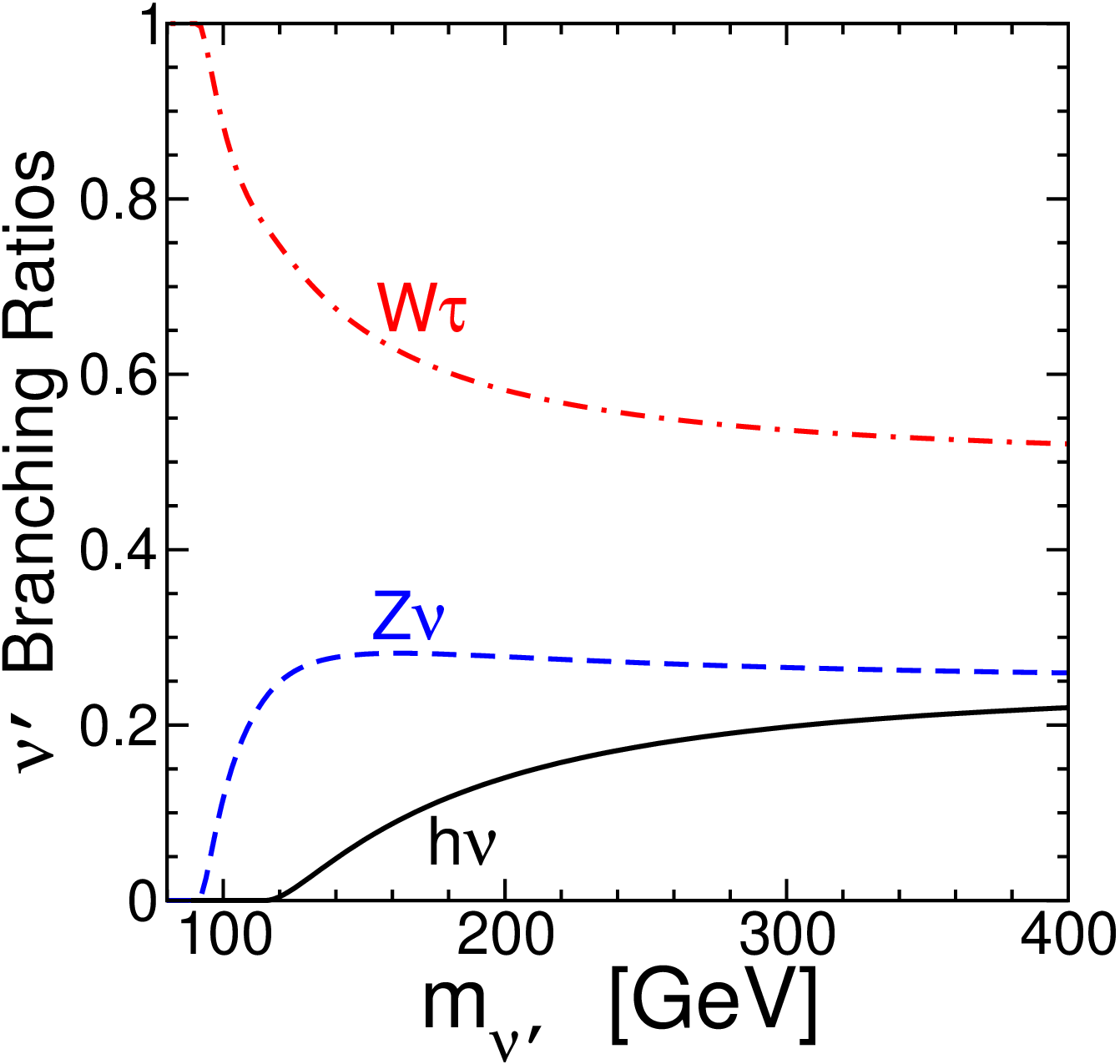}
\end{flushright}
\end{minipage}
\caption{\label{fig:BRLND}
The branching ratios of the lightest new quark $b'$ (left panel) and
the lightest new lepton $\nu_1'$ (right panel) in the LND model.
The $\nu_1'$ results assume that $\epsn \gg \epse$; if instead
$\epse \gg \epsn$ then $BR(\nu_1' \rightarrow W\tau) = 1$ (not shown).}
\end{figure}

First consider the decays of $b'$.
The mass matrix for the down-type quarks 
resulting from eqs.~(\ref{eq:WMSSM}),
(\ref{eq:WLND}), and (\ref{eq:WLNDmix}) is:
\beq
{\cal M}_d = \begin{pmatrix}
     M_D & 0 \cr
     \epsd v_d & y_b v_d \cr
     \end{pmatrix} ,
\label{eq:WLNDmassmix}
\eeq
with eigenstates $b$ and $b'$.
The $b'$ decay
can take place only through the $\epsd$ coupling, 
to final states $Wt$, $Zb$, and $h^0b$. Formulas for these decay widths 
are given in \AppendixB. 
To leading order, the branching ratios only depend on the mass of the 
$b'$, and the results are graphed in Figure \ref{fig:BRLND}.%
Note that in the limit of large $m_{b'}$, the branching ratios 
are ``democratic" between charged and neutral currents,
approaching $0.5$, $0.25$, and $0.25$ for 
$Wt$, $Zb$, and $h^0b$ 
respectively, in accord with
the Goldstone boson equivalence theorem. 
However, for smaller masses, kinematic suppression reduces the
$Wt$ branching ratio, so that, for example,
the three final states have comparable branching ratios for
$m_{b'}$ in the vicinity of 300 to 400 GeV. 

The LHC signals 
include $pp \rightarrow b_1' \bar b_1' \rightarrow
W^+ W^- t \bar t \rightarrow W^+ W^- W^+ W^- b \bar b$. When two 
same-charge $W$'s decay leptonically
and the other two $W$'s decay hadronically, this leads to a same-charge 
dilepton plus multi-jets 
(including two $b$ jets) plus missing transverse energy signal,
with a total branching ratio as high as 25\%. 
This signal is also the basis for the current Tevatron bound $m_{b'} > 325$ 
GeV,
but this assumes BR$(b' \rightarrow Wt) = 100$\%; 
since the actual branching ratio predicted by the LND model for that
mass range is more than a factor of 3 smaller, 
the model prediction for the signal 
in the channel that was searched is more than an order of magnitude 
smaller, and decreases sharply for lower $m_{b_1'}$. 
In over half of the other $b_1' \bar b_1'$ production events, there will 
be four or more $b$ jets, coming mostly from events with $h^0 b 
\rightarrow bb\bar b$ decays but also from $Zb \rightarrow b b \bar b$. 
The Tevatron limit \cite{CDF8737} of $m_{b'} > 268$ GeV from assuming 
BR($b' \rightarrow Zb) = 100$\% is in a mass range where the actual 
branching ratio is about $0.55$, so the actual predicted signal from the 
LND model is more than a factor of 3 smaller. The limit 
of $m_{b'} > 295$ GeV from 
\cite{CDFbp211}, a search which is motivated in part by 
\cite{Choudhury:2001hs,Bjorken:2002vt}, is based on the idealized large 
mass limit ``democratic" branching, but in the relevant mass range the 
model prediction has BR$(b' \rightarrow W t)$ more than a factor of 2 
smaller, and decreasing very rapidly for smaller $m_{b'}$, due to the 
kinematic suppression. The neutral current decays, including $Z \rightarrow 
\ell^+\ell^-$, could also play an important role at the LHC, see for 
example \cite{AguilarSaavedra:2009es} for a similar case.

The decay of $\nu_1'$ in the LND model is dependent on two
different mixing Yukawa couplings $\epsn$ and $\epse$.
The mass matrix for the neutral leptons in the 
$(L, N, \ell_3, \overline L, \overline N)$ basis 
resulting from 
eqs.~(\ref{eq:WMSSM}), 
(\ref{eq:WLND}) and
(\ref{eq:WLNDmix})
is 
\beq
\begin{pmatrix} 0 & {\cal M}_\nu \cr {\cal M}^T_\nu & 0 \end{pmatrix}, 
\quad
\mbox{where}
\quad
{\cal M}_\nu^T = \begin{pmatrix}
     M_L & h_N v_d & 0\cr
     k_N v_u & M_N & \epsn v_U \cr
     \end{pmatrix} ,
\label{eq:Nmassmix}
\eeq
with the masses of the Standard Model neutrinos neglected. The 
corresponding mass eigenstates are a Standard Model neutrino $\nu$ and two 
extra massive neutrino states $\nu_1'$ and $\nu_2'$. The mass matrix for 
the charged leptons is
\beq
{\cal M}_e = \begin{pmatrix}
             -M_L & \epse v_d \cr
             0 & y_\tau v_d
             \end{pmatrix}.
\label{eq:Emassmix}
\eeq
Formulas for the resulting 
decay widths for $\nu_1' \rightarrow W\tau$ and
$Z\nu$ and $h^0 \nu$ are given in \AppendixB. 
If one assumes that $\epse \gg \epsn$,
then the decay $\nu_1' \rightarrow W \tau$ has a nearly 100\% branching ratio. 
If the opposite limit applies,
$\epsn \gg \epse$, 
then the branching ratios as a function of $m_{\nu_1'}$ are as shown
in the right panel of Figure \ref{fig:BRLND}. Note that in the limit of large 
$m_{\nu_1'}$, the branching ratios for $W\tau$, $Z\nu$, 
and $h^0 \nu$ asymptote
to 0.5, 0.25, 0.25 respectively when $\epsn$ dominates, 
again in accordance with Goldstone boson equivalence with equal charged
and neutral currents.
So, depending on which Yukawa coupling dominates, one could have interesting
hadron collider signatures from $\nu_1' \bar \nu_1'$ production, 
such as $W^+ W^- \tau^+ \tau^-$, and 
$h^0 h^0 + \missET$, and $ZZ + \missET$, and 
$Wh^0 + \missET$ and $Zh^0 + \missET$. 
So far, there are no 
published limits specifically
on $m_{\nu'}$ based on collider pair production with these final states.
If $M_L \lsim M_N$ in this model, then $\tau'$ will be not
much heavier than $\nu_1'$, and so there will be additional contributions
to the signal from $\tau'\nu_1'$ production and 
$\tau^{\prime +}\tau^{\prime -}$ production, followed by 
$\tau' \rightarrow W^{(*)} \nu'$.
It should also be noted that production of $\nu_{1,2}'$ and $\tau_1'$
might well be dominated by cascade decays from heavier strongly interacting
superpartners.

\subsection{The QUE model}
\label{subsec:QUEdecays}

In the QUE model, the lightest of the new quarks is always the 
charge $2/3$ quark $t_1'$. After being pair-produced at hadron colliders,
it can decay due to mixing with the
Standard Model fermions through the superpotential
\beq
W = 
  \epsu H_u q_3 \overline U 
+ \epsup H_u Q \overline u_3 
- \epsd H_d Q \overline d_3 
,
\label{eq:WQUEmix}
\eeq
where $\epsu$, $\epsup$, and $\epsd$ 
are new Yukawa couplings that are assumed here to be small enough to treat as
perturbations compared to other entries in the mass matrices.
The resulting mass matrices for the up-type quarks and down-type quarks are 
\beq
{\cal M}_u = \begin{pmatrix}
     M_Q & k_U v_u & \epsup v_u \cr
     h_U v_d & M_U & 0\cr
     0 &  \epsu v_u & y_t v_u \cr
\end{pmatrix}
,
\qquad\>
{\cal M}_d = \begin{pmatrix}
    -M_Q & \epsd v_d \cr
     0 & y_b v_d \cr
\end{pmatrix},
\label{eq:QUEmassmix}
\eeq
with mass eigenstates $t, t_1', t_2'$ and $b, b'$ respectively.
Formulas for the resulting decay widths for $t_1'$ to $Wb$, $Zt$, and 
$h^0t$ are presented in \AppendixB. I will concentrate on the three cases 
where one of 
the mixing Yukawa couplings 
in eq.~(\ref{eq:WQUEmix}) dominates over the other two. 
The branching ratios depend on the mass of $t_1'$ and on the type of 
mixing. If $\epsd$ 
provides the dominant effect, then the decays are 
dominantly charged-current, or ``$W$-philic", with 
BR$(t_1' \rightarrow Wb) = 1$. This is the scenario for which the 
Tevatron limit is now $m_{t'} > 311$ GeV \cite{CDF9446}. If instead 
$\epsup$ dominates, then the decays are 
dominantly neutral-current, or ``$W$-phobic"; in the limit of 
large $m_{t'}$, the branching ratios asymptote to BR$(t_1' \rightarrow 
Wb) 
= 0$ and BR$(t_1' \rightarrow Zt) =$ BR$(t_1' \rightarrow h^0 t) = 0.5$. 
Finally, if $\epsu$ dominates, the the decays are ``democratic", with 
branching ratios for $Wb$, $Zt$, and $h^0$ approaching 0.5, 0.25, and 
0.25 respectively in the large $m_{t_1'}$ limit. 
\begin{figure}[!tp]
\begin{minipage}[]{0.47\linewidth}
\begin{flushright}
\includegraphics[width=7.8cm,angle=0]{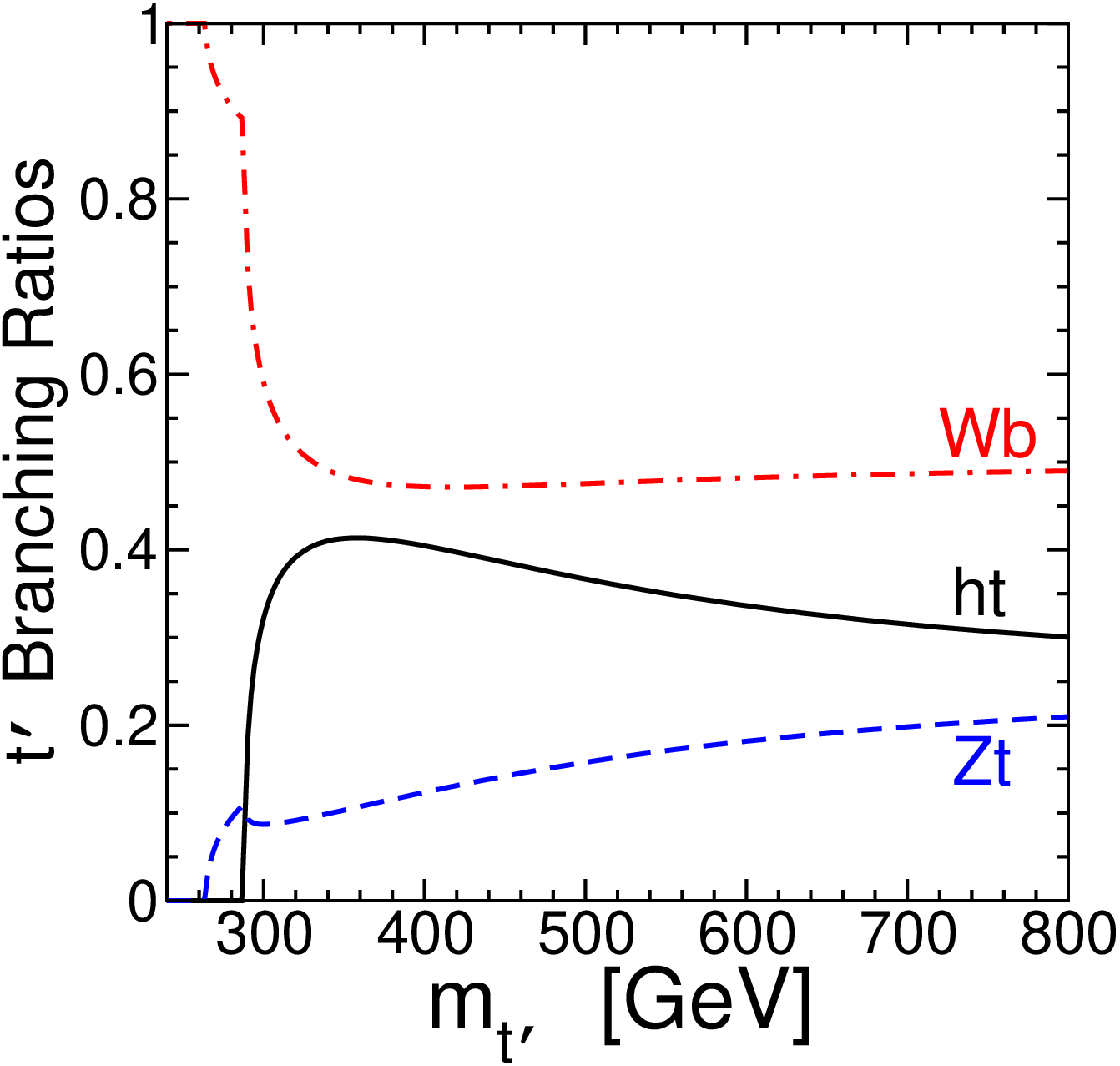}
\end{flushright}
\end{minipage}
\begin{minipage}[]{0.52\linewidth}
\begin{flushright}
\includegraphics[width=7.8cm,angle=0]{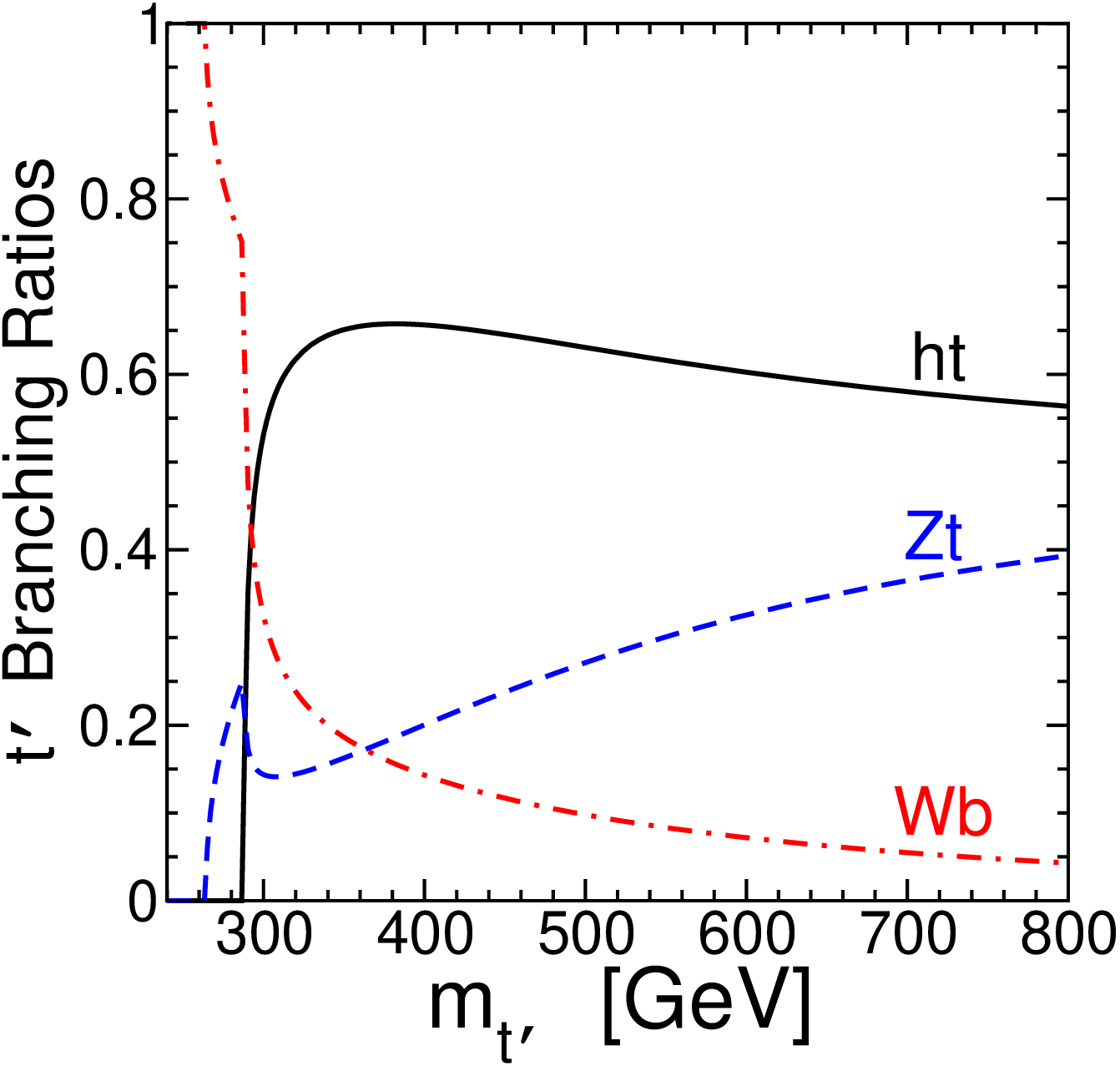}
\end{flushright}
\end{minipage}
\caption{\label{fig:BRQUE}
Branching ratios for the lightest extra quark, $t_1'$, in the QUE model
with $M_Q = M_U$,
to final states $Wb$, $Zt$, 
and $h^0 t$, as a function of $m_{\tilde t_1}$.
The left panel shows the ``democratic" case that arises when $\epsu$ 
dominates (with equal charged and neutral currents),
and the right panel shows the ``$W$-phobic" (mostly neutral current) 
case that 
arises when $\epsup$ dominates.
In the ``$W$-philic" case that arises when $\epsd$ dominates, then 
$BR(t'\rightarrow Wb) = 1$ (not shown).}
\end{figure}
Numerical results are shown 
in Figure \ref{fig:BRQUE} as a function of $m_{t_1'}$, for the case
that $k_U$ is at its fixed point value, and $h_U=0$, and $M_Q = M_U$.
(The results are only mildly sensitive to the last two assumptions.)
By taking the different mixing Yukawa couplings $\epsu$, $\epsup$, and 
$\epsd$ to be comparable, one can get essentially any result one wants 
for the branching ratios, but it seems reasonable to assume that one of 
the individual mixing Yukawa couplings dominates in the absence of some 
organizing principle. So the possible signatures will include $W^+W^-b\bar b$,
(similar to the Standard Model $t\bar t$ signature, but with larger 
invariant masses; see \cite{
Holdom:2006mr,
Arik:1996qd,
AguilarSaavedra:2005pv,
Skiba:2007fw,
AguilarSaavedra:2009es} for recent studies of comparable signals), and
$ZZt \bar t$ and $h^0 h^0 t \bar t$, etc.
If $M_Q \lsim M_U$ in this model, then the $b'$ will be not much heavier
than the $t'$, and one should expect an additional component of the signal
from $b'\bar t'$ and $b' \bar b'$ production, 
followed by $b' \rightarrow W^{(*)} t'$.
 
\begin{figure}[!tp]
\begin{minipage}[]{0.47\linewidth}
\caption{\label{fig:BRtaup}
Branching ratios for $\tau'$ decays to $W\nu$, $Z\tau$, and $h^0 \tau$ in the 
QUE and QDEE models, as a function of $m_{\tau'}$.}
\end{minipage}
\begin{minipage}[]{0.52\linewidth}
\begin{flushright}
\includegraphics[width=8.0cm,angle=0]{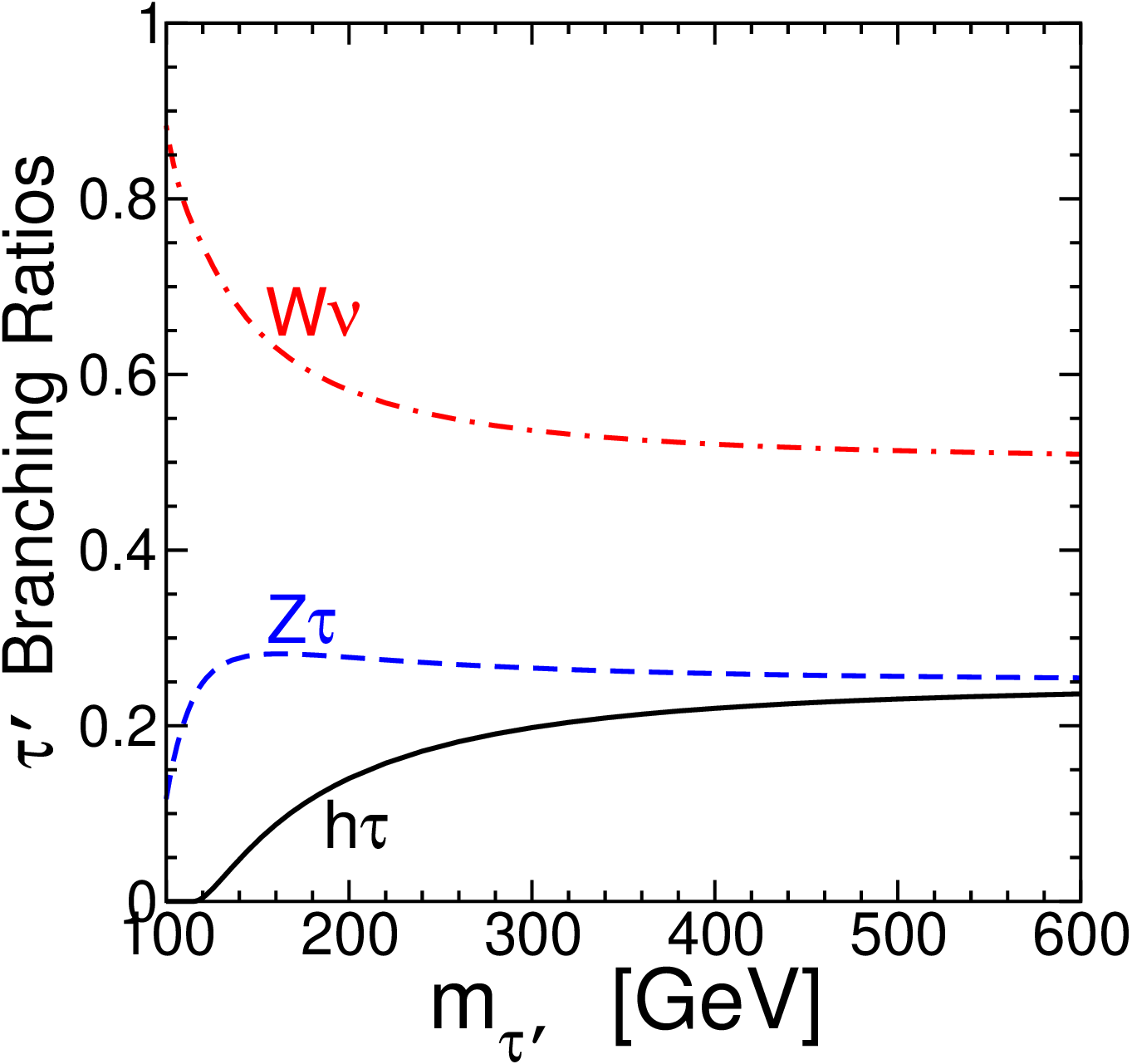}
\end{flushright}
\end{minipage}
\end{figure}

The $\tau'$ in the QUE model mixes with the Standard Model $\tau$ lepton
through a superpotential term:
\beq
W = -\epse H_d \ell_3 \overline E.
\label{eq:WQUEmixE}
\eeq
The mass matrix for the charged leptons resulting from this and 
eqs.~(\ref{eq:WMSSM}) and (\ref{eq:WQUE}) is:
\beq
{\cal M}_e = \begin{pmatrix}
     M_E & 0 \cr
     \epse v_d & y_\tau v_d \cr
     \end{pmatrix} ,
\label{eq:QUEmassmatE}
\eeq
with mass eigenstates $\tau$ and $\tau'$.
It follows that $\tau'$ can decay to $W\nu$, $Z\tau$, and $h^0 \tau$, 
with decay widths that are
computed in \AppendixB. 
Because there is only one relevant Yukawa mixing term, the branching 
ratios depend only on $m_{\tau'}$. They are shown in Figure 
\ref{fig:BRtaup}, assuming $m_{h^0} = 115$ GeV.
The largest branching ratio for $\tau'$ is always to $W\nu$, and in the 
large 
$m_{\tau'}$ limit, Goldstone boson equivalence provides that the $W\nu$, 
$Z\tau$, and $h^0\tau$ branching ratios approach $0.5$, $0.25$, and 
$0.25$, respectively. The most immediately relevant searches at 
hadron colliders will be in the mass range of $m_{\tau'}$ just above 
100 GeV, where the electroweak pair-production cross-section can be 
sufficiently large, and limits do not presently exist.
However, note that the appearance of $\tau_1'$
could easily be dominated by cascade decays from heavier strongly interacting
superpartners.

\subsection{The QDEE model}
\label{subsec:QDEEdecays}

\begin{figure}[!tp]
\begin{minipage}[]{0.47\linewidth}
\begin{flushright}
\includegraphics[width=8.0cm,angle=0]{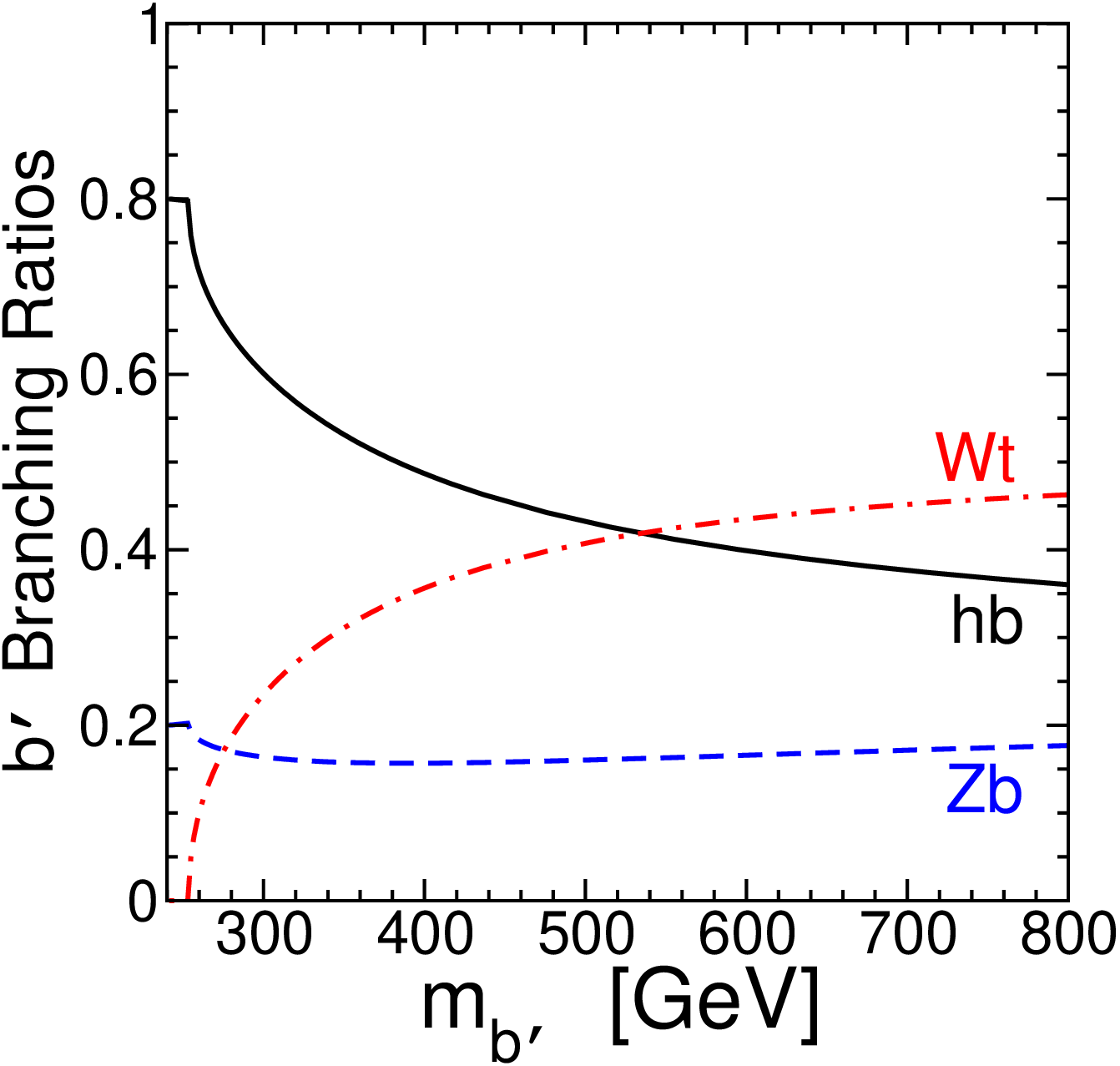}
\end{flushright}
\end{minipage}
\begin{minipage}[]{0.52\linewidth}
\begin{flushright}
\includegraphics[width=8.0cm,angle=0]{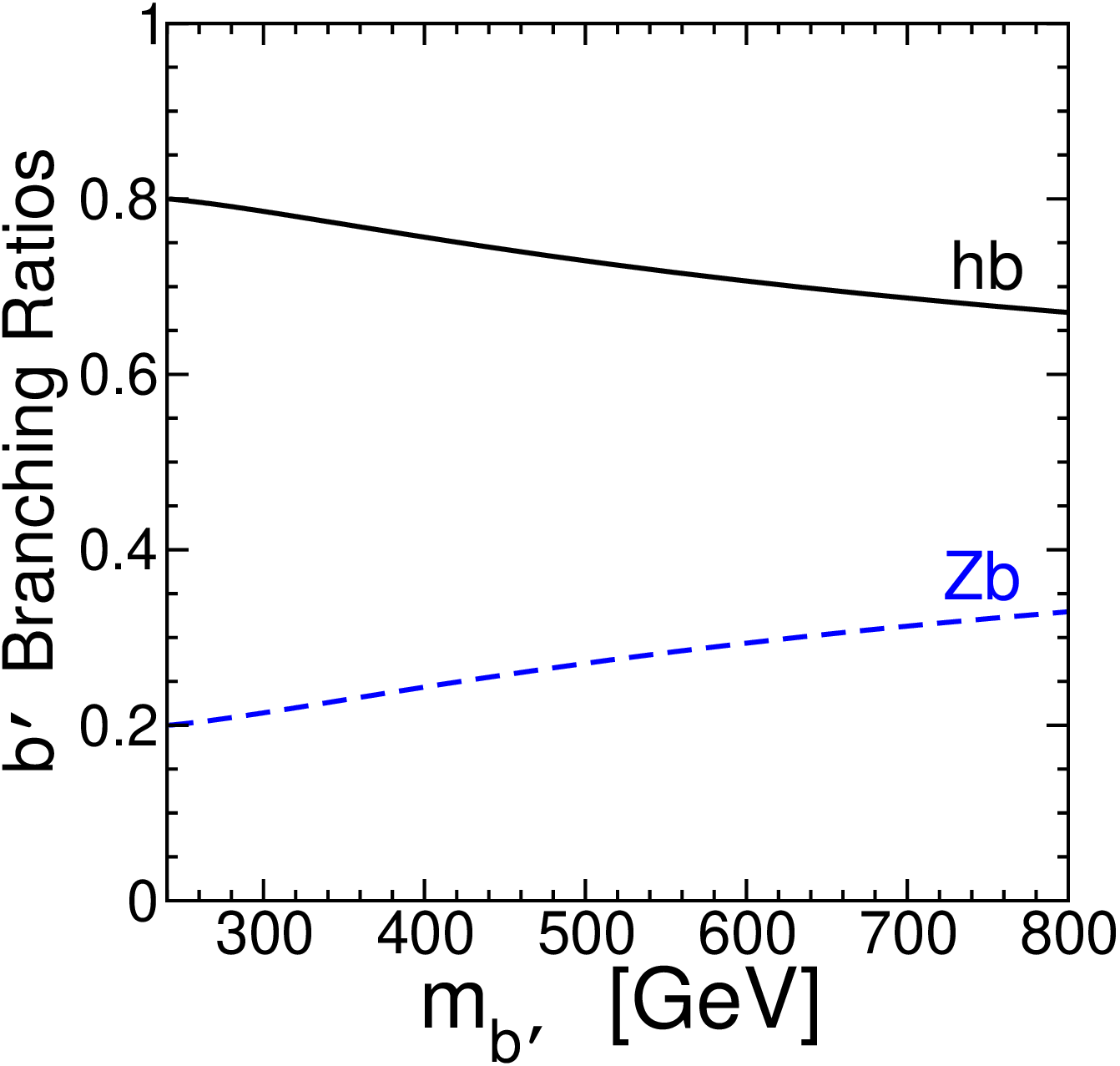}
\end{flushright}
\end{minipage}
\caption{\label{fig:BRQDEE}
Branching ratios for the lightest extra quark, $b_1'$, in the QDEE model
with $M_Q = M_D$, to final states $Wt$, $Zb$, and $h^0 b$.
The left panel shows the ``democratic" case that $\epsd$ dominates,
and the right panel shows the ``$W$-phobic" case that $\epsdp$ dominates,
leading to mostly neutral-current decays.
In the ``$W$-philic" case that $\epsu$ dominates leading to mostly charged-current decays, then
$BR(b_1' \rightarrow Wt) = 1$ (not shown).}
\end{figure}

In the QDEE model, the new fermions consist of a $b_1'$, $b_2'$, $t'$, 
and $\tau_1'$, $\tau_2'$. 
In this model, the lighter charge $-1/3$ quark $b_1'$ is
always lighter than the $t'$. 
The decays of $b'_1$ in the QDEE model are brought about by
superpotential mixing terms with third-family quarks:
\beq
W = 
  -\epsd H_d q_3 \overline D 
-\epsdp H_d Q \overline d_3 
+ \epsu H_u Q \overline u_3.
\label{eq:WQDEEmix}
\eeq
In the gauge eigenstate basis, the resulting mass matrices for the down-type quarks and up-type quarks are 
\beq
{\cal M}_d = \begin{pmatrix}
     M_Q & k_D v_u & 0 \cr
     h_D v_d & M_D & \epsd v_d \cr
     \epsdp v_d & 0 & y_b v_d \cr
\end{pmatrix}
,
\qquad\>
{\cal M}_u = \begin{pmatrix}
    -M_Q & 0 \cr
     \epsu v_u & y_t v_u \cr
\end{pmatrix}.
\label{eq:QDEEmassmix}
\eeq
with mass eigenstates $b, b_1', b_2'$ and $t, t'$ respectively. Formulas 
for the resulting decay widths for $b_1'$ to $Wt$, $Zb$, and $h^0b$ 
are given in \AppendixB. As in the case of the QUE model, I will 
consider the three cases where one of the mixing Yukawa couplings in 
eq.~(\ref{eq:WQDEEmix}) dominates over the other two. 
Then the branching ratios depend on 
the mass of $b_1'$ and on the type of mixing. If $\epsu$ provides the 
dominant effect, then the decays are dominantly charged-current, or
``$W$-philic", with BR$(b_1' 
\rightarrow Wt) = 1$ provided that it is kinematically allowed. 
The resulting signal at hadron colliders will be 
$b_1' \bar b_1' \rightarrow W^+W^-t \bar t \rightarrow 
W^+W^+W^-W^-b\bar b$.
This is 
the scenario for which the Tevatron limit is presently $m_{b'} > 325$ GeV 
\cite{CDF9759}, based on the same-charge dilepton plus $b$-jets signal 
already mentioned above for the LND model. If instead $\epsdp$ is 
dominant, then the decays are dominantly neutral-current, or
``$W$-phobic", with BR$(b_1' \rightarrow 
Wt) = 0$; in the limit of large $m_{b'}$, the branching ratios slowly 
approach BR$(b_1' \rightarrow Zb) =$ BR$(b_1' \rightarrow h^0 b) = 0.5$, 
but with $h^0b$ larger for finite masses. Finally, if $\epsd$ is 
dominant, the the decays are ``democratic", with branching ratios for 
$Wt$, $Zb$, and $h^0b$ approaching 0.5, 0.25, and 0.25 respectively in 
the large $m_{b_1'}$ limit. 
The predicted branching ratios are shown in Figure 
\ref{fig:BRQDEE} as a function of $m_{b_1'}$ for the latter two cases,
assumings $k_D$ is at its fixed point value, and $h_D=0$ and
$M_Q = M_D$. 
(However, it should be noted that, unlike in the QUE model case, 
the results shown are somewhat sensitive to
the last of these assumptions.)
Note that in the ``democratic" case, the branching ratios are similar to 
what one obtains for the $b'$ of the LND model. 
The CDF limit $m_{b'} > 295$ GeV 
was obtained in the idealized case
of branching ratios obtained in the high mass limit, but for finite 
$m_{b'}$, the actual
BR$(b' \rightarrow Wt)$ is much smaller and BR$(b' \rightarrow h^0 b)$ is 
larger. 
In contrast, the 
same-charge dilepton signal from $b_1' b_1' \rightarrow W^+W^-t\bar t$ is 
turned off in the ``$W$-phobic" case, where the largest overall branching 
ratio is typically to six b quarks, 
yielding the interesting possible signal 
$b_1' b_1' \rightarrow h^0 h^0b\bar b 
\rightarrow bbb \bar b \bar b \bar b$.
Decays to leptons through $Z$ bosons are unfortunately suppressed by both
small BR$(Z \rightarrow \ell^+\ell^-)$ and small BR$(b_1' \rightarrow Z b)$
in this case.
If $M_Q \lsim M_D$ in this model, then the $t'$ will be not much heavier
than the $b'$, and one should expect an additional component of the signal
from $t'\bar b'$ and $t' \bar t'$ production, 
followed by $t' \rightarrow W^{(*)} b'$.

For $\tau'_1$ in the QDEE model, the 
branching ratio situation is essentially 
the same as for the QUE model as discussed above.

\section{Outlook}
\label{sec:outlook}

In this paper, I have studied supersymmetric models that have vector-like
fermions that are consistent with perturbative gauge coupling unification
and have large Yukawa couplings that can significantly raise the Higgs mass
in supersymmetry. Some of the more important features found for these models are:
\begin{itemize}
\item There are three types of models consistent with perturbative gauge coupling unification and all
new particles near the TeV scale. The first type (LND) contains 
up to three copies of the ${\bf 5}+{\bf \overline {5}}$
of $SU(5)$. The second type (QUE) contains a 
${\bf 10} + {\bf \overline{10}}$ of $SU(5)$. The third
type (QDEE) is not classifiable in terms of complete representations of 
$SU(5)$, but consists of the
fields $Q,D,E,E$ and their conjugates.
\item A complete vector-like family (i.e. a ${\bf 16} + {\bf \overline{16}}$
of $SO(10)$) could also be entertained, 
but was not considered here because a multi-loop renormalization group analysis shows that this would forfeit perturbative unification and 
high-scale control
unless (at least some of) the new particles are much heavier than 1 TeV.
\item The constraints imposed by oblique 
corrections to electroweak observables
are rather mild, especially in comparison to the corresponding 
constraints on a chiral fourth family, and are easily accommodated by 
present data as long as the new quarks with Yukawa couplings 
are heavier than about 400 GeV, and perhaps considerably lower.
\item The model framework is consistent with the hypothesis that 
gaugino masses dominate soft supersymmetry breaking near the unification 
scale, without problems from sleptons being too light as is the case in 
so-called 
mSUGRA models.
\item The lightest Higgs mass can be substantially raised in the QUE and 
QDEE models if the Yukawa couplings are near their fixed points.
However,
the extent of this is limited if one takes seriously the prediction for
the fixed point behavior of the scalar trilinear couplings, which limits
the mixing in the new squark sector.
For example, if the new quarks are at $M_F = 400$ GeV, and their
scalar partners have an average mass of $M_S = 1000$ GeV, then
one finds an increase in $m_{h^0}$ of up to about 15 GeV
(see Figure \ref{fig:deltamh}).
For larger $M_S$, this contribution increases, but at the expense 
of apparently more severe fine-tuning of the electroweak scale. 
\item Despite the sizable positive contribution to  the lightest Higgs, 
the contributions to the
$\mu$ parameter are also raised, so it is difficult to make any 
unambiguous claim for 
an improvement in
the supersymmetric little hierarchy problem.
\item The new fermions can decay through any mixture of neutral and charged
currents to third-family fermions and $W,Z,h^0$ weak bosons, 
but with different combinations correlated to the possible 
superpotential couplings that mix the new
fermions with the Standard Model ones. 
\item Existing bounds from direct searches at
the Tevatron do not significantly constrain the parameter space of these models
after precision electroweak constraints are taken into account.
\end{itemize} 
The collider phenomenology of the MSSM augmented by the new particles in 
these models should be both rich and confusing, leading to a difficult 
challenge at the LHC and beyond in deciphering the new discoveries.


\section*{Appendix A: Contributions to precision
electroweak parameters}
\label{appendixPEW} \renewcommand{\theequation}{A.\arabic{equation}}
\setcounter{equation}{0}
\setcounter{footnote}{1}

This Appendix gives formulas for the contributions of the new chiral
supermultiplets to the Peskin-Takeuchi precision electroweak parameters 
\cite{Peskin:1991sw}.
For convenience I will follow the notations 
and conventions of \cite{Martin:2004id}. 
The oblique parameters $S$ and $T$ are
defined in terms of electroweak vector boson self-energies by
\beq
\frac{\alpha S}{4 s_W^2 c_W^2} &=& \Bigl [
\Pi_{ZZ}(M_Z^2) - \Pi_{ZZ}(0) 
- \frac{c_{2W}}{c_W s_W} \Pi_{Z\gamma}(M_Z^2) - \Pi_{\gamma\gamma}(M_Z^2)
\Bigr ]/M_Z^2
,
\\
\alpha T &=& \Pi_{WW}(0)/M_W^2 - \Pi_{ZZ}(0)/M_Z^2
.
\eeq 
In the following, the one-loop integral functions $G(x)$, $H(x,y)$, 
$B(x,y)$, and $F(x,y)$
are as defined in ref.~\cite{Martin:2004id}, 
and particle names should be understood to stand for the 
squared mass when used as an argument of one of these functions, which
also have an implicit argument $s$ which is identified with
the invariant mass of the self-energy function in which they appear.

\begin{center}{\em 1.~Corrections to electroweak vector boson 
self-energies in 
the LND model}
\end{center}

For the LND model, define the gauge eigenstate new 
neutral lepton mass matrix by
\beq
{\cal M}_\nu = \begin{pmatrix} M_L & k_N v_u \cr  
                    h_N v_d & M_N \cr 
    \end{pmatrix}
,
\eeq
and unitary mixing matrices $L$ and $R$ by
\beq
L^* {\cal M}_\nu R^{\dagger} = {\rm diag}(m_{\nu'_1}, m_{\nu'_2}),
\eeq
and note $m_{\tau'} = M_L$.
Then the $(\nu_1', \nu_2', \tau')$
fermion contributions to the electroweak vector boson
self-energies are:
\beq
\Delta \Pi_{\gamma\gamma} &=& 
-\frac{N_c}{16 \pi^2} 2 g^2 s_W^2 e_e^2 G(\tau'),
\\
\Delta \Pi_{Z\gamma} &=& 
-\frac{N_c}{16 \pi^2} g s_W 
e_e (g^{Z}_{\tau' \tau^{\prime\dagger}}  - 
g^{Z}_{\bar \tau' \bar \tau^{\prime\dagger}}) G(\tau')  ,
\\
\Delta \Pi_{ZZ} &=& 
-\frac{N_c}{16 \pi^2} \biggl [
(|g^{Z}_{\tau' \tau^{\prime\dagger}}|^2 + 
|g^{Z}_{\bar \tau' \bar \tau^{\prime\dagger}}|^2) G(\tau') 
\nonumber \\ && + 
\sum_{i,j=1}^2 (|g^{Z}_{\nu_i' \nu_j^{\prime\dagger}}|^2 + 
|g^{Z}_{\bar \nu_i' \bar \nu_j^{\prime\dagger}}|^2) H(\nu'_i, \nu'_j) 
-4 {\rm Re}(
g^{Z}_{\nu_i' \nu_j^{\prime\dagger}} 
g^{Z}_{\bar \nu_i' \bar \nu_j^{\prime\dagger}}) 
m_{\nu'_i} m_{\nu'_j} B(\nu'_i, \nu'_j)
\biggr ]
,
\\
\Delta \Pi_{WW} &=& 
-\frac{N_c}{16 \pi^2} \sum_{i=1,2} \left [
(|g^{W}_{\nu_i' \tau^{\prime\dagger}}|^2 + 
|g^{W}_{\bar \nu_i' \bar \tau^{\prime\dagger}} |^2) H(\tau', \nu'_i) 
 -4 {\rm Re}(g^{W}_{\nu_i' \tau^{\prime\dagger}} 
 g^{W}_{\bar \nu_i' \bar \tau^{\prime\dagger}}) 
 m_{\tau'} m_{\nu'_i} B(\tau', \nu_i')
\right ]
,\phantom{xxx}
\eeq
where $N_c = 1$ and $e_e = -1$ and 
the massive vector boson couplings with the new leptons are:
\beq
g^{Z}_{\nu_i' \nu_j^{\prime\dagger}} &=& 
\frac{g}{2 c_W}  L_{i1}^* L_{j1} ,
\qquad\qquad
g^{Z}_{\bar \nu_i' \bar \nu_j^{\prime\dagger}} 
=
-\frac{g}{2 c_W} R_{i1}^* R_{j1},
\\ 
g^{Z}_{\tau' \tau^{\prime\dagger}} &=& 
-g^{Z}_{\bar \tau' \bar \tau^{\prime\dagger}}  =
\frac{g}{c_W} \left (-\frac{1}{2} - e_e s_W^2 \right ) ,
\\ 
g^{W}_{\nu_i' \tau^{\prime\dagger}} &=& g L_{i1}^*/\sqrt{2},
\qquad\qquad
g^{W}_{\bar \nu_i' \bar \tau^{\prime\dagger}} 
=
-g R_{i1}^* /\sqrt{2}.
\eeq

To obtain the $(\tilde \nu_{1,2,3,4}', \tilde \tau_{1,2}')$
scalar contribution, consider the new sneutrino
squared-mass matrix:
\beq
M^2_{\tilde \nu} = M^2_\nu + 
\begin{pmatrix}
m^2_L + \Delta_{\frac{1}{2},0} 
   & 0 & b_L^* & a_{k_N}^* v_u - \mu k_N v_d \cr
0 & m^2_{N}  & 
   a_{h_N}^* v_d -  \mu h_N v_u & b_N^* \cr
b_L & a_{h_N} v_d -  \mu^* h_N v_u & m^2_{\overline L} + 
\Delta_{-\frac{1}{2},0}  & 0 
\cr
a_{k_N} v_u - \mu^* k_N v_d & b_N & 0 & m^2_{\overline N} 
\cr
\end{pmatrix}
,
\eeq
where the supersymmetric part (also equal to the fermion squared-mass matrix) is:
\beq
M^2_\nu = 
\begin{pmatrix}
{\cal M}_\nu {\cal M}_\nu^\dagger & 0 \cr
0 & {\cal M}_\nu^\dagger {\cal M}_\nu
\end{pmatrix} 
.
\eeq
Also, the new charged slepton squared-mass matrix is given by:
\beq
M^2_{\tilde e} =
\begin{pmatrix}
M_L^2 + m^2_L + \Delta_{-\frac{1}{2},-1} &  -b_L^*\cr 
-b_L & M_L^2 + m^2_{\overline L} + \Delta_{\frac{1}{2},1}\cr
\end{pmatrix}
.
\eeq
Now define unitary
scalar mixing matrices $U$ and $V$ by:
\beq
U M^2_{\tilde \nu} U^{\dagger} &=& 
{\rm diag}(m^2_{\tilde \nu'_1}, 
m^2_{\tilde \nu'_2},m^2_{\tilde \nu'_3},m^2_{\tilde \nu'_4})
,
\qquad\qquad
V M^2_{\tilde e} V^{\dagger} 
=
{\rm diag}(m^2_{\tilde \tau'_1}, 
m^2_{\tilde \tau'_2}).
\eeq
Then the scalar contributions to the vector boson self-energies are:
\beq
\Delta \Pi_{\gamma\gamma} &=& 
\frac{N_c}{16 \pi^2} g^2 s_W^2 
e_e^2 \sum_{i=1}^2 F(\tilde \tau'_i, \tilde \tau'_i),
\\
\Delta \Pi_{Z\gamma} &=& 
\frac{N_c}{16 \pi^2} g s_W e_e \sum_{i=1}^2 
g^{Z}_{\tilde \tau_i' \tilde \tau_i^{\prime *}} 
F(\tilde \tau'_i, \tilde \tau'_i)
,
\\
\Delta \Pi_{ZZ} &=& 
\frac{N_c}{16 \pi^2} \left [ 
 \sum_{i,j=1}^4 |g^{Z}_{\tilde \nu_i' \tilde \nu_j^{\prime *}} |^2  
 F(\tilde \nu'_i, \tilde \nu'_j)
+\sum_{i,j=1}^2 |g^{Z}_{\tilde \tau_i' \tilde \tau_j^{\prime *}} |^2  
F(\tilde \tau'_i, \tilde \tau'_j)
\right ]
,
\\
\Delta \Pi_{WW} &=& 
\frac{N_c}{16 \pi^2} \sum_{i=1}^2 \sum_{j=1}^4
|g^{W}_{\tilde \tau_i' \tilde \nu_j^{\prime *}}|^2 
F(\tilde \tau'_i, \tilde \nu'_j),
\eeq 
where the vector boson couplings with the new sleptons are:
\beq
g^{Z}_{\tilde \nu_i' \tilde \nu_j^{\prime *}} &=& 
\frac{g}{2c_W} (U_{i1}^* U_{j1} + U_{i3}^* U_{j3}) 
,
\qquad\qquad
g^{Z}_{\tilde \tau_i' \tilde \tau_j^{\prime *}} =  
\frac{g}{c_W} \left (
-\frac{1}{2} - e_e s_W^2 \right )\delta_{ij}
,
\\
g^{W}_{\tilde \tau_i' \tilde \nu_j^{\prime *}} &=& 
g ( V_{i1}^* U_{j1} - V_{i2}^* U_{j3})/\sqrt{2}.
\eeq

\begin{center}{\em 2.~Corrections to electroweak vector boson self-energies in 
the QUE model}
\end{center}

For the QUE model, the gauge eigenstate new 
up-type quark mass matrix is:
\beq
{\cal M}_u = \begin{pmatrix} M_Q & k_U v_u \cr  
                    h_U v_d & M_U \cr 
    \end{pmatrix}
,
\eeq
with unitary mixing matrices $L$ and $R$ defined by
\beq
L^* {\cal M}_u R^{\dagger} = {\rm diag}(m_{t'_1}, m_{t'_2}),
\eeq
and $m_{b'} = M_Q$.
Then the $(t_1', t_2', b')$
fermion contributions to the electroweak vector boson
self-energies are:
\beq
\Delta \Pi_{\gamma\gamma} &=& 
-\frac{N_c}{16 \pi^2} 2 g^2 s_W^2 \biggl [ 
e_u^2 \sum_{i=1,2} G(t'_i)  + e_d^2 G(b') \biggr ],
\\
\Delta \Pi_{Z\gamma} &=& 
-\frac{N_c}{16 \pi^2} g s_W \biggl [ 
e_u \sum_{i=1,2} (g^{Z}_{t_i' t_i^{\prime\dagger}} - 
g^{Z}_{\bar t_i' \bar t_i^{\prime\dagger}}) G(t'_i)  
+ e_d (g^{Z}_{b'b^{\prime\dagger}}  - 
g^{Z}_{\bar b' \bar b^{\prime\dagger}}) G(b')  
\biggr ],
\\
\Delta \Pi_{ZZ} &=& 
-\frac{N_c}{16 \pi^2} \biggl [
(|g^{Z}_{b'b^{\prime\dagger}}|^2 + 
|g^{Z}_{\bar b' \bar b^{\prime\dagger}}|^2) G(b') 
\nonumber \\ && + 
\sum_{i,j=1}^2 (|g^{Z}_{t_i' t_j^{\prime\dagger}}|^2 + 
|g^{Z}_{\bar t_i' \bar t_j^{\prime\dagger}}|^2) H(t'_i, t'_j) 
-4 {\rm Re}(
g^{Z}_{t_i' t_j^{\prime\dagger}} g^{Z}_{\bar t_i' \bar t_j^{\prime\dagger}}) 
m_{t'_i} m_{t'_j} B(t'_i, t'_j)
\biggr ]
,
\\
\Delta \Pi_{WW} &=& 
-\frac{N_c}{16 \pi^2} \sum_{i=1,2} \left [
(|g^{W}_{t_i' b^{\prime\dagger}}|^2 + |g^{W}_{\bar t_i' \bar b^{\prime\dagger}} |^2) H(b', t'_i) 
 -4 {\rm Re}(g^{W}_{t_i' b^{\prime\dagger}} 
 g^{W}_{\bar t_i' \bar b^{\prime\dagger}}) m_{b'} m_{t'_i} B(b', t_i')
\right ]
,\phantom{xxx}
\eeq
where $N_c = 3$ and $e_u = 2/3$ and $e_d = -1/3$ and 
the massive vector boson couplings with the new quarks are:
\beq
g^{Z}_{t_i' t_j^{\prime\dagger}} &=& 
\frac{g}{c_W} \left (\frac{1}{2} L_{i1}^* L_{j1} 
- e_u s_W^2 \delta_{ij}\right ),
\qquad\qquad
g^{Z}_{\bar t_i' \bar t_j^{\prime\dagger}} = \frac{g}{c_W} \left ( 
-\frac{1}{2} R_{i1}^* R_{j1} + e_u s_W^2 \delta_{ij}\right ),
\\ 
g^{Z}_{b'b^{\prime\dagger}} &=& -g^{Z}_{\bar b' \bar b^{\prime\dagger}}  =
\frac{g}{c_W} \left (-\frac{1}{2} - e_d s_W^2 \right )
,
\phantom{xxx}
\\ 
g^{W}_{t_i' b^{\prime\dagger}} &=& g L_{i1}^*/\sqrt{2},
\qquad\qquad
g^{W}_{\bar t_i' \bar b^{\prime\dagger}} 
=
-g R_{i1}^* /\sqrt{2}.
\eeq

To obtain the $(\tilde t_{1,2,3,4}', \tilde b_{1,2}')$
scalar contribution, consider the up-type squark
squared-mass matrix:
\beq
M^2_{\tilde u} = M^2_u + 
\begin{pmatrix}
m^2_Q + \Delta_{\frac{1}{2},\frac{2}{3}} 
& 0 & b_Q^* & a_{k_U}^* v_u - \mu k_U v_d \cr
0 & m^2_{U} + \Delta_{0,\frac{2}{3}} & 
a_{h_U}^* v_d -  \mu h_U v_u & b_U^* \cr
b_Q & a_{h_U} v_d -  \mu^* h_U v_u & m^2_{\overline Q} + 
\Delta_{-\frac{1}{2},-\frac{2}{3}}  & 0 
\cr
a_{k_U} v_u - \mu^* k_U v_d & b_U & 0 & m^2_{\overline U} 
+ \Delta_{0,-\frac{2}{3}} \cr
\end{pmatrix}
,
\eeq
where the supersymmetric part (also equal to the fermion squared-mass matrix) is:
\beq
M^2_u = 
\begin{pmatrix}
{\cal M}_u {\cal M}_u^\dagger & 0 \cr
0 & {\cal M}_u^\dagger {\cal M}_u
\end{pmatrix} 
.
\eeq
Also, the down-type squark mass matrix is
\beq
M^2_{\tilde d} =
\begin{pmatrix}
M_Q^2 + m^2_Q + \Delta_{-\frac{1}{2},-\frac{1}{3}} &  -b_Q^*\cr -b_Q & 
M_Q^2 + m^2_{\overline Q} + \Delta_{\frac{1}{2},\frac{1}{3}}\cr
\end{pmatrix}
.
\eeq
Now define unitary
scalar mixing matrices $U$ and $V$ by
\beq
U M^2_{\tilde u} U^{\dagger} &=& 
{\rm diag}(m^2_{\tilde t'_1}, m^2_{\tilde t'_2},
m^2_{\tilde t'_3},m^2_{\tilde t'_4})
,
\qquad\qquad
V M^2_{\tilde d} V^{\dagger} 
=
{\rm diag}(m^2_{\tilde b'_1}, m^2_{\tilde b'_2}).
\eeq
Then the scalar contributions to the vector boson self-energies are:
\beq
\Delta \Pi_{\gamma\gamma} &=& 
\frac{N_c}{16 \pi^2} g^2 s_W^2 \left [
       e_u^2 \sum_{i=1}^4 F(\tilde t'_i, \tilde t'_i)
      +e_d^2 \sum_{i=1}^2 F(\tilde b'_i, \tilde b'_i)
\right ],
\\
\Delta \Pi_{Z\gamma} &=& 
\frac{N_c}{16 \pi^2} g s_W \left [ 
 e_u \sum_{i=1}^4 g^{Z}_{\tilde t_i' \tilde t_i^{\prime *}} 
 F(\tilde t'_i, \tilde t'_i)
+e_d \sum_{i=1}^2 g^{Z}_{\tilde b_i' \tilde b_i^{\prime *}} 
 F(\tilde b'_i, \tilde b'_i)
\right ],
\\
\Delta \Pi_{ZZ} &=& 
\frac{N_c}{16 \pi^2} \left [ 
 \sum_{i,j=1}^4 |g^{Z}_{\tilde t_i' \tilde t_j^{\prime *}} |^2  F(\tilde t'_i, \tilde t'_j)
+\sum_{i,j=1}^2 |g^{Z}_{\tilde b_i' \tilde b_j^{\prime *}} |^2  F(\tilde b'_i, \tilde b'_j)
\right ]
,
\\
\Delta \Pi_{WW} &=& 
\frac{N_c}{16 \pi^2} \sum_{i=1}^2 \sum_{j=1}^4
|g^{W}_{\tilde b_i' \tilde t_j^{\prime *}}|^2 F(\tilde b'_i, \tilde t'_j),
\eeq 
where the vector boson couplings with the new squarks are:
\beq
g^{Z}_{\tilde t_i' \tilde t_j^{\prime *}} &=& \frac{g}{c_W} \left [
\frac{1}{2} (U_{i1}^* U_{j1} + U_{i3}^* U_{j3}) 
- e_u s_W^2 \delta_{ij}\right ]
,
\qquad\qquad
g^{Z}_{\tilde b_i' \tilde b_j^{\prime *}} =  \frac{g}{c_W} \left (
-\frac{1}{2} - e_d s_W^2 \right )\delta_{ij}
,
\phantom{xxx}
\\
g^{W}_{\tilde b_i' \tilde t_j^{\prime *}} &=& 
g ( V_{i1}^* U_{j1} - V_{i2}^* U_{j3})/\sqrt{2}.
\eeq

\begin{center}
{\em 3.~Corrections to electroweak vector boson self-energies in 
the QDEE model}
\end{center}

For the QDEE model, define the gauge eigenstate new 
down-type quark mass matrix by:
\beq
{\cal M}_d = \begin{pmatrix} M_Q & k_D v_u \cr  
                    h_D v_d & M_D \cr 
    \end{pmatrix}
\eeq
and unitary mixing matrices $L$ and $R$ by:
\beq
R^* {\cal M}_d L^{\dagger} = {\rm diag}(m_{b'_1}, m_{b'_2}),
\eeq
and note $m_{t'} = M_Q$.
Then the $(b_1', b_2', t')$
fermion contributions to the electroweak vector boson
self-energies are:
\beq
\Delta \Pi_{\gamma\gamma} &=& 
-\frac{N_c}{16 \pi^2} 2 g^2 s_W^2 \biggl [ 
e_d^2 \sum_{i=1,2} G(b'_i)  + e_u^2 G(t') \biggr ],
\\
\Delta \Pi_{Z\gamma} &=& 
-\frac{N_c}{16 \pi^2} g s_W \biggl [ 
e_d \sum_{i=1,2} (g^{Z}_{b_i' b_i^{\prime\dagger}} - 
g^{Z}_{\bar b_i' \bar b_i^{\prime\dagger}}) G(b'_i)  
+ e_u (g^{Z}_{t't^{\prime\dagger}}  - 
g^{Z}_{\bar t' \bar t^{\prime\dagger}}) G(t')  
\biggr ],
\\
\Delta \Pi_{ZZ} &=& 
-\frac{N_c}{16 \pi^2} \biggl [
(|g^{Z}_{t't^{\prime\dagger}}|^2 + 
|g^{Z}_{\bar t' \bar t^{\prime\dagger}}|^2) G(t') 
\nonumber \\ && + 
\sum_{i,j=1}^2 (|g^{Z}_{b_i' b_j^{\prime\dagger}}|^2 + 
|g^{Z}_{\bar b_i' \bar b_j^{\prime\dagger}}|^2) H(b'_i, b'_j) 
-4 {\rm Re}(
g^{Z}_{b_i' b_j^{\prime\dagger}} g^{Z}_{\bar b_i' \bar b_j^{\prime\dagger}}) 
m_{b'_i} m_{b'_j} B(b'_i, b'_j)
\biggr ]
,
\\
\Delta \Pi_{WW} &=& 
-\frac{N_c}{16 \pi^2} \sum_{i=1,2} \left [
(|g^{W}_{b_i' t^{\prime\dagger}}|^2 + |g^{W}_{\bar b_i' \bar t^{\prime\dagger}} |^2) H(t', b'_i) 
 -4 {\rm Re}(g^{W}_{b_i' t^{\prime\dagger}} 
 g^{W}_{\bar b_i' \bar t^{\prime\dagger}}) m_{t'} m_{b'_i} B(t', b_i')
\right ]
,\phantom{xxx}
\eeq
where $N_c = 3$ and $e_u = 2/3$ and $e_d = -1/3$ and 
the massive vector boson couplings with the new quarks are:
\beq
g^{Z}_{b_i' b_j^{\prime\dagger}} &=& 
\frac{g}{c_W} \left (-\frac{1}{2} L_{i1}^* L_{j1} 
- e_d s_W^2 \delta_{ij}\right ),
\qquad\qquad
g^{Z}_{\bar b_i' \bar b_j^{\prime\dagger}} = \frac{g}{c_W} \left ( 
\frac{1}{2} R_{i1}^* R_{j1} + e_d s_W^2 \delta_{ij}\right ),
\\ 
g^{Z}_{t't^{\prime\dagger}} &=& -g^{Z}_{\bar t' \bar t^{\prime\dagger}}  =
\frac{g}{c_W} \left (\frac{1}{2} - e_u s_W^2 \right )
,
\\ 
g^{W}_{b_i' t^{\prime\dagger}} &=& -g L_{i1}^*/\sqrt{2},
\qquad\qquad
g^{W}_{\bar b_i' \bar t^{\prime\dagger}} 
= g R_{i1}^* /\sqrt{2}.
\eeq

To obtain the $(\tilde b_{1,2,3,4}', \tilde t_{1,2}')$
scalar contribution, start with the down-type squark
squared-mass matrix:
\beq
M^2_{\tilde d} = M^2_d + 
\begin{pmatrix}
m^2_{\overline Q} + \Delta_{\frac{1}{2},\frac{1}{3}} 
& 0 & b_Q^* & a_{k_D}^* v_u - \mu k_D v_d 
\cr
0 & m^2_{\overline D} + \Delta_{0,\frac{1}{3}} & 
a_{h_D}^* v_d -  \mu h_D v_u & b_D^* 
\cr
b_Q & a_{h_D} v_d -  \mu^* h_D v_u & m^2_{Q} + 
\Delta_{-\frac{1}{2},-\frac{1}{3}}  & 0 
\cr
a_{k_D} v_u - \mu^* k_D v_d & b_D & 0 & m^2_{D} 
+ \Delta_{0,-\frac{1}{3}} 
\cr
\end{pmatrix}
,
\eeq
where the supersymmetric part 
(also equal to the fermion squared-mass matrix) is:
\beq
M^2_d = 
\begin{pmatrix}
{\cal M}_d {\cal M}_d^\dagger & 0 \cr
0 & {\cal M}_d^\dagger {\cal M}_d
\end{pmatrix} 
.
\eeq
Also, the up-type squark squared-mass matrix is given by:
\beq
M^2_{\tilde u} =
\begin{pmatrix}
M_Q^2 + m^2_{\overline Q} + \Delta_{-\frac{1}{2},-\frac{2}{3}} &  -b_Q^*
\cr 
-b_Q & M_{Q}^2 + m^2_{Q} + \Delta_{\frac{1}{2},\frac{2}{3}}
\end{pmatrix}
.
\eeq
Now define unitary
scalar mixing matrices $U$ and $V$ by:
\beq
U M^2_{\tilde d} U^{\dagger} &=& 
{\rm diag}(m^2_{\tilde b'_1}, m^2_{\tilde b'_2},m^2_{\tilde b'_3},
m^2_{\tilde b'_4})
,
\qquad\qquad
V M^2_{\tilde u} V^{\dagger} 
=
{\rm diag}(m^2_{\tilde t'_1}, 
m^2_{\tilde t'_2}).
\eeq
Then the scalar contributions to the vector boson self-energies are:
\beq
\Delta \Pi_{\gamma\gamma} &=& 
\frac{N_c}{16 \pi^2} g^2 s_W^2 \left [
       e_d^2 \sum_{i=1}^4 F(\tilde b'_i, \tilde b'_i)
      +e_u^2 \sum_{i=1}^2 F(\tilde t'_i, \tilde t'_i)
\right ],
\\
\Delta \Pi_{Z\gamma} &=& 
\frac{N_c}{16 \pi^2} g s_W \left [ 
-e_d \sum_{i=1}^4 g^{Z}_{\tilde b_i^{\prime *} \tilde b_i'} 
 F(\tilde b'_i, \tilde b'_i)
-e_u \sum_{i=1}^2 g^{Z}_{\tilde t_i^{\prime *}\tilde t_i' } 
 F(\tilde t'_i, \tilde t'_i)
\right ],
\\
\Delta \Pi_{ZZ} &=& 
\frac{N_c}{16 \pi^2} \left [ 
 \sum_{i,j=1}^4 |g^{Z}_{\tilde b_i^{\prime *}\tilde b_j' } |^2  
 F(\tilde b'_i, \tilde b'_j)
+\sum_{i,j=1}^2 |g^{Z}_{\tilde t_i^{\prime *}\tilde t_j' } |^2  
F(\tilde t'_i, \tilde t'_j)
\right ]
,
\\
\Delta \Pi_{WW} &=& 
\frac{N_c}{16 \pi^2} \sum_{i=1}^2 \sum_{j=1}^4
|g^{W}_{\tilde t_i^{\prime *} \tilde b_j'}|^2 F(\tilde t'_i, \tilde b'_j),
\eeq 
where the vector boson couplings with the new squarks are
\beq
g^{Z}_{\tilde b_i^{\prime *}\tilde b_j'} &=& \frac{g}{c_W} \left [
\frac{1}{2} (U_{i1}^* U_{j1} + U_{i3}^* U_{j3}) 
+ e_d s_W^2 \delta_{ij}\right ]
,
\qquad\qquad
g^{Z}_{\tilde t_i^{\prime *}\tilde t_j'} =  \frac{g}{c_W} \left (
-\frac{1}{2} + e_u s_W^2 \right )\delta_{ij}
,
\phantom{xxx}
\\
g^{W}_{\tilde t_i^{\prime *} \tilde b_j' } &=& 
g ( V_{i1}^* U_{j1} - V_{i2}^* U_{j3})/\sqrt{2}
.
\eeq

\section*{Appendix B: Formulas for decay widths of new quarks and leptons}
\label{appendixBR} \renewcommand{\theequation}{B.\arabic{equation}}
\setcounter{equation}{0}
\setcounter{footnote}{1}
\setcounter{subsubsection}{0}

This Appendix gives formulas for the decay widths of the lightest of the new quarks and
leptons to Standard Model states. These decays are assumed to be mediated by
Yukawa couplings that provide small mass mixings that can be treated as 
perturbations compared to the other entries in the mass matrices. In the following, 
$\lambda(x,y,z) = x^2 + y^2 + z^2 - 2 x y - 2 x z - 2 y z$.

\begin{center}
{\em 1.~Decays of $b'$ in the LND model}
\end{center}

In the LND model, the lightest quark $b'$ can decay to Standard Model
states because of the mixing Yukawa parameter $\epsd$ in 
eq.~(\ref{eq:WLNDmix}). In terms of
the mass matrix ${\cal M}_d$ in eq.~(\ref{eq:WLNDmassmix}),
define unitary mixing matrices $L$ and $R$ by:
\beq
L^* {\cal M}_d R^{\dagger} = {\rm diag}(m_b, m_{b'}).
\eeq
The relevant couplings of $b'$ to Standard Model particles are
\beq
g^{W}_{b' t^\dagger} &=& g L^*_{22}/\sqrt{2} 
,
\qquad\qquad
g^{Z}_{b' b^\dagger} 
= 
-\frac{g}{2 c_W} L^*_{22} L_{12}
,
\\
y^{h^0}_{b' \bar b} &=& 
-\sin(\alpha) (y_b R_{12} + \epsd R_{11}) L_{22}/\sqrt{2}
,
\\
y^{h^0}_{\bar b' b} &=& 
-\sin(\alpha) (y_b R_{22} + \epsd R_{21}) L_{12}/\sqrt{2}
.
\eeq
It follows that the decay widths of $b'$ are:
\beq
\Gamma(b' \rightarrow W t) &=& 
\frac{m_{b'}}{32 \pi} |g^{W}_{b' t^\dagger}|^2
\lambda^{1/2}(1, r_W, r_t) (1 + r_t - 2 r_W + (1-r_t)^2/r_W)
,
\\
\Gamma(b' \rightarrow Z b) &=& 
\frac{m_{b'}}{32 \pi} |g^{Z}_{b' b^\dagger}|^2 
(1 - r_Z)^2 (2 + 1/r_Z)
,
\\
\Gamma(b' \rightarrow h^0 b) &=& \frac{m_{b'}}{32 \pi}
\left (
|y^{h^0}_{b' \bar b}|^2 + |y^{h^0}_{\bar b' b}|^2
\right )(1 - r_{h^0})^2
,\phantom{xxx}
\eeq
where $m_{b}$ is neglected for kinematic purposes and
$r_i = m_i^2/m^2_{b'}$ for $i = Z, W, h^0$.

\begin{center}
{\em 2.~Decays of $\nu_1'$ in the LND model}
\end{center}

Consider the decays of $\nu_1'$, the lighter new neutral lepton in the
LND model, brought about by the superpotential mixing terms 
$\epsn$ and $\epse$ in 
eq.~(\ref{eq:WLNDmix}). 
Define unitary mixing matrices $L$ $(3\times 3)$ and $R$ $(2 \times 2)$ 
in terms of the neutral lepton mass matrix in eq.~(\ref{eq:Nmassmix}) by:
\beq
R^* {\cal M}_\nu^T L^{\dagger} &=&
\begin{pmatrix}
0 & m_{\nu_1'} & 0 \cr
0 & 0 & m_{\nu_2'} \cr
\end{pmatrix}
\eeq
where we are neglecting the tau neutrino mass. Also define
unitary matrices $L'$ and $R'$ in terms of the charged lepton mass 
matrix in eq.~(\ref{eq:Emassmix}) by:
\beq
L^{\prime *} {\cal M}_e R^{\prime \dagger} &=& {\rm diag}(m_{\tau}, m_{\tau'}).
\eeq
Then the relevant couplings of $\nu_1'$ to Standard Model particles are:
\beq
g^{W}_{\nu_1' \tau^\dagger} &=& 
g (L_{21}^* L_{11}' + L_{23}^* L_{12}' )/\sqrt{2}
\qquad\quad
g^{W}_{\bar \nu_1' \bar \tau^\dagger} 
= 
g R_{11}^* R_{11}'/\sqrt{2}
\\
g^Z_{\nu_1' \nu^\dagger} &=& \frac{g}{2 c_W} (L_{21}^* L_{11} + L_{23}^* L_{13})
\\
y^{h^0}_{\bar \nu_1' \nu } &=&
\frac{\cos\alpha}{\sqrt{2}} (\epsn L_{13} + k_N L_{11}) R_{12}
-\frac{\sin\alpha}{\sqrt{2}} h_N L_{12} R_{11} .
\eeq
It follows that the decay widths of $\nu_1'$ are:
\beq
\Gamma(\nu_1' \rightarrow W \tau) &=& \frac{m_{\nu_1'}}{32 \pi} 
(1 - r_W)^2 (2 + 1/r_W) (|g^{W}_{\nu_1' \tau^\dagger}|^2 + 
|g^{W}_{\bar \nu_1' \bar \tau^\dagger}|^2)
,
\\
\Gamma(\nu_1' \rightarrow Z \nu_\tau) &=& \frac{m_{\nu_1'}}{32 \pi } 
(1 - r_Z)^2 (2 + 1/r_Z) |g^Z_{\nu_1' \nu^\dagger} |^2
,
\\
\Gamma(\nu_1' \rightarrow h^0 \nu_\tau) &=& \frac{m_{\nu_1'}}{32 \pi}
(1 - r_{h^0})^2
|y^{h^0}_{\bar \nu_1' \nu}|^2
,\phantom{xxx}
\eeq
where $m_{\tau}$ and $m_{\nu_\tau}$ are neglected for kinematic purposes
and
$r_i = m_i^2/m^2_{\nu_1'}$ for $i = Z, W, h^0$.

\begin{center}
{\em 3.~Decays of $t'_1$ in the QUE model}
\end{center}

Consider the decays of $t'_1$, the lightest new quark 
in the QUE model, brought about by the
superpotential mixing terms in eq.~(\ref{eq:WQUEmix}).
Define unitary mixing matrices  $L$, $R$, $L'$, $R'$
in terms of the mass matrices in eq.~(\ref{eq:QUEmassmix})
by:
\beq
L^* {\cal M}_u R^{\dagger} = {\rm diag}(m_t, m_{t'_1}, m_{t'_2}), 
\qquad\qquad
L^{\prime *} {\cal M}_d R^{\prime \dagger} = {\rm diag}(m_b, m_{b'}).
\eeq
Then the relevant couplings of $t_1'$ to Standard Model particles are:
\beq
g^W_{t_1'b^\dagger} &=& 
g (L_{21}^* L_{11}^{\prime} + L_{23}^* L_{12}^{\prime})/\sqrt{2}
,
\qquad\qquad
g^W_{\bar t_1'\bar b^\dagger} 
=
g  R_{21}^* R'_{11}/\sqrt{2}
,
\\
g^{Z}_{t_1't^\dagger} &=& \frac{g}{2 c_W} (L_{21}^* L_{11} + L_{23}^* L_{13}) 
,
\qquad\qquad
g^Z_{\bar t_1' \bar t^\dagger} 
= -\frac{g}{2 c_W} R_{21}^* R_{11}
,
\\
y^{h^0}_{t_1' \bar t} &=& \frac{\cos\alpha}{\sqrt{2}} \left (
\epsu L_{23} R_{12} + \epsup L_{21} R_{13} + k_U L_{21} R_{12} + y_t L_{23} R_{13}
\right ) - \frac{\sin\alpha}{\sqrt{2}} h_U L_{22} R_{11}
,
\\
y^{h^0}_{\bar t_1' t} &=& \frac{\cos\alpha}{\sqrt{2}} \left (
\epsu L_{13} R_{22} + \epsup L_{11} R_{23} + k_U L_{11} R_{22} + y_t L_{13} R_{23}
\right ) - \frac{\sin\alpha}{\sqrt{2}} h_U L_{12} R_{21}
.
\phantom{xxx}
\eeq
It follows that the decay widths of $t_1'$ are:
\beq
\Gamma(t_1' \rightarrow W b) &=&
\frac{m_{t_1'}}{32 \pi} 
(1 - r_W)^2 (2 + 1/r_W)
(|g^W_{t_1'b^\dagger}|^2 + |g^W_{\bar t_1'\bar b^\dagger}|^2) 
,
\\
\Gamma(t_1' \rightarrow Z t) &=&
\frac{m_{t_1'}}{32 \pi} \lambda^{1/2}(1, r_Z, r_t)
\Bigl [
(1 + r_t - 2 r_Z + (1 - r_t)^2/r_Z)
(|g^{Z}_{t_1't^\dagger}|^2 + |g^Z_{\bar t_1' \bar t^\dagger} |^2)
\nonumber \\ &&
+ 12 \sqrt{r_t} 
{\rm Re} (g^{Z}_{t_1't^\dagger} g^Z_{\bar t_1' \bar t^\dagger} ) 
\Bigr ]
,
\\
\Gamma(t_1' \rightarrow h^0 t) &=&
\frac{m_{t_1'}}{32 \pi} \lambda^{1/2}(1, r_{h^0}, r_t)
\left [
(1 + r_t - r_{h^0}) 
(|y^{h^0}_{t_1' \bar t}|^2 + |y^{h^0}_{\bar t_1' t}|^2) 
+ 4 \sqrt{r_t} {\rm Re}(y^{h^0}_{\bar t_1' t} y^{h^0}_{t_1' \bar t} ) 
\right ],\phantom{xxxx}
\eeq
where the bottom
quark is treated as massless for purposes of kinematics
and $r_i = m_i^2/m^2_{t_1'}$ for $i = t, Z, W, h^0$.


\begin{center}
{\em 4.~Decays of $b'_1$ in the QDEE model}
\end{center}

Consider the decays of $b_1'$, the lightest new quark in the QDEE model,
brought about by the superpotential mixing terms in 
eq.~(\ref{eq:WQDEEmix}).
Define unitary mixing matrices  $R$, $L$, $R'$, $L'$ 
in terms of the mass matrices in eq.~(\ref{eq:QDEEmassmix}) by:
\beq
R^* {\cal M}_d L^{\dagger} = {\rm diag}(m_b, m_{b'_1}, m_{b'_2}), 
\qquad\qquad
R^{\prime *} {\cal M}_u L^{\prime \dagger} = {\rm diag}(m_t, m_{t'}).
\eeq
Then the relevant couplings of $b_1'$ to Standard Model particles are:
\beq
g^W_{b_1' t^\dagger} &=& g ( 
L_{21}^* L_{11}^{\prime} + L_{23}^* L_{12}^{\prime})/\sqrt{2}
,
\qquad\qquad
g^W_{\bar b_1' \bar t^\dagger} 
= 
g R_{21}^* R'_{11}/\sqrt{2} 
,
\\
g^Z_{b_1' b^\dagger} &=& -\frac{g}{2 c_W} (L_{21}^* L_{11} + L_{23}^* L_{13}) 
,
\qquad\qquad
g^Z_{\bar b_1' \bar b^\dagger} 
= 
\frac{g}{2 c_W} R_{21}^* R_{11} 
,
\\
y^{h^0}_{b_1' \bar b} &=& -\frac{\sin\alpha}{\sqrt{2}} \left (
\epsd L_{23} R_{12} + \epsdp L_{21} R_{13} + h_D L_{21} R_{12} + y_b L_{23} R_{13}
\right ) + \frac{\cos\alpha}{\sqrt{2}} k_D L_{22} R_{11}
,
\\
y^{h^0}_{\bar b_1' b}  &=& -\frac{\sin\alpha}{\sqrt{2}} \left (
\epsd L_{13} R_{22} + \epsdp L_{11} R_{23} + h_D L_{11} R_{22} + y_b L_{13} R_{23}
\right ) + \frac{\cos\alpha}{\sqrt{2}} k_D L_{12} R_{21}
.
\phantom{xxx}
\eeq
It follows that the decay widths of $b_1'$ are:
\beq
\Gamma(b_1' \rightarrow W t) &=&
\frac{m_{b_1'}}{32 \pi} \lambda^{1/2}(1, r_W, r_t) \Bigl[
(1 + r_t - 2 r_W + (1 - r_t)^2/r_W) (|g^W_{b_1' t^\dagger}|^2 + 
|g^W_{\bar b_1' \bar t^\dagger}|^2)
\phantom{xxx}
\nonumber \\ &&
+ 12 \sqrt{r_t} {\rm Re}(g^W_{b_1' t^\dagger} g^W_{\bar b_1' \bar t^\dagger}) 
\Bigr ]
,
\\
\Gamma(b_1' \rightarrow Z b) &=&
\frac{m_{b_1'}}{32 \pi} (1- r_Z)^2 (2 + 1/r_Z)
(|g^Z_{b_1' b^\dagger}|^2 + |g^Z_{\bar b_1' \bar b^\dagger}|^2)
,
\\
\Gamma(b_1' \rightarrow h^0 b) &=&
\frac{m_{b_1'}}{32 \pi} (1 - r_{h^0})^2
(|y^{h^0}_{b_1' \bar b} |^2 + |y^{h^0}_{\bar b_1' b} |^2),
\eeq
where the bottom quark is treated as massless for purposes of kinematics 
and $r_i = m_i^2/m^2_{b_1'}$ for $i = t, Z, W, h^0$.

\begin{center}
{\em 5.~Decays of $\tau'$ in the QUE and QDEE models}
\end{center}

Consider the decays of $\tau'$ in the QUE model, 
brought about by the superpotential
mixing term $\epse$ in eq.~(\ref{eq:WQUEmixE}).
In terms of the mass matrix
eq.~(\ref{eq:QUEmassmatE}),
define unitary mixing matrices $L$ and $R$ by:
\beq
L^* {\cal M}_e R^{\dagger} = {\rm diag}(m_\tau, m_{\tau'}).
\eeq
Then the relevant couplings of $\tau'$ to Standard Model particles are:
\beq
g^W_{\tau' \nu^\dagger} &=& g L_{22}^*/\sqrt{2}
,
\qquad\qquad
g^Z_{\tau' \tau^\dagger} 
=
-\frac{g}{2c_W} L_{22}^* L_{12}
,
\\
y^{h^0}_{\tau'\bar \tau} &=&
-\sin(\alpha) L_{22} (y_\tau R_{12} + \epse R_{11})/\sqrt{2}
,
\\
y^{h^0}_{\bar \tau'\tau} &=&
-\sin(\alpha) L_{12} (y_\tau R_{22} + \epse R_{21})/\sqrt{2}
.
\eeq
It follows that the decay widths of $\tau'$ are:
\beq
\Gamma(\tau' \rightarrow W \nu) &=& \frac{m_{\tau'}}{32 \pi} 
(1 - r_W)^2 (2 + 1/r_W) |g^W_{\tau' \nu^\dagger}|^2
,
\\
\Gamma(\tau' \rightarrow Z \tau) &=& \frac{m_{\tau'}}{32 \pi } 
(1 - r_Z)^2 (2 + 1/r_Z) |g^Z_{\tau' \tau^\dagger}|^2
,
\\
\Gamma(\tau' \rightarrow h^0 \tau) &=& \frac{m_{\tau'}}{32 \pi}
(1 - r_{h^0})^2 (|y^{h^0}_{\tau'\bar \tau}|^2 + 
|y^{h^0}_{\bar \tau'\tau}|^2)
,\phantom{xxx}
\eeq
where $r_i = m_i^2/m^2_{\tau'}$ for $i = Z, W, h^0$, and $m_{\tau}$ is 
neglected for kinematic purposes. In the QDEE model, the same calculation 
holds, provided that $M_E$ is replaced by $M_{E_1}$ corresponding to the 
lighter
mass eigenstate $m_{\tau'}$.

\bigskip \noindent 
{\it Acknowledgments:} I am indebted to James Wells
for useful comments.
This work was supported in part by the National Science Foundation grant 
number PHY-0757325.

\noindent 
Note added: shortly after
the present paper, one with some related
subject matter appeared \cite{Graham:2009gy}.



\begin{thebibliography}{90}
\baselineskip=11.9pt

\bibitem{primer}
For a review of supersymmetry at the TeV scale, see
  S.P.~Martin,
  ``A supersymmetry primer,''
  [hep-ph/9709356] (version 5, December 2008).

\bibitem{Frampton:1999xi}
  P.H.~Frampton, P.Q.~Hung and M.~Sher,
  Phys.\ Rept.\  {\bf 330}, 263 (2000)
  [hep-ph/9903387],
  and references therein.

\bibitem{Maltoni:1999ta}
  M.~Maltoni, V.A.~Novikov, L.B.~Okun, A.N.~Rozanov and M.I.~Vysotsky,
  Phys.\ Lett.\  B {\bf 476}, 107 (2000)
  [hep-ph/9911535].

\bibitem{He:2001tp}
  H.J.~He, N.~Polonsky and S.f.~Su,
  Phys.\ Rev.\  D {\bf 64}, 053004 (2001)
  [hep-ph/0102144].

\bibitem{Novikov:2002tk}
  V.A.~Novikov, L.B.~Okun, A.N.~Rozanov and M.I.~Vysotsky,
  JETP Lett.\  {\bf 76}, 127 (2002)
  [hep-ph/0203132].

\bibitem{Holdom:2006mr}
  B.~Holdom,
  JHEP {\bf 0608}, 076 (2006)
  [hep-ph/0606146],
  JHEP {\bf 0703}, 063 (2007)
  [hep-ph/0702037],
  JHEP {\bf 0708}, 069 (2007)
  [hep-ph/0705.1736].

\bibitem{Kribs:2007nz}  
G.D.~Kribs, T.~Plehn, M.~Spannowsky and T.M.P.~Tait,
  Phys.\ Rev.\  D {\bf 76}, 075016 (2007)
  [hep-ph/0706.3718].

\bibitem{Hung:2007ak}
  P.Q.~Hung and M.~Sher,
  Phys.\ Rev.\  D {\bf 77}, 037302 (2008)
  [hep-ph/0711.4353].

\bibitem{Ozcan:2008qk}
  V.E.~Ozcan, S.~Sultansoy and G.~Unel,
  ``Search for 4th family quarks with the ATLAS detector,''
  [hep-ex/0802.2621].

\bibitem{Fok:2008yg}
  R.~Fok and G.D.~Kribs,
  Phys.\ Rev.\  D {\bf 78}, 075023 (2008)
  [hep-ph/0803.4207].

\bibitem{Murdock:2008rx}
  Z.~Murdock, S.~Nandi and Z.~Tavartkiladze,
  Phys.\ Lett.\  B {\bf 668}, 303 (2008)
  [hep-ph/0806.2064].

\bibitem{Ibrahim:2008gg}
  T.~Ibrahim and P.~Nath,
  Phys.\ Rev.\  D {\bf 78}, 075013 (2008)
  [hep-ph/0806.3880].

\bibitem{Burdman:2008qh}
  G.~Burdman, L.~Da Rold, O.~Eboli and R.~Matheus,
  Phys.\ Rev.\  D {\bf 79}, 075026 (2009)
  [hep-ph/0812.0368].

\bibitem{Dobrescu:2009vz}
  B.A.~Dobrescu, K.~Kong and R.~Mahbubani,
  JHEP {\bf 0906}, 001 (2009)
  [hep-ph/0902.0792].

\bibitem{Bobrowski:2009ng}
  M.~Bobrowski, A.~Lenz, J.~Riedl and J.~Rohrwild,
  Phys.Rev.D79:113006,2009,''
  [hep-ph/0902.4883].

\bibitem{Chanowitz:2009mz}
For example, see
  M.S.~Chanowitz,
  Phys.\ Rev.\  D {\bf 79}, 113008 (2009)
  [hep-ph/0904.3570].

\bibitem{Novikov:2009kc}
  V.A.~Novikov, A.N.~Rozanov and M.I.~Vysotsky,
  ``Once more on extra quark-lepton generations and precision measurements,''
  [hep-ph/0904.4570].

\bibitem{Holdom:2009rf}
  B.~Holdom, W.S.~Hou, T.~Hurth, M.L.~Mangano, S.~Sultansoy and G.~Unel,
  ``Four Statements about the Fourth Generation,''
  [hep-ph/0904.4698].

\bibitem{Liu:2009cc}
  C.~Liu,
  ``Supersymmetry and Vector-like Extra Generation,''
  [hep-ph/0907.3011].

\bibitem{Litsey:2009rp}
  S.~Litsey and M.~Sher,
  ``Higgs Masses in the Four Generation MSSM,''
  [hep-ph/0908.0502].

\bibitem{Moroi:1991mg}
  T.~Moroi and Y.~Okada,
  Mod.\ Phys.\ Lett.\  A {\bf 7}, 187 (1992).

\bibitem{Moroi:1992zk}
  T.~Moroi and Y.~Okada,
  Phys.\ Lett.\  B {\bf 295}, 73 (1992).

\bibitem{Babu:2004xg}
  K.S.~Babu, I.~Gogoladze and C.~Kolda,
  ``Perturbative unification and Higgs boson mass bounds,''
  [hep-ph/0410085].

\bibitem{Babu:2008ge}
  K.S.~Babu, I.~Gogoladze, M.U.~Rehman and Q.~Shafi,
  Phys.\ Rev.\  D {\bf 78}, 055017 (2008)
  [hep-ph/0807.3055].

\bibitem{betas:1}
D.R.T.~Jones,
  Nucl.\ Phys.\  B {\bf 87}, 127 (1975). 
D.R.T.~Jones and L.~Mezincescu,
  Phys.\ Lett.\  B {\bf 136}, 242 (1984).
P.C.~West,
  Phys.\ Lett.\  B {\bf 137}, 371 (1984).
A.~Parkes and P.C.~West,
  Phys.\ Lett.\  B {\bf 138}, 99 (1984).

\bibitem{betas:2}
S.P.~Martin and M.T.~Vaughn,
  Phys.\ Lett.\  B {\bf 318}, 331 (1993)
  [hep-ph/9308222],
  Phys.\ Rev.\  D {\bf 50}, 2282 (1994)  
  [Erratum-ibid.\  D {\bf 78}, 039903 (2008)]
  [hep-ph/9311340].
Y.~Yamada,
  Phys.\ Rev.\  D {\bf 50}, 3537 (1994)
  [hep-ph/9401241].
I.~Jack and D.R.T.~Jones,
  Phys.\ Lett.\  B {\bf 333}, 372 (1994) 
  [hep-ph/9405233].
I.~Jack et al,
  Phys.\ Rev.\  D {\bf 50}, 5481 (1994)
  [hep-ph/9407291].

\bibitem{betas:3}
I.~Jack, D.R.T.~Jones and C.G.~North,
  Phys.\ Lett.\  B {\bf 386}, 138 (1996)
  [hep-ph/9606323].
I.~Jack and D.R.T.~Jones,
  Phys.\ Lett.\  B {\bf 415}, 383 (1997)
  [hep-ph/9709364].

\bibitem{Kolda:1996ea}  
C.F.~Kolda and J.~March-Russell,
  Phys.\ Rev.\  D {\bf 55}, 4252 (1997)
  [hep-ph/9609480].
  
\bibitem{AguilarSaavedra:2005pw}
  J.A.~Aguilar-Saavedra {\it et al.},
  Eur.\ Phys.\ J.\  C {\bf 46}, 43 (2006)
  [hep-ph/0511344].

\bibitem{Kane:1998im}
  G.L.~Kane and S.F.~King, 
  Phys.\ Lett.\  B {\bf 451}, 113 (1999)
  [hep-ph/9810374].

\bibitem{LEPEWWG}
  ALEPH Collaboration and DELPHI Collaboration and L3 Collaboration and 
  OPAL Collaboration and SLD Collaboration and LEP Electroweak 
  Working Group and SLD Electroweak Group and SLD Heavy Flavour Group,
  Phys.\ Rept.\  {\bf 427}, 257 (2006)
  [hep-ex/0509008].

\bibitem{LEPEWWG2}  
J.~Alcaraz {\it et al.}  [ALEPH Collaboration and DELPHI Collaboration and
                  L3 Collaboration and OPAL Collaboration and LEP Electroweak Working Group],
  ``A Combination of preliminary electroweak measurements and constraints on
  the standard model,''
  [hep-ex/0612034],
ALEPH Collaboration and CDF Collaboration and D0 Collaboration 
and DELPHI Collaboration and L3 Collaboration and OPAL Collaboration and 
SLD Collaboration and LEP Electroweak Working Group and 
Tevatron Electroweak Working Group and SLD Electroweak Working Group and Heavy 
Flavour Group,
``Precision Electroweak Measurements and Constraints on the Standard Model,''
  [hep-ex/0811.4682].

\bibitem{RPP}
  C.~Amsler {\it et al.}  [Particle Data Group],
  ``Review of particle physics,''
  Phys.\ Lett.\  B {\bf 667}, 1 (2008).

\bibitem{Peskin:1991sw}
  M.E.~Peskin and T.~Takeuchi,
  Phys.\ Rev.\  D {\bf 46}, 381 (1992).

\bibitem{Wmass}
CDF Collaboration and D0 Collaboration,
  ``Combination of CDF and D0 results on the $W$ boson mass and width,''
  [hep-ex/0808.0147],
  
\bibitem{Tevatrontop}
   Tevatron Electroweak Working Group and CDF Collaboration 
   and D0 Collaboration,
  ``Combination of CDF and D0 Results on the Mass of the Top Quark,''
  [hep-ex/0903.2503].

\bibitem{Awramik:2006uz}
  M.~Awramik, M.~Czakon and A.~Freitas,
  JHEP {\bf 0611}, 048 (2006)
  [hep-ph/0608099].

\bibitem{Awramik:2003rn}
  M.~Awramik, M.~Czakon, A.~Freitas and G.~Weiglein,
  Phys.\ Rev.\  D {\bf 69}, 053006 (2004)
  [hep-ph/0311148].

\bibitem{Ferroglia:2002rg}
  A.~Ferroglia, G.~Ossola, M.~Passera and A.~Sirlin,
  Phys.\ Rev.\  D {\bf 65}, 113002 (2002)
  [hep-ph/0203224].

\bibitem{Lavoura:1992np}
  L.~Lavoura and J.P.~Silva,
  Phys.\ Rev.\  D {\bf 47}, 2046 (1993).

\bibitem{Maekawa:1995ha}
  N.~Maekawa,
  Phys.\ Rev.\  D {\bf 52}, 1684 (1995).

\bibitem{CDF9446} 
  J.~Conway et al, 
  CDF Public Note 9446,
  ``Search for Heavy Top $t' \rightarrow Wq$ In Lepton Plus Jet Events 
  in 2.8 fb$^{-1}$" 
  (unpublished).

\bibitem{CDF9759} 
  M.~Hickman et al, 
  CDF Public Note 9759,
  ``Search for fermion-pair decays 
  $Q \bar Q \rightarrow (tW^\pm) (\bar t W^\mp)$
  in same-charge dilepton events with 2.7 fb$^{-1}$", 
  (unpublished).

\bibitem{CDF8737} 
  T.~Aaltonen {\it et al.}  [CDF Collaboration],
  Phys.\ Rev.\  D {\bf 76}, 072006 (2007)
  [hep-ex/0706.3264].

\bibitem{CDFbp211} 
  C.~Wolfe et al, 
  ``Search for Heavy, Right Handed Quarks in Dilepton + Jets + Large $H_T$", 
  \begin{verbatim}http://www-cdf.fnal.gov/physics/exotic/r2a/20070810.heavy_obj_dilepX_wolfe/\end{verbatim}
  (unpublished).

\bibitem{Acosta:2002ju}
  D.E.~Acosta {\it et al.}  [CDF Collaboration],
  Phys.\ Rev.\ Lett.\  {\bf 90}, 131801 (2003)
  [hep-ex/0211064].

\bibitem{CDF7244} 
  A.L.~Scott, D.~Stuart, et al, 
  CDF Public Note 7244,
  ``Search for Long-Lived Parents of the $Z^0$ Boson", 
  (unpublished).

\bibitem{Aaltonen:2009kea}
  T.~Aaltonen {\it et al.}  [CDF Collaboration],
  Phys.\ Rev.\ Lett.\  {\bf 103}, 021802 (2009)
  [hep-ex/0902.1266].

\bibitem{Culbertson:2000am}
  R.L.~Culbertson {\it et al.}  [Tevatron Run II Study SUSY Working Group],
  ``Low scale and gauge mediated supersymmetry breaking at the Fermilab
  Tevatron Run II,''  [hep-ph/0008070].

\bibitem{Kraan:2004tz}  
  A.C.~Kraan,  
  Eur.\ Phys.\ J.\  C {\bf 37}, 91 (2004)
  [hep-ex/0404001].  

\bibitem{Fairbairn:2006gg}
  M.~Fairbairn et al, 
  Phys.\ Rept.\  {\bf 438}, 1 (2007)
  [hep-ph/0611040].

\bibitem{Aad:2009wy}
  G.~Aad {\it et al.}  [The ATLAS Collaboration],
  ``Expected Performance of the ATLAS Experiment - Detector, Trigger and
  Physics,''
  [hep-ex/0901.0512].


\bibitem{CTEQ5}
  H.L.~Lai {\it et al.}  [CTEQ Collaboration],
  Eur.\ Phys.\ J.\  C {\bf 12}, 375 (2000)
  [hep-ph/9903282].


\bibitem{Arik:1996qd}
  E.~Arik {\it et al.},
  Phys.\ Rev.\  D {\bf 58}, 117701 (1998).

\bibitem{AguilarSaavedra:2005pv}
  J.~A.~Aguilar-Saavedra,
  Phys.\ Lett.\  B {\bf 625}, 234 (2005)
  [Erratum-ibid.\  B {\bf 633}, 792 (2006)]
  [hep-ph/0506187].

\bibitem{Mehdiyev:2006tz}
  R.~Mehdiyev, S.~Sultansoy, G.~Unel and M.~Yilmaz,
  Eur.\ Phys.\ J.\  C {\bf 49}, 613 (2007)
  [hep-ex/0603005].

\bibitem{Skiba:2007fw}
  W.~Skiba and D.~Tucker-Smith,
  Phys.\ Rev.\  D {\bf 75}, 115010 (2007)
  [hep-ph/0701247].

\bibitem{AguilarSaavedra:2009es}
  J.A.~Aguilar-Saavedra,
  ``Identifying top partners at LHC,''
  [hep-ph/0907.3155].


\bibitem{Han:2003wu}
  T.~Han, H.E.~Logan, B.~McElrath and L.T.~Wang,
  Phys.\ Rev.\  D {\bf 67}, 095004 (2003)
  [hep-ph/0301040].

\bibitem{Perelstein:2003wd}
  M.~Perelstein, M.E.~Peskin and A.~Pierce,
  Phys.\ Rev.\  D {\bf 69}, 075002 (2004)
  [hep-ph/0310039].

\bibitem{Azuelos:2004dm}
  G.~Azuelos {\it et al.},
  Eur.\ Phys.\ J.\  C {\bf 39S2}, 13 (2005)
  [hep-ph/0402037].

\bibitem{Dennis:2007tv}
  C.~Dennis, M.~Karagoz Unel, G.~Servant and J.~Tseng,
  ``Multi-W events at LHC from a warped extra dimension with custodial
  symmetry,''
  [hep-ph/0701158].

\bibitem{Contino:2008hi}
  R.~Contino and G.~Servant,
  JHEP {\bf 0806}, 026 (2008)
  [hep-ph/0801.1679].

\bibitem{Atre:2008iu}
  A.~Atre, M.~Carena, T.~Han and J.~Santiago,
  Phys.\ Rev.\  D {\bf 79}, 054018 (2009)
  [hep-ph/0806.3966].

\bibitem{Mrazek:2009yu}
  J.~Mrazek and A.~Wulzer,
  ``A Strong Sector at the LHC: Top Partners in Same-Sign Dileptons,''
  [hep-ph/0909.3977].


\bibitem{Choudhury:2001hs}
  D.~Choudhury, T.M.P.~Tait and C.E.M.~Wagner,
  Phys.\ Rev.\  D {\bf 65}, 053002 (2002)  
  [hep-ph/0109097].

\bibitem{Bjorken:2002vt}
  J.D.~Bjorken, S.~Pakvasa and S.F.~Tuan,
  Phys.\ Rev.\  D {\bf 66}, 053008 (2002)
  [hep-ph/0206116].

\bibitem{Martin:2004id}
  S.P.~Martin, K.~Tobe and J.D.~Wells,
  Phys.\ Rev.\  D {\bf 71}, 073014 (2005)
  [hep-ph/0412424].

\bibitem{Graham:2009gy}
  P.W.~Graham, A.~Ismail, S.~Rajendran and P.~Saraswat,
  ``A Little Solution to the Little Hierarchy Problem: A Vector-like
  Generation,''
  [hep-ph/0910.3020].


\end{thebibliography}
\end{document}